# User Association and Resource Allocation in 5G (AURA-5G): A Joint Optimization Framework

Akshay Jain, Elena Lopez-Aguilera, and Ilker Demirkol

***Abstract*—5G wireless networks will be extremely dense, given the projected increase in the number of users and access points (APs), as well as heterogeneous, given the different types of APs and the applications being accessed by the users. In such challenging environments, efficient mobility management (MM), and specifically user association, will be critical to assist the 5G networks in provisioning the Quality of Service (QoS) of diverse applications 5G targets to serve. Whilst determining the most suitable AP for the users, multiple constraints such as available backhaul capacity, link latency, etc., will need to be accommodated for. Hence, to provide an optimal user association solution, in this paper we present a joint optimization framework, namely AURA-5G. Under this framework we formulate our user association strategy as a Mixed Integer Linear Program (MILP) that aims to maximize the total sum rate of the network whilst optimizing the bandwidth assignment and access point selection. We analyze multiple active application profiles simultaneously, i.e. enhanced Mobile Broadband (eMBB) and massive Machine Type Communication (mMTC), in the network and study the performance of AURA-5G. Additionally, we provision a novel study on the multiple dual connectivity modes, wherein the user can be connected to either one macro cell and a possible small cell, or with any two favorable candidate access points. Utilizing the AURA-5G framework, we perform a novel comparative study of all the considered scenarios on the basis of total network throughput, performance against baseline scenario and system fairness. We show that the AURA-5G optimal solutions improve the different network scenarios in terms of total network throughput as compared to the baseline scenario, which is a conventional user association solution. Further, we also present a fidelity analysis of the AURA-5G framework based on the backhaul utilization, latency compliance, convergence time distribution and solvability. And since, a given network cannot always guarantee to satisfy the future network loads and application constraints, we show that AURA-5G can be utilized by the operators/vendors to evaluate the myriad network re-dimensioning approaches for attaining a feasible and optimal solution. Henceforth, we then explore the possibility of network re-dimensioning and study its impact on system performance for scenarios where the performance of AURA-5G is severely impacted due to the extremely strict nature of the constraints imposed in the MILP.***

***Index Terms*—5G, User Association, Optimization, Mobility Management, HetNets, Bandwidth Allocation**

This work has been supported by the EU Horizon 2020 research and innovation programme under grant agreement No. 675806 (5GAuRA).

We would also like to thank Dr. H. Birkan Yilmaz, now a professor at Bogazici University in Turkey, for providing important insights and comments during the development of this work.

Akshay Jain and Elena Lopez-Aguilera are with the Department of Network Engineering, Universitat Politècnica de Catalunya BarcelonaTECH, Barcelona 08034, ES. Ilker Demirkol is with the Dept. of Mining, Industrial and ICT Systems Engineering, Universitat Politècnica de Catalunya BarcelonaTECH, Barcelona 08034, ES. E–mail: akshay.jain@upc.edu, elopez@entel.upc.edu, and ilker.demirkol@upc.edu. Akshay Jain is the corresponding author.

## I. Introduction

The upcoming 5G networks will be characterized by an extremely dense and heterogeneous amalgamation of access points (AP) with different Radio Access Technologies (RATs), users as well as application types [1]. Such a network environment will present significant challenges to the complex task of mobility management (MM) [2]. As part of the MM objective for 5G networks, seamless mobility with extremely low latency will need to be provisioned. However, to guarantee such latency and reliability characteristics, 5G MM mechanisms will be broadly required to ensure fast handover methods, efficient signaling during mobility events, optimal and fast user to AP associations, reliable and fast path re-configuration as well as service migration [3].

Specifically, to ensure that the quality of service (QoS) requested by an application on a given user is met, a user does not experience frequent handovers (FHO) as well as the network capabilities and capacities are respected, efficient user association techniques will be critical. And given the exponentially increasing number of users that 5G networks will cater to [4], finding the optimal user equipment (UE)-AP association will present a significant challenge for these user association techniques. This will be further exacerbated by the fact that 5G networks will be constrained by a multitude of requirements imposed for ensuring application QoS as well as limitations with regards to technological capabilties.

To elaborate, in [5], 3GPP established that to be able to provision services such as Virtual Reality (VR)/Augmented Reality (AR) among others, a minimum rate requirement of 100 Mbps would be required for enhanced Mobile Broadband (eMBB) services. Further, it was also determined that such services would necessitate anywhere between 5-10 ms latency (or round trip delay) [5]. Alongside these requirements, massive machine type communication (mMTC) services would need to be serviced anytime and anywhere, even though they do not communicate as regularly as the eMBB services. Further, the network would have to accomodate very high density of mMTC devices that will be prevalent in 5G networks, e.g. 24000 users per $km^2$ according to [6]. Moreover, the ultra-reliable low latency communication (URLLC) services will require latency within the range of 1-3 ms as well as extreme network reliability [5].

Coupled with these aforesaid requirements, 5G networks will also be challenged by the amount of available resources. Concretely, while mmWave small cells (SCs) will help resolve the lack of resources in current sub-6 GHz access network, the corresponding backhaul links will become increasingly strained. Further, the availability of APs with links that satisfy the latency requirements will be critical.

TABLE I
ABBREVIATIONS LIST

| | |
|---|---|
| RSSI | Received Signal Strength Indicator |
| AP | Access Point |
| MILP | Mixed Integer Linear Program |
| QoS | Quality of Service |
| HetNet | Heterogeneous Networks |
| MM | Mobility Management |
| FHO | Frequent Handover |
| RAT | Radio Access Technology |
| SDN | Software Defined Networking |
| UE | User Equipment |
| VR | Virtual Reality |
| AR | Augmented Reality |
| eMBB | enhanced Mobile Broadband |
| mMTC | massive Machine Type communication |
| URLLC | Ultra-reliable Low Latency communication |
| SC | Small Cells |
| MC | Macro Cells |
| SNR | Signal to Noise Ratio |
| SINR | Signal to Interference and Noise Ratio |
| MIH | Media Independent Handover |
| TOPSIS | Technique for Order of Preference by Similarity to Ideal Solution |
| MADM | Multi Attribute Decision Making |
| DC | Dual Connectivity |
| mmWave | Millimeter Wave |
| eNB | Evolved NodeB |
| HPP | Homogeneous Poisson Point process |
| LTE | Long Term Evolution |
| MHz | MegaHertz |
| 3GPP | Third Generation Partnership Project |
| RRC | Radio Resource Control |
| MRT | Minimum Rate |
| CPL | Constrained Path Latency |
| CB | Constrained Backhaul |
| gNB | next generation NodeB |
| dB | Decibels |
| LOS | Line of Sight |
| NLOS | Non Line of Sight |
| SF | Shadow Fading |

Henceforth, these aforesaid requirements and technological challenges will make the problem of user association alongside resource allocation in 5G heterogeneous networks (HetNets) important to explore and address. Consequently, in this work we aim to do the same. Notably, a broad spectrum of strategies/methods to accomplish the task of user association in 5G HetNets have been discussed in the literature, which we have also taken cognizance of in Section II. However, certain gaps still exist with regards to the aforesaid problem. And so, we elaborate them as follows:

1) Most of the research works discussed in the literature present a user association method that allows the user to connect to only one AP at most [25]–[28]. Certain works such as [7], [8], etc., discuss the problem of user association in a Dual Connectivity (DC) scenario. However, the analysis is limited to scenarios where SCs are tightly coupled to macro cells (MCs). Concretely, this means that the choice of an SC is governed by the choice of the MC. While this is inline with the current 3GPP DC standards in Release-15 [9], it is in general a very restrictive choice. Henceforth, we state that a gap exists here where none of the works in the state of the art, to the best of our knowledge, consider that an independent choice of MC and SC, or even two SCs or MCs to serve a user can be made. Note that a relatively tangential work to ours in [10] discusses such a possibility for slice level mobility in 5G networks. However, they do not present any concrete methodology or analysis for the purpose of user association.
2) None of the works present in the literature provide an application aware strategy [12]–[21], [25]–[28]. Concretely, they do not consider or analyze the impact of the prevalence of eMBB and mMTC services together. This is critical for user association in 5G HetNets, since the presence of different services will lead to different bottlenecks within the same system as shown in this study, thus making the process of finding an optimal association even more complicated.
3) Delving deeper into the analysis presented in the literature, it is evident that for the computation of the signal quality in terms of Signal to Interference and Noise ratio ($SINR$) an isotropic transmission and reception model is assumed. However, none of the user association works explore the impact of transmit and receive beamforming on the overall system performance, as well as for the complexity of computation of the $SINR$.
4) While certain research efforts discuss the computational complexity of the non-linear optimization framework for the user association problem, e.g. [22], [24], [33], [34], none of these works provision a detailed analysis with regards to the computation time, solvability, as well as other network parameters such as achieved latency and backhaul utilization.

Given the aforesaid deficiencies, to the best of our knowledge, we present the very first study in literature with regards to application aware user association in 5G HetNets in this paper. Concretely, we have explored the prevalence of multiple services and their impacts on the user association problem. Henceforth, for our study we have considered the scenarios where there are only eMBB services, and where both eMBB and mMTC services co-exist. We characterize the performance of our Joint Optimization framework, i.e., AURA-5G, in both these setups and present insights, which currently are not provided by any other research effort. Note that, we leave the study involving URLLC services as part of a follow-up work to this paper. Additionally, for our evaluation process we utilize realistic network scenarios and parameters. This consequently, helps establish the efficacy of the AURA-5G framework. However, a detailed discussion with regards to the scenarios and parameters is deferred until Sections IV and V.

Furthermore, we also consider the DC scenarios wherein we explore the futuristic trends of having *independent choices of MC and SC, i.e., the choice of SC is not geo-restricted to the coverage of the chosen MC* and *the possibility of selecting either two SCs or two MCs*. In addition to the DC scenarios, we also study the single association strategies (which most research works in the state of the art consider) and a baseline strategy (discussed in detail in Section IV). These latter scenarios provide us a basis for comparison for the *AURA-5G* framework, that we have developed.

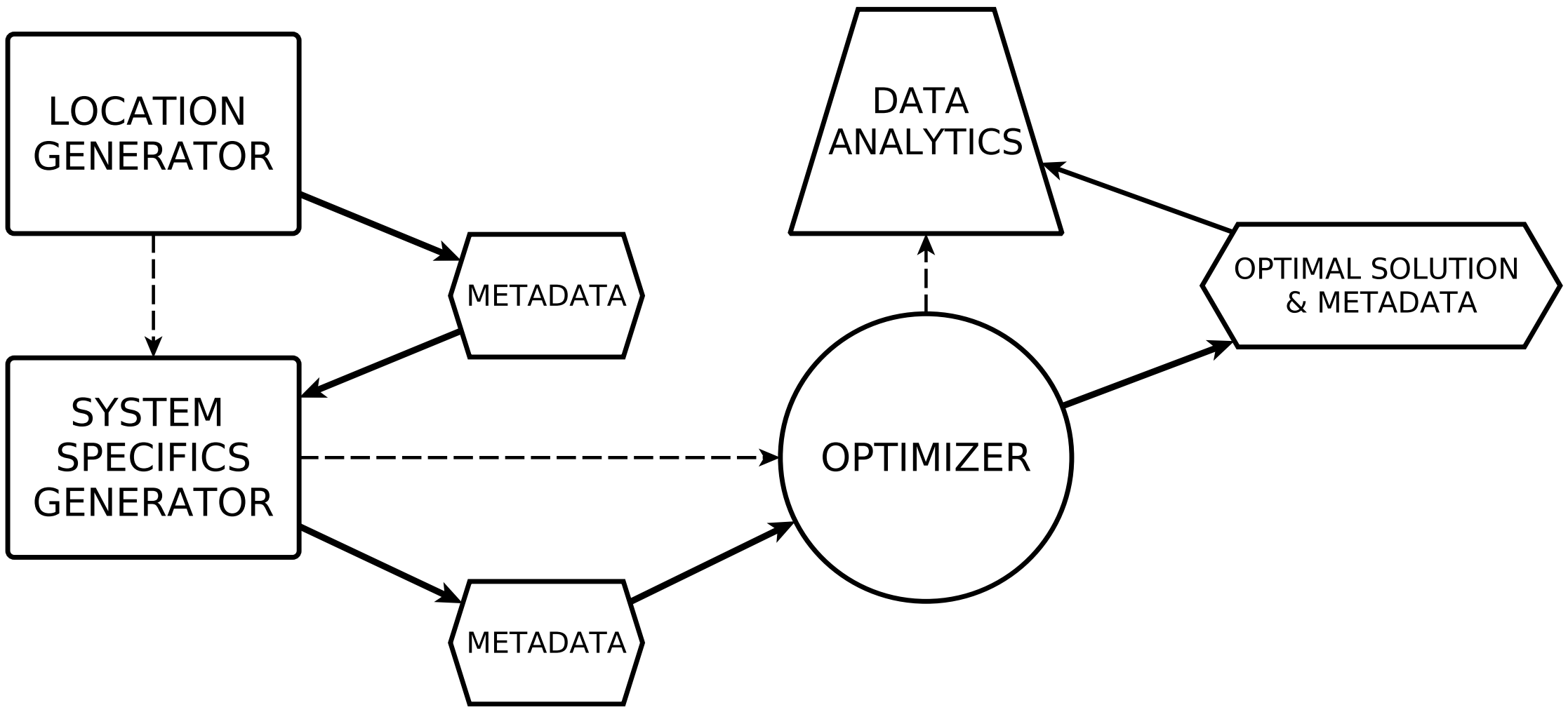

Fig. 1. AURA-5G Framework. The logical flow, i.e. flow of control, within the developed tool is depicted using dashed arrows, whilst solid arrows indicate the data flow in the program.

Moreover, in this paper we have also presented a detailed study into the performance of such joint optimization strategies when the environment is interference limited due to omni-directional antennas or when beamforming is utilized. The motivation behind exploring the aforesaid scenarios, is the fact that while most of the current day radio antennas do not utilize beamforming, networks such as 5G and beyond 5G will utilize massive MIMO setups that will support beamforming. Hence, it becomes imperative to study both these scenarios, by virtue of the algorithm being deploy-able irrespective of the infrastructural setup.

Next, as part of the contributions of this paper, we also emphasize on the AURA-5G software framework which we have developed and published[1] to obtain the aforementioned deep insights into the user association problem. As can be seen from the framework diagram in Fig. 1, AURA-5G is basically composed of four building blocks. The very first block, i.e. *location generator* block, takes care of the generation of location specific information for the users and APs to be utilized during the analysis. Concretely, it generates the location coordinates for the users and APs within the topology. Next, the *system specifics generator* block creates the backhaul link based details for the system, computes the SINR matrix for the system (according to whether we are in an interference limited scenario analysis or a beamforming based scenario analysis) and saves the necessary metadata that would be required by the subsequent *optimizer* block. For the *optimizer* block we utilize the **Gurobi** toolbox [11] and solve the Mixed Integer Linear Programming formulation (MILP), discussed in Section III, to compute the optimal solution. Note that, in the *optimizer* block we also specify the particular scenario (described in Section IV) that has to be evaluated on the system specifics generated by the preceding framework boxes. The *optimizer* block then saves the optimal solution and supporting metadata, which are consequently utilized by the *data representation/analytics* box for gaining insights into the obtained solution for the given scenario.

The rest of this paper is organized as follows: In Section II we present the state-of-the-art strategies/techniques employed with regards to the user association mechanisms for 5G networks. In Section III, we discuss in detail the optimization framework, i.e., the MILP framework, for computing the optimal user association given the objective and corresponding constraints. We also discuss the aspects of *Linearization* and challenges involved in developing the *AURA-5G* framework in this section. In Section IV, we present the various scenarios that have been studied in this paper. We then elaborate on the evaluation framework that has been adopted in this paper, in Section V. The results from the evaluation and the corresponding insights have then been articulated in Section VI and VII. Specifically, in Section VII we discuss how AURA-5G can provide insights to the operators and help them re-dimension their networks for improved system performance. Finally, we conclude this paper in Section VIII.

## II. State of the Art

User association has been an area of major interest ever since wireless networks came into existence. Over the course of those many years, multiple intriguing algorithms and methodologies have been proposed by academia and industry, aiming to resolve the incremental challenges that user association presents as wireless networks continue to evolve. Concretely, while 2G and Wi-Fi networks utilized an Required Signal Strength Indicator (RSSI)/ Signal-to-Noise Ratio (SNR) based criterion for AP selection, the increasing complexity of the networks has driven the growth of algorithms that utilize TOPSIS [12], Fuzzy logic [13]–[15], Genetic Algorithms

[1]The complete framework has been developed using Python. It can be found at: *the github repository link will be provided after the review process.*

[13], Multi Attribute Decision Making (MADM) [16]–[21] and Optimization theory [22]–[24] among others as tools to accomplish an optimal user association. Moreover, they not only aim to provision an optimal association, based on maximizing sum rate of users for example, but they also try to optimize other system parameters such as interference, bandwidth allocation, interference reduction, etc.

Specifically, the authors in [25] consider an SDN enabled LTE network and utilize the global knowledge of the SDN controller so as to be able to distribute information about the backhaul load of neighboring APs as well as the bottleneck backhaul link bandwidth for an AP of interest. Such a mechanism consequently allows for a more backhaul and AP load aware load-balancing mechanism. Next, in [26] the authors propose a distributed user association policy wherein they explore multiple dimensions to the user association problem. These dimensions explore the association policy while trying to optimize rate, throughput, minimize file transfer delays and load balancing. They do so by varying a factor, which characterizes each of these dimensions. Further, the distributed nature encompasses the fact that part of the optimal solution search is performed on the UE and a part of it is performed on the AP. Further, authors in [27] utilize the k-nearest neighbors principle and the azimuth angle to determine the next AP to associate to in a mobile environment. They analyze the pattern associated with a given k-value and then utilize pattern recognition methods to determine the optimal APs to associate to given an ultra-dense topology and severe interference scenarios. Moreover, in [28], the authors utilize Media Independent Handover (MIH) and its corresponding services so as to be able to execute a MADM algorithm. They compare it with other simplistic counterparts, such as Additive Weighting strategies, to project the benefits of MIH based MADM. While these discussed strategies consider multiple constraints and propose effective solutions, they limit the number of APs that a UE can connect/choose to just one.

As a consequence, authors in [7] proposed a Dual Connectivity (DC) based User Association solution, wherein they formulate an optimization problem based on maximizing the sum throughput. Since, the problem is ridden with non-linearities and is NP-hard, they propose a tractable solution to achieve the optimal solution. Also, in [8], the authors study the problem of user association with the objective of maximizing the weighted sum rate with user rate (minimum and maximum) constraints as well as with another objective of maximizing the proportional fairness of the system. Note that, in [8], for DC, the authors consider that a UE can be associated with an MC and an SC that is within the coverage of this selected MC. Next, the authors of [29] consider the problem of user association for the uplink with power allocation optimization being another objective. They consider user minimum rate requirements as well as backhaul limitations at the pico BSs as the constraints and aim to find an association that reduces the overall cost for the network while demands of all the users are satisfied.

Moreover in [30] the authors consider the downlink aspect of dual connectivity as well as the mmWave aided HetNets. Correspondingly, the authors develop a two stage iterative algorithm for user association where their objective is to maximize the total network throughput subject to the fact that the overall access-fairness amongst the users has to be improved. However, for the analysis the authors consider a very small representative scenario, which in essence does not represent the real world densities of SCs, MCs and users. In addition, in [31] the authors propose an opportunistic method for cell selection in a DC scenario, wherein the load characteristics of the network are taken into account. However, according to the network model considered for simulations, their analysis is only limited to the sub-6 GHz scenarios. Another significant study in the direction of user association for 5G networks is in [32]. The authors of this work analyze the impact of wireless backhaul on user association. The optimal association is determined such that it maximizes the overall network throughput whilst not exceeding the backhaul capacity limits.

Lastly, multiple studies such as [22], [24], [33], [34] propose a joint optimization approach towards user association. In these studies the objective function is maximized/minimized, depending on the utility, and an optimal association strategy that provisions the same is determined. Additionally, they also incorporate, within their optimization approach, a search for the association that will lead to an optimal system interference/energy consumption regime/spectrum allocation system, etc. Such mechanisms are termed as *joint optimization* approaches, and through the use of binary decision variables they introduce another dimension of non-linearity to an already non-linear optimization problem. However, a relaxation of these decision variables or a decomposition into simpler sub problems is in general possible. And so, in [22], [24], [33], [34] such techniques have been utilized and a discussion on the optimal solution obtained has been presented.

## III. The Optimization Framework: Mathematical Formulation and Solver Implementation

The discussion with regards to the state of the art studies, in Section II, and the contributions of this work, in Section I, provides us with the platform to now present the mathematical formulation of our joint optimization strategy. Subsequently, we also discuss the challenges that need to be addressed for an efficient implementation of the AURA-5G framework.

And so, we consider a wireless heterogeneous network scenario, wherein 4G-LTE evolved NodeBs (eNBs) provide the role of MCs and the 5G next generation NodeBs (gNBs) function as the SCs. We denote the set of MCs as $\mathcal{M}$ and the set of SCs as $\mathcal{N}$. We consider that the SCs are connected to the MCs via a backhaul link. This backhaul link can be either wired (fiber based) or wireless (mmWave based). Further, the MCs have backhaul links to the core network. However, these backhaul links are only wired (fiber based) in nature. We denote these aforementioned backhaul links as, $B_m$ and $B_n$ (where $m \in \mathcal{M}$ and $n \in \mathcal{N}$) for the $m^{th}$ MC and $n^{th}$ SC, respectively. Further, the capacity of each of the backhaul links in the considered scenario is denoted as $C_m$ and $C_n$ for the $m^{th}$ MC and $n^{th}$ SC, respectively. Given, the heterogeneous characteristic of the backhaul technologies, their correspond-ing capacities will also be different. We elaborate more on this in Section V, wherein we describe the system model in detail.

Next, we specify the delay imposed by the backhaul links as $D_t$, where $t = 1 \dots d$. Here, $d$ is the number of links in the considered scenario. Further, $d > (|\mathcal{M}| + |\mathcal{N}|)$, where $|\cdot|$ denotes the cardinality of the set, because each MC is defined with a backhaul network that has one or more hops to the core network. However, for the purpose of backhaul utilization analysis, the multiple hops from any given MC can be considered together as a single link. This is so because, all the wired hops from MCs are defined to have the same capacity. In addition, for the SCs there is an additional hop (link), i.e., the connecting link to the MC, which may be wired or wireless depending on the operator deployment strategy. Similar to the MC backhaul hops, the wired SC to MC links also have the same capacity but less than that of the MC to CN links. We provide numerical details with regards to this in our system model description in Section V.

The users within the HetNet are deployed using a homogeneous poisson point process (HPP), and are denoted as $U_f$, where $f = 1 \dots u$, and $u$ being the total number of users within the scenario. For the ease of understanding, we introduce Table II, which contains a list of all the variables, constants and notations that have been utilized for this work. Given these preliminaries, we state the objective of our user association strategy. Concretely, our objective is to maximize the overall system throughput in the downlink, whilst adhering to the various constraints that the 5G HetNets will impose. It is imperative to state here that, the *AURA-5G* framework is also applicable to the uplink. Specifically, in this work we consider the backhaul capacity, minimum required rate and path latency (one-way downlink delay) as the constraints for our joint optimization problem. Thus, we frame our user association strategy as a MILP problem as follows:

$$\max_{x_{ij}, g_{ijk}} \sum_i \sum_j \sum_k x_{ij} g_{ijk} w_k log_2(1 + \Psi_{ij}) \tag{1}$$

$$\text{s.t.} \sum_i \sum_k x_{ij} g_{ijk} w_k \leq W_j \qquad \forall j \tag{2}$$

$$\sum_j x_{ij} \leq 2 \qquad \forall i \tag{3}$$

$$\sum_k g_{ijk} \leq 1 \qquad \forall i, j \tag{4}$$

$$\sum_j \sum_k x_{ij} r_{ijk} \geq R_i \qquad \forall i \tag{5}$$

$$\sum_i \sum_k x_{ij} r_{ijk} \leq C_j \qquad \forall j \in \mathcal{N} \tag{6}$$

$$\sum_i \sum_k x_{ij} r_{ijk} + \sum_t \xi_{tj} \leq C_j \qquad \forall j \in \mathcal{M} \tag{7}$$

$$p_j x_{ij} \leq l_i \qquad \forall i \tag{8}$$

$$x_{ij}, g_{ijk} \in \{0, 1\} \qquad \forall i, j, k \tag{9}$$

where, $x_{ij}$ indicates the association of user $i$ to AP $j$. A value of 1 signifies an active association and 0 defines that there is no association. Further, $g_{ijk}$ defines the bandwidth assignment to a user $i$ at AP $j$, which has $k$ different available bandwidth options. A value of 1 for any given $i, j$ and $k$ combination defines the fact that the bandwidth option $k$ at AP $j$ for user $i$ has been selected, while a value of 0 for the same defines vice versa. In eq (1), which defines our total sum rate maximization objective, $w_k$ is a constant value that indicates the actual bandwidth resource in *MHz* for the option $k$. Next, the constraint defined in eq (2) specifies that the total bandwidth resources allocated to all the users associated with AP $j$ cannot exceed the total available bandwidth $W_j$ at AP $j$. In eq (3), we define the dual connectivity constraint wherein a user can select a maximum of two APs. As we will see later in this section, we modify this constraint to study both the single and dual connectivity scenarios. Subsequently, the constraint in eq (4) guarantees that no more than one bandwidth option can be chosen by a user $i$ at an AP $j$.

TABLE II
DEFINITIONS LIST FOR NOTATIONS, VARIABLES AND CONSTANTS

| | |
|---|---|
| $i, j, k, f, t$ | Index variables |
| $\mathcal{M}$ | Set of Macro-cells |
| $\lvert\mathcal{M}\rvert$ | Total Macro-cells in the system |
| $\mathcal{N}$ | Set of Small-cells |
| $\lvert\mathcal{N}\rvert$ | Total Small-cells in the system |
| $B_m$ | Backhaul link for Macro-cell $m$ |
| $B_n$ | Backhaul link for Small-cell $n$ |
| $C_m$ | Capacity of Backhaul link for Macro-cell $m$ |
| $C_n$ | Capacity of Backhaul link for Small-cell $n$ |
| $D_t$ | Delay imposed by link $t$ where $t \in (1, \dots, d)$ |
| $d$ | Total number of links in the scenario |
| $U_f$ | User $f$ where $f \in (1, \dots, u)$ |
| $u$ | Total number of users in the scenario |
| $x_{ij}$ | Binary variable indicating association of user $i$ with AP $j$ |
| $g_{ijk}$ | Binary variable indicating selection of bandwidth option $k$ at AP $j$ for user $i$ |
| $w_k$ | Bandwidth option $k$ at an AP |
| $\Psi_{ij}$ | SINR registered by user $i$ for AP $j$ |
| $W_j$ | Total available bandwidth at AP $j$ |
| $r_{ijk}$ | Composite variable defining the rate offered by AP $j$ to user $i$ with bandwidth option $k$ |
| $\xi_{tj}$ | Backhaul resource consumption by Small-cell $t$ associated to Macro-cell $j$ |
| $p_j$ | collection of links defining a path to the core network from AP $j$ |
| $R_i$ | Minimum Rate constraint for user $j$ |
| $C_j$ | Collection of backhaul capacities for the Macro and Small cells |
| $l_i$ | Maximum bearable downlink latency [delay] for a user $i$ |
| $\Gamma_{ijk}$ | Binary variable, introduced for linearization (Section II.A), indicating association of user $i$ with AP $j$ where bandwidth option $k$ has been assigned. |
| $V_{ijk}$ | A constant, representing $w_k log_2(1 + \Psi_{ij})$ |

Next in eq (5) we specify the minimum rate constraint wherein, the sum rate for a user $i$ from the APs and the corresponding bandwidth options it selects at those APs has to be greater than minimum rate requirement $R_i$, for every user in the scenario. Here, for the sake of brevity, we also define a composite variable $r_{ijk}$, computed as $g_{ijk} w_k log_2(1 + \Psi_{ij})$, which corresponds to the rate AP $j$ offers to user $i$ at bandwidth option $k$ in case there is an active association between them, i.e. $x_{ij} = 1$. Note that, $R_i$ will depend on the type of application a user accesses, i.e., eMBB, mMTC and URLLC applications will have different minimum rate requirements. In this work, we only consider the eMBB and mMTC applications, and utilize the 3GPP 5G specifications [35] and other literature works such as [36] for their minimum rate requirements.

In eqs. (6) and (7), we introduce the backhaul capacity constraint, wherein the total allocated link rate to all the users associated to a given AP $j$ cannot exceed the available link bandwidth $C_j$. It is important to state here that, the backhaul capacity constraints for the SCs (eq. (6)) and MCs (eq. (7)) are characteristically different. This is so because, in the considered scenario, an SC always has a backhaul link, either wired or wireless, to an MC. Henceforth for the MC, it is mandatory that we consider the contribution of the SCs as well in order to ensure that the backhaul capacity constraint is not violated. Consequently in our MILP formulation, in eq. (7) we introduce the term $\xi_{tj}$, which specifies the rate consumption by SC $t$ at MC $j$. It is expressed by the left hand side of eq. (6), and is equivalent to the capacity of the backhaul link utilized by all the users associated with SC $t$.

Next, in eq (8) we introduce the path latency constraint for each user $i$. We define $l_i$ as the downlink latency (delay) that an application on user $i$ can permit, based on its QoS requirements. We also introduce an additional system variable, $p_j$, which specifies the cumulative latency offered by the links that connect AP $j$ to the core network. Here, by a link we specifically mean a wired/wireless hop towards the core network from the AP $j$. Hence, and as we will observe in further detail in Section V, different APs will offer different latency (delay) as is the case in real networks. Consequently, the constraint in eq (8) will assist the algorithm in selecting an association for all applications in the system that assures that their latency requirements are satisfied.

Lastly, in eq (9), we state that $x_{ij}$ and $g_{ijk}$ are both binary variables. Henceforth, this preceding discussion concretizes our joint optimization objective, wherein we not only aim to find the right user-AP association, i.e. $x_{ij}$, but also the possible bandwidth allocation through $g_{ijk}$. However, it is important to note here that the multiplication of $x_{ij}$ and $g_{ijk}$ in our objective function, i.e., eq (1) and subsequently in the constraints in eqs. (2), (5), (6) and (7), introduces a non-linearity. To resolve this we perform a linearization operation, which is detailed in the following text.

### A. Linearization

To avoid the non-linearity introduced by the multiplicative term involving two binary decision variables, i.e. $x_{ij}$ and $g_{ijk}$, in our optimization problem formulated in eqs. (1)-(9), we perform a simplistic linearization operation that will enable us to apply our proposed user association strategy as a MILP.

Firstly, we introduce the linearization term in eq. (10) wherein we replace the multiplicative quantity by a single binary variable.

$$\Gamma_{ijk} = x_{ij} g_{ijk} \forall i, j, k \tag{10}$$

where $\Gamma_{ijk} \in \{0, 1\}$. A value of 1 denotes the active association of a user $i$ with AP $j$ with bandwidth option $k$ allocated at this AP, while 0 indicates vice versa. Subsequently, we replace $x_{ij} g_{ijk}$ in eqs. (1)-(9) with $\Gamma_{ijk}$. In order to make this linearization functional, we will also need additional constraints that establish a relationship between $\Gamma_{ijk}$, $x_{ij}$ and $g_{ijk}$. These additional constraints are as follows:

$$\Gamma_{ijk} \leq g_{ijk} \qquad \forall i, j, k \tag{11}$$

$$\Gamma_{ijk} \leq x_{ij} \qquad \forall i, j, k \tag{12}$$

$$\Gamma_{ijk} \geq g_{ijk} + x_{ij} - 1 \qquad \forall i, j, k \tag{13}$$

The aforesaid equations establish the necessary relationship required between the linearizing variable, and the variables comprising the term that is being linearized. Henceforth, we now present our modified MILP formulation, as a result of the aforesaid linearization, in eqs. (14)-(25) as follows:

$$\max_{\Gamma_{ijk}} \sum_i \sum_j \sum_k \Gamma_{ijk} w_k log_2(1 + \Psi_{ij}) \tag{14}$$

$$\text{s.t.} \sum_i \sum_k \Gamma_{ijk} w_k \leq W_j \qquad \forall j \tag{15}$$

$$\sum_j x_{ij} \leq 2 \qquad \forall i \tag{16}$$

$$\sum_k g_{ijk} \leq 1 \qquad \forall i, j \tag{17}$$

$$\sum_j \sum_k \Gamma_{ijk} V_{ijk} \geq R_i \qquad \forall i \tag{18}$$

$$\sum_i \sum_k \Gamma_{ijk} V_{ijk} \leq C_j \qquad \forall j \in N \tag{19}$$

$$\sum_i \sum_k \Gamma_{ijk} V_{ijk} + \sum_t \xi'_{tj} \leq C_j \qquad \forall j \in M \tag{20}$$

$$p_j x_{ij} \leq l_i \qquad \forall i \tag{21}$$

$$\Gamma_{ijk} \leq g_{ijk} \qquad \forall i, j, k \tag{22}$$

$$\Gamma_{ijk} \leq x_{ij} \qquad \forall i, j, k \tag{23}$$

$$\Gamma_{ijk} \geq g_{ijk} + x_{ij} - 1 \qquad \forall i, j, k \tag{24}$$

$$x_{ij}, g_{ijk}, \Gamma_{ijk} \in \{0, 1\} \qquad \forall i, j, k \tag{25}$$

where $V_{ijk} = w_k log_2(1 + \Psi_{ij})$, and is a constant since the values for both $w_k$ and $log_2(1 + \Psi_{ij})$ are defined/computed beforehand. Further, $\xi'_{tj}$ represents the modified variable for expressing the contribution of the SCs towards the backhaul utilization to an MC. We introduce this modified variable to account for the linearization operation, since the computation of $\xi_{tj}$ in eq. (7) involves the multiplicative term $x_{ij} g_{ijk}$, as observed from our discussions regarding eq. (6) and eq. (7).

### B. Solver Implementation Challenges

While we may have linearized the system of equations for our optimization framework in Section III.A, unfortunately non-linearity still exists given that the variables $x_{ij}$, $g_{ijk}$, and consequently $\Gamma_{ijk}$ are binary in nature. However, we establish that a simplistic approach, wherein we – a) relax the binary nature of the aforesaid variables to bounded constraints, and

b) threshold the solution values of these integral variables; can help us avoid such non-linearities. Moreover, solvers such as Gurobi allow the users to program optimization problems, such as ours, and solve them using LP relaxation, branch-and-bound and other advanced mixed integer programming techniques [11]. And so, we utilize this powerful characteristic of Gurobi to solve our optimization framework, and consequently, determine the optimal user association strategy.

In addition, and as mentioned earlier in Section I, we have developed an implementation framework named *AURA-5G* that also undertakes the tedious task of computing the link SINR matrix. The complexity of this process is highlighted by the fact that in scenarios where there is transmit and receive beamforming, the computation of the link SINR matrix will require the system to know beforehand the beam directions of all the APs. Concretely, for an UE of interest, all the other UE-AP associations must be known so as to be able to compute the interference from the APs other than the AP of interest. Note that an AP will only create interference at the UE when the transmit beam of the AP and receive beam of the UE are aligned with each other, whole or in part.

And so, in the following text, through a hypothetical scenario, we show the complexity of computing the aforementioned link SINR matrix. Let us consider the scenario where there is a UE and $Z$ possible APs to which this UE can attach to in any receive beam direction. Let us also define a binary variable $\delta \in \{0, 1\}$ which indicates whether an AP, through its transmit beam, creates interference for the AP of interest at the UE under observation, i.e., $\delta = 1$, or it does not, i.e., $\delta = 0$. Thus, the total number of combinations of interfering beams (APs) that needs to be explored to determine the value of SINR, for an AP of interest, is given as $2^{(Z-1)}$.

Next, and for the sake of simplicity, we quantize the number of possible receiver beam directions as $\Phi$. As a consequence, number of computations required to determine the vector of SINR values for a given UE can be expressed as:

$$[(2^{Z-1} + 2)Z]\Phi \tag{26}$$

where the additive term of 2 indicates an addition operation for computing *interference plus noise* term and a division operation for ultimately computing the *SINR*. Thus from eq. (26), it can be seen that the number of computations, and hence the number of combinations that need to be explored, grows exponentially with the number of candidate APs $Z$, in scenarios where there is receive and transmit beamforming. This validates our earlier claim regarding the tediousness of computing the SINR matrix.

Certain works in literature, such as [37]–[39], provide insights as to how a statistical estimate for the SINR can be obtained in a beamforming scenario given a user and multiple candidate APs. However, these works do not account for the possibility of multiple users in the vicinity of the user of interest. Thus, we utilize a simplified process to determine the SINR at any given UE in a beamformed regime, wherein we only consider the receiver beamforming at the UE and allow the APs to transmit in an omnidirectional manner. This reduces the number of computations significantly, because now the remaining $Z - 1$ APs will create an interference for the AP of interest. Hence, for $\Phi$ quantized receive beam directions, the number of operations required are:

$$[(Z + 1)Z]\Phi \tag{27}$$

Comparing eqs. (26) and (27), we establish that for a given UE our method utilizes significantly less number of operations to compute the SINR, and hence overcome the earlier said challenge of computing the SINR matrix in a beamformed environment. However, it must be stated that the computed SINR estimate will be a lower bound on the actual SINR value. This is so because, we do not consider the transmit beamforming on the APs. Consequently, we increase the number of interferers in our computation compared to those where both transmit and receive beamforming is utilized. Notably though, the efficacy of our analysis for the optimization framework is further enhanced as it utilizes the lower bound, i.e. worst case scenario, for the SINR according to the preceding discussions.

Following this optimization framework, in the next section we introduce the various scenarios that have been explored in this work. We also introduce the necessary modifications to the constraints in the MILP framework to study the corresponding scenarios.

## IV. Scenarios Evaluated

The optimization framework developed in Section III presented the objective and the multiple real-network constraints that will be utilized when deciding the most optimal AP selection for a given set of users and their corresponding locations. Based on this framework, in this section we introduce the myriad scenarios that have been explored in this work. We also present the necessary modifications, if required, in our optimization framework to study the corresponding scenarios. Table III illustrates all the scenarios that have been discussed.

### A. Deployment Strategies

For the analyzed scenarios, we generate a set of topologies by deploying the MCs, SCs and users based on the parameters defined in Table V. Of specific interest amongst these is the deployment of SCs within the scenario map. While MCs are at fixed locations, governed by the scenario map size and the MC inter-site distance, the SCs are distributed based on an HPP around each MC. The density is defined in [6] based on the Metis-II project guidelines.

Given these deployment characteristics, we undertake a study on scenarios where these SCs are deployed in a circle of radius $0.5 \times ISDMC$ (see Table V), termed as *Circular Deployment* from here on, and scenarios where they are deployed in a *Square Deployment*. In the latter scenario, the SCs are deployed in a square whose center is at the MC location and the length of each edge is equal to the MC inter-site distance. Note that, while actual deployments will vary depending on operator requirements, *Circular Deployment* provides a realistic and simple deployment strategy for the SCs. Further, and as we will see in Section V (Fig. 5), a *Circular Deployment* strategy will lead to areas around MC edges where there will be no coverage via SCs. Hence, to circumvent this issue, we also explore the *Square Deployment* scenario.

TABLE III
ANALYZED SCENARIOS

| Composite Scenario Name | Circular Deploy-ment | Square Deploy-ment | AnyDC | MCSC | Interference Limited | Beamformed | eMBB | eMBB + mMTC |
|---|---|---|---|---|---|---|---|---|
| CABE | ✓ | – | ✓ | – | – | ✓ | ✓ | – |
| CMBE | ✓ | – | – | ✓ | – | ✓ | ✓ | – |
| CAIE | ✓ | – | ✓ | – | ✓ | – | ✓ | – |
| CMIE | ✓ | – | – | ✓ | ✓ | – | ✓ | – |
| SABE | – | ✓ | ✓ | – | – | ✓ | ✓ | – |
| SMBE | – | ✓ | – | ✓ | – | ✓ | ✓ | – |
| SAIE | – | ✓ | ✓ | – | ✓ | – | ✓ | – |
| SMIE | – | ✓ | – | ✓ | ✓ | – | ✓ | – |
| CABEm | ✓ | – | ✓ | – | – | ✓ | – | ✓ |
| CMBEm | ✓ | – | – | ✓ | – | ✓ | – | ✓ |
| CAIEm | ✓ | – | ✓ | – | ✓ | – | – | ✓ |
| CMIEm | ✓ | – | – | ✓ | ✓ | – | – | ✓ |
| SABEm | – | ✓ | ✓ | – | – | ✓ | – | ✓ |
| SMBEm | – | ✓ | – | ✓ | – | ✓ | – | ✓ |
| SAIEm | – | ✓ | ✓ | – | ✓ | – | – | ✓ |
| SMIEm | – | ✓ | – | ✓ | ✓ | – | – | ✓ |

The goal of including these deployment strategies into our study is to give the operators an insight as to how different deployment characteristics can impact the system performance whilst defining a UE to AP association map. This will allow them to understand the benefits and drawbacks of each of these deployment strategies with regards to the joint user association and resource allocation problem. Moreover, it also provides a framework for the operators to introduce their own custom deployments, and analyze the behavior of the user association strategy.

### B. Service Classes

5G, as has been discussed in Section I, will cater to the multiple service classes, i.e., eMBB, mMTC and URLLC [35]. As a consequence, in this work, we study the performance of *AURA-5G* in the presence of only *eMBB service* requests, as well as for the case where *eMBB and mMTC service requests* are generated simultaneously within the topology under study, to show the impact of provisioning of diverse services. Note that, while the eMBB services will request significantly higher throughput (we study the impact of the minimum rate requirements of eMBB services in 5G, as detailed later), mMTC services due to their relatively higher density but low individual data rates will create bottlenecks in the access and the backhaul networks for the eMBB service requests. However, in this work, for mMTC devices we consider the guard band mode of operation. Hence, the mMTC devices do not consume resources in the access network and just contribute towards the consumption of BH resources. In addition to these services, URLLC services present the unique challenge of ensuring not only low latency but also significantly high levels of reliability from the network. Nevertheless, we postpone our discussion with regards to URLLC services to a sequel of this work.

### C. Directivity Regimes

In Section III. B, we briefly commented upon the fact that in this article two scenarios, depending on whether beamforming is used, are explored. Concretely, we consider one scenario wherein all transmit and receive antennas have a $360^{\circ}$ transmit and receive pattern, respectively. As we will detail next, this will be an *Interference Limited* regime. And so, we also study the behavior of AURA-5G in *beamformed regimes*, wherein, and for the sake of simplicity (See Section III.B for details), we only consider receiver beamforming at the UE. We elaborate further on the aforesaid *directivity regimes* as follows:

- **Beamformed Regime:** In the scenarios where there is beamforming we consider only receive beamforming at the UE, and utilize it for the purpose of calculating the values of the SINR, i.e. $\Psi_{ij}$, for all user $i$ and small cell AP $j$ pairs[2]. For the macro-cells we employ sectorization, details of which are specified in our system model in Section V, and allow the UEs to have isotropic reception for the frequencies at which the MCs operate.
- **Interference Limited Regime:** Without beamforming, the environment is flooded with multiple interfering signals which can deeply degrade the performance of the system. Concretely, for this scenario neither the SC and MC APs employ any sort of beamforming/sectorization nor do the UEs employ any receiver beamforming. Thus, we also evaluate our MILP framework for user association in such a challenging network environment.

### D. Dual Connectivity Modes

With the current 5G standardization, i.e., Release-15, EN-DC and MR-DC have been formalized as one of the critical

[2]Note that, transmit beamforming at the Small Cell APs can also be considered here. However, this complicates the computation of the SINR as discussed in Section III.B

features for provisioning higher data rates for users. However, current standards constrain the choice of small cells severely, by limiting them to the secondary cell group (SCG) specified by the MC [40]. Hence, in this paper we outline two DC strategies that build on the current standards and explore the possibility of either – a) MCSC: having an MC and a possible SC (not geo-restricted by the choice of MC), or b) AnyDC: choosing any two possible APs. We elaborate further on the aforementioned DC connectivity modes, as follows:

- **MCSC:** In this mode of DC, a UE is required to attach to an MC and select at most one SC. The choice of MC does not geo-restrict the choice of SC, as we attempt to go beyond the existing standards on SCG. Additionally, it must be reiterated here that, a connection does not guarantee bandwidth allocation as it is subject to the available physical resources at that moment. Such a scenario can be seen as equivalent to the one where a UE is in a *RRC connected inactive state* at that AP [41].

  And so for this DC mode, the dual connectivity constraint [eq. (16)] in our optimization framework in Section III is modified to:

$$\sum_{j \in \mathcal{M}} x_{ij} == 1 \qquad \forall i \qquad (28)$$

$$\sum_{j \in \mathcal{N}} x_{ij} \leq 1 \qquad \forall i \qquad (29)$$

  recall that $\mathcal{M}$ and $\mathcal{N}$, as shown in Table II, represent the set of MCs and SCs, respectively. Concretely, eq. (28) ensures that each UE selects an MC, and eq. (29) enables them to select at most one SC.

- **AnyDC:** This mode for DC will permit the users to select any two APs irrespective of the fact that whether they are an MC or an SC. Consequently, we also incorporate this scenario in our study, which, as we will see in Section VI, rightly points towards the potential for improved performance but at a higher computational cost as compared to *MCSC* scenarios. By higher computational costs here we refer to the convergence time to the optimal solution, which we study later in Section VI.E.

  Hence, to study the *AnyDC* scenario, the dual connectivity constraint in our optimization framework in Section III is modified as follows:

$$\sum_{j \in \mathcal{M} \cup \mathcal{N}} x_{ij} == 2 \qquad \forall i \qquad (30)$$

### *E. Baseline and Single Association*

In our study, we also analyze the *single association* and *baseline association* scenarios as benchmark solutions. Consequently, we conduct a performance comparison of our user association and resource allocation strategy, i.e., *AURA-5G*, with these scenarios based on the obtained performance metrics from the DC modes discussed in Section IV.D.

We further elaborate on these two association strategies in the text that follows.

- **Single Association:** As the name suggests, in this scenario we enable the UE to connect to at most one AP. This is in essence what current day wireless networks offer. And so, we modify the dual connectivity constraint in eq. (16) to:

$$\sum_{j \in \mathcal{M} \cup \mathcal{N}} x_{ij} \leq 1 \qquad \forall i \qquad (31)$$

  Note that, the Single Association (SA) scenarios along side the DC mode scenarios are of significant interest since they will be prevalent in situations where it is not possible for the network to allocate resources on two APs for a given UE. Henceforth, we observe and analyze their performance along side the DC mode scenarios and compare it with the baseline scenario, which we elaborate upon next.

- **Baseline Association:** For the baseline scenario, we adopt the user association strategy that is being used by current day mobile networks, Wi-Fi, etc. Concretely, we utilize Algorithm 1, wherein we first compute the SNR that the users would observe from each AP. Based on this observed SNR, we associate the users to the AP with the best SNR. However, to compute the achieveable data rate we utilize the SINR ($\Psi(i, j)$) at the UE for the chosen AP. Given these UE-AP pairs, the bandwidth ($B$) at any given AP is then divided equally amongst all the UEs associated to it.

**Algorithm 1** Baseline Scenario Generation

**procedure** BASELINEGENERATOR
  $N_User \leftarrow$ Number of Users
  $N_APs \leftarrow$ Number of Access Points
  $R \leftarrow$ Vector of data rates for all users
  $N_{AP} \leftarrow$ Vector for number of users per AP
  $m_id \leftarrow$ Index of the AP with the highest SNR for user i
  $N_{AP} \leftarrow zeros(N_APs)$
  $R \leftarrow zeros(N_User)$
  $i, j \leftarrow 1$
  $iter_user \leftarrow N_User$
  $iter_AP \leftarrow N_APs$
  **for** $i < iter_user$ **do**
    **for** $j < iter_AP$ **do**
      $SNR(i, j) \leftarrow$ SNR of user i from AP j
      $\Psi(i, j) \leftarrow$ SINR of user i from AP j
    $m_id \leftarrow find(max(SNR(i, :)))$
    $N_{AP}(m_id) \leftarrow N_{AP}(m_id) + 1$
  **for** $i < iter_user$ **do**
    $idx \leftarrow find(max(SNR(i, :)))$
    $R(i) \leftarrow (\frac{B}{N_{AP}(idx)}) log_2(1 + \Psi(i, idx))$

### *F. Constraint Based Scenarios*

In addition to different combinations of topology, DC mode, Directivity and Service Class based scenarios, in our study we introduce an amalgamation of different network constraints, listed in Table IV, as well. These myriad combination of constraints are then combined with the scenarios in Table

III, following which they are optimized and analyzed by the *AURA-5G* framework for user association and resource allocation.

TABLE IV
CONSTRAINT COMBINATIONS FOR SCENARIOS

| Constraint Combination | Description |
|---|---|
| MRT | Minimum Rate Constraint for eMBB services. |
| CB | The wired backhaul link capacity is capped. For SCs, we cap the capacity of the backhaul to the MC at 1 Gbps, while for the MC to the CN it is 10 Gbps [42]. |
| CPL | eMBB applications will also have latency constraints, although not as strict as the URLLC applications. However, taking the requirements into account, we also explore the impact of constrained path latency. |
| MRT + CB | Minimum Rate Requirements and Constrained Backhaul together. |
| MRT + CPL | Minimum Rate and Constrained Path Latency constraints together. |
| MRT + CPL + CB | Minimum Rate Constraint, Constrained Backhaul and Path Latency constraint, all need to be satisfied simultaneously. |
| CB + CPL | Backhaul and Path Latency constraints are employed together. |

As we will see from our observations in Section VI, different combinations of these constraints on the scenarios analyzed have significant impact on the performance metrics. Further, interesting insights that can be utilized by the operator to enhance the performance for the purpose of UE-AP association as well as resource allocation, have been outlined.

## V. EVALUATION FRAMEWORK

In this section we establish the evaluation framework that we have considered for analyzing the multiple scenarios, described in Section IV, by utilizing the optimization framework developed in Section III. Concretely, we consider the topology, as shown in Fig. 2, with a geographical area of $600m \times 600m$. The scenario under investigation consists of a heterogeneous multi-layered radio access deployment. For this, we consider the 4G-LTE eNodeB as the MCs operating at the sub-6GHz frequency range, specifically at $3.55GHz$ with an inter-site distance of $200m$ [6]. Further, we deploy SCs utilizing a homogeneous poisson point process (HPP) within the vicinity of each MC. Note that, we repeatedly generate the location coordinates for the SCs, using the aforementioned HPP, until they have a minimum of $20m$ inter-site distance [6]. In addition, they operate on the mmWave frequency range of $27GHz$ for the access network, i.e. from SC to user, and at $73GHz$ for the possible wireless backhaul to the MC in accordance with [43], [44].

These SCs are then connected to an MC either via a wireless or a wired backhaul link. We utilize the fact that SCs operating in the mmWave frequency range, due to the operational characteristics of mmWave, i.e. high atmospheric absorption and severe blockage from the various objects in the transmission (TX) path, will have a significantly reduced TX range as compared to the MCs [45]. Henceforth, we specify a breakpoint distance of $25m$ from the MC, beyond which SCs are linked to the MC using a wired backhaul link. This in turn implies that the SCs within the aforementioned distance of the MC connect to it via a wireless link. Further, we specify an out of band operation regime for these wireless backhaul links, wherein the total available bandwidth is divided equally amongst the SCs attached to a given MC. Thus, to compute the available capacity on this link, we utilize the *Shannon-Hartley* theorem specified in eq (32).

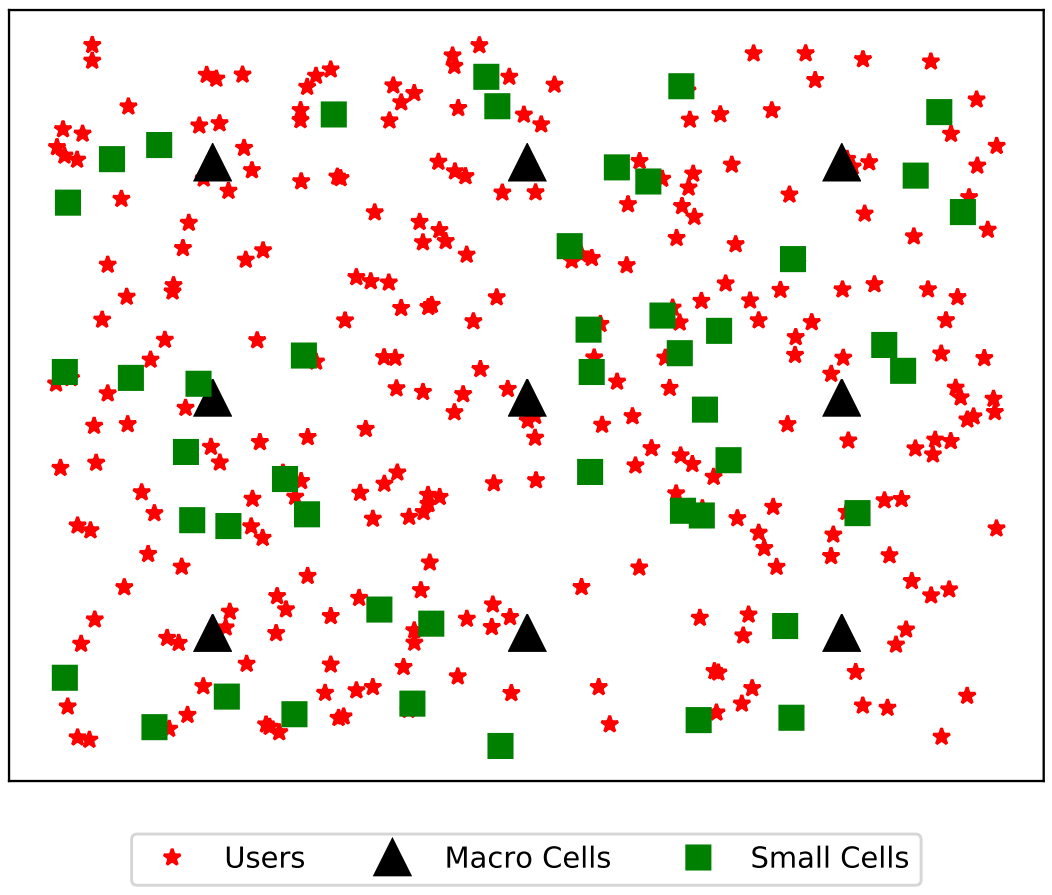

Fig. 2. Illustrative example of the network topology under study

$$C = W \times log_2(1 + \Psi') \quad (32)$$

where $C$ is the channel capacity, $W$ is the transmission bandwith and $\Psi'$ is the calculated signal to noise ratio between the SC and MC. Additionally, for the backhaul network we consider relevant wired technologies as specified in [42] and deploy a 10Gbps capacity fiber link from the MC to the core network. The wired backhaul link between SC and MC has a capacity of 1 Gbps. Note that, we dimension the backhaul link capacities such that an MC is able to serve all the SCs connected to it.

Next, we specify that each MC is connected to the core network via wired links. The number of hops for a given MC to the core network is chosen from a discrete uniform distribution over 1 to 4 hops. Further, each of the wired links within our defined topology imposes a delay of 1ms [42]. Additionally, since the SCs can have a wireless backhaul link to the MC, we define that a wireless link also imposes a 1ms delay [46].

We then deploy the users in the scenario area by utilizing a HPP. As specified in our discussions in Section IV, we consider scenarios where either only eMBB devices or both eMBB and mMTC devices exist. Hence, for the purpose of analysis we consider the various user densities for both eMBB and mMTC devices, as listed in Table V. We simplify our evaluation framework by utilizing the fact that mMTC devices operate in the guard band mode. Hence, they consume only backhaul resources in our evaluation framework. Further, in [36], it is stated that mMTC devices generate traffic between 1 Kbps and 1 Mbps mostly. Consequently, we consider a uniform distribution between 1-1000 Kbps and utilize it to compute the backhaul network resources consumed by the mMTC devices.

TABLE V
EVALUATION PARAMETERS

| Parameter Description | Value | Parameter Description | Value |
|---|---|---|---|
| Speed of Light ($c$) | $3\times10^8$ m/s | LTE eNB (Macro Cell) operating frequency | 3.55 GHz |
| Small-cell access network operating frequency | 27 GHz | Small-cell backhaul operating frequency | 73 GHz |
| Height of user | 1.5 m | Height of Small-cell | 10 m |
| Height of Macro-cell | 25 m | Simulation Area | $0.36\times10^6$m$^2$ |
| Number of eMBB users | [150,175,200,225,250,275] | Density of mMTC users | 24000 per MC |
| Transmit Antenna Gain for MCBS | 17 dBi | Transmit Antenna Gain for SCBS | 30 dBi |
| Macro-cell Transmit Power | 49 dBm | Small-cell Transmit Power | 23 dBm |
| UE Receive Gain for Small-cell | 14 dBi | UE Receive Gain for Macro-cell | 0 dBi |
| Small-Cell Bandwidth for access network | 1 GHz | Macro-cell Bandwidth for access network | 80 MHz |
| Number of hops from MC to core network | $U \in [1,4]$ | White Noise Power | −174 dBm/Hz |
| Break point distance for wireless backhaul between SC and MC | 25 m | Macro-cell Intersite distance (ISDMC) | 200 m |
| Small-cell Intersite distance | 20 m | Minimum Rate for eMBB users | 100 Mbps |
| Wired Backhaul Capacity from SC to MC | 1 Gbps | Wired Backhaul Capacity from MC to core network | 10 Gbps |
| Wireless link delay | 1 ms | Wired link delay | 1 ms |
| Maximum Permissible latency for eMBB services | 3 ms | Number of Iterations for evaluation | 100 |
| Small-Cell bandwidth options for users (BWSC) | [50 MHz, 100 MHz, 200 MHz] | Macro-cell bandwidth options for users (BWMC) | [1.5 MHz, 3 MHz, 5 MHz, 10 MHz, 20 MHz] |
| Data rate range for mMTC services | $U \in [1,1000]$ Kbps | UE receive beam Half Power Beamwidth (HPBW) | $45^{\circ}$ |
| Pathloss Exponent (Small Cell and LOS condition) | 2.1 | SF Std. deviation (Small Cell and LOS condition) | 4.4 |
| Pathloss Exponent (Macro Cell and LOS condition) | 2.0 | SF Std. deviation (Macro Cell and LOS condition) | 2.4 |
| Pathloss Exponent (Small Cell and NLOS condition) | 3.2 | SF Std. deviation (Small Cell and NLOS condition) | 8.0 |
| Pathloss Exponent (Macro Cell and NLOS condition) | 2.9 | SF Std. deviation (Macro Cell and NLOS condition) | 5.7 |
| Minimum Rate Requirement (eMBB services) | 100 Mbps | Latency Requirement (eMBB services) | 3 ms |
| Optimizer Cutoff Time | 600 seconds | Number of SCs per MC | $U \in [3,10]$ |

We also specify the bandwidth options that a UE has from a given SC as well as an MC [47]. In addition, for the channel model we adopt the NYU CI model [43], [48], which is expressed as follows:

$$PL = FSPL + 10nlog_{10}(d/d_0) + X_\sigma \tag{33}$$

where $PL$ defines the pathloss in dB, $FSPL$ [in dB] is computed as $20log_{10}\left(\frac{4\pi f d_0\times10^9}{c}\right)$, $d_0$ is 1m, and $X_\sigma$ denotes the shadow fading component with a standard deviation of $\sigma$. Based on the experiments carried out in [43] we adopt the pathloss coefficient, i.e., the value of $n$, and the standard deviation $\sigma$ for shadowing. These values have been specified in Table V. Note that, we consider both the Urban Micro (U-Mi Street Canyon) and the Urban Macro (U-Ma) scenarios in [43], [48] as they are reflective of the scenarios that will prevail for SC and MC, respectively, in a dense urban environment. Moreover, we also take into account the possibility of encountering obstacles in such dense urban environments by simulating the LOS-NLOS probability models for U-Mi Street Canyon and U-Ma scenarios as specified by 3GPP in [49]. To illustrate here, for U-Mi Street Canyon the LOS probability model is expressed as shown in eq. (34), where $d_{2D}$ is the

$$P_{LOS}^{SC} = \begin{cases} 1 & , d_{2D} \leq 18m \\ \frac{18}{d_{2D}} + exp^{(-\frac{d_{2D}}{36})(1-\frac{18}{d_{2D}})} & , 18m < d_{2D} \end{cases} \tag{34}$$

$$P_{LOS}^{MC} = \begin{cases} 1 & , d_{2D} \leq 18m \\ \left[\frac{18}{d_{2D}} + exp^{(-\frac{d_{2D}}{63})(1-\frac{18}{d_{2D}})}\right]\left[1 + C'(h_{UT})\frac{5}{4}\left(\frac{d_{2D}}{100}\right)^2 exp^{(-\frac{d_{2D}}{150})}\right] & , 18m < d_{2D} \end{cases} \tag{35}$$

$$C'(h_{UT}) = \begin{cases} 0 & , h_{UT} \leq 13m \\ \left(\frac{h_{UT}-13}{10}\right)^{1.5} & , 13m < h_{UT} \leq 23m \end{cases} \tag{36}$$

two dimensional distance between transmitter and receiver. Further, the U-Ma LOS probability model has been expressed, in a manner similar to the U-Mi model, in eqs. (35) and (36), where $h_{UT}$ represents the height of the user terminal, i.e. UE.

Lastly, we also provision parameters such as the MC height, SC height, UE height, transmit and receive gains, intersite distances as well as QoS requirements for the services discussed in this paper (minimum rate and latency). These parameters have been derived utilizing 5GPPP project proposals [6], 3GPP specifications [50] and other relevant research efforts [44].

Given the setup detailed so far, we perform 100 Monte Carlo runs for each scenario and constraint combination. These Monte Carlo trials help us attain a certain measure of confidence over our observations. Additionally, we also define a cutoff period of 600 seconds for our optimizer to determine a solution for the user association problem. The reason for such a cutoff timer being, in any dynamic network environment such a time period would be more than sufficient to determine an optimal association. And so, we now list all the other simulation parameters, along side the parameters discussed thus far in this section, and their corresponding description and values in Table V.

With this background, in the next section we evaluate the performance of the AURA-5G framework based on *Total network throughput*, *System fairness*, *Backhaul utilization*, *Latency compliance*, *Convergence time* and *Solvability*. It is important to state here that, the scenarios, detailed in Section III, along side the parameters, elaborated upon earlier in this section, provision a very realistic scenario. As a consequence, this accentuates the efficacy of our framework in provisioning a realistic and implementable framework for industry and academia.

## VI. Results and Discussions

Based on the evaluations performed by utilizing the AURA-5G framework, in this section and Section VII we consolidate and discuss our findings in detail. We structure our discussion into two phases wherein at first (in Section VI) we consider the setup that utilizes the evaluation parameters in Table V as is. The observations for this setup considers scenarios with only eMBB users and, eMBB and mMTC users together. Secondly, based on crucial observations from the first phase, in Section VII we perform network re-dimensioning and then present details on how AURA-5G can assist the operator in gaining insights that will ultimately lead to improved system performance through re-dimensioning.

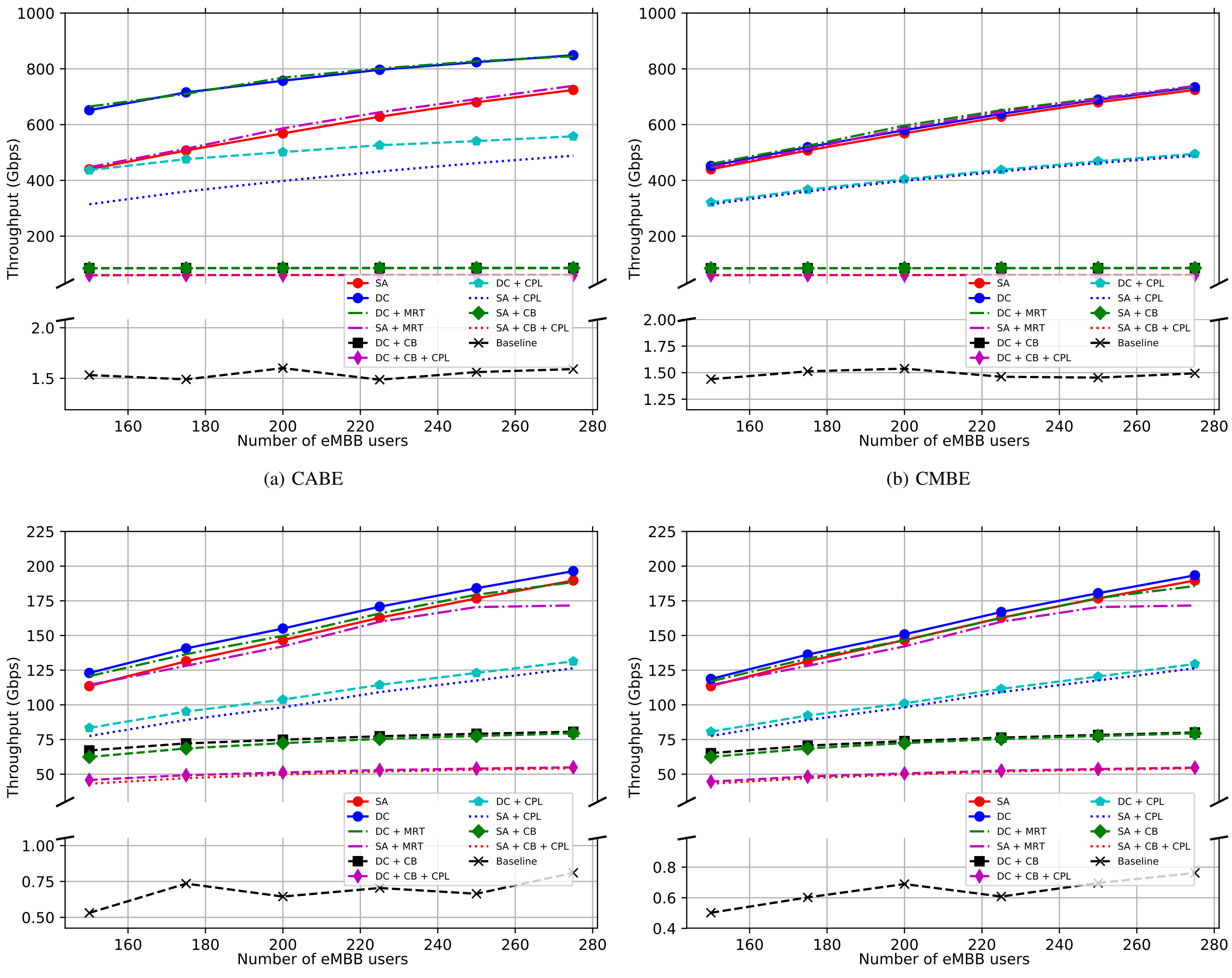

(a) CABE (b) CMBE (c) CAIE (d) CMIE

Fig. 3. Total Network Throughput for multiple combination of constraints being employed on (a) CABE, (b) CMBE, (c) CAIE and(d) CMIE scenarios.

We now proceed towards our discussion, and reiterate that we utilize the notations presented in Tables III and IV for the myriad scenarios explored in this paper.

### A. Total Network Throughput

*1) eMBB services based scenarios:* For the scenarios where users with only eMBB services are considered, we present our observations with regards to the total network throughput in Figs. 3 and 4. For the purpose of comparison, we also include the observations from the baseline scenario in Figs. 3 and 4. Note that, the observations presented have been obtained after averaging the total network throughput over 100 Monte-Carlo trials. In addition, the Minimum Rate (MRT) constraint (see Table IV for description), due to its strict nature, can lead to circumstances where either an optimal solution does not exist or it takes too long (greater than 600 seconds) for the optimizer to search for one. In either case, we consider these simulation trials to be unsuccessful with regards to finding an optimal solution, and hence, exclude them from the evaluation of the AURA-5G framework for the *total network throughput*, *system fairness*, *backhaul utilization* and *latency compliance* metrics. We refer the reader to Section VI.F and VII, wherein a more detailed discussion with regards to the issue of *solvability* and how it is addressed has been provided. Henceforth, for the *total network throughput* analysis, we now evaluate the *CABE*, *CMBE*, *CAIE* and *CMIE* scenarios in Figs. 3(a)-(d), where multiple combination of constraints (specified in Table IV), beamformed regime and circular deployment have been considered.

From Figs. 3(a)-(d), firstly we deduce that the AURA-5G

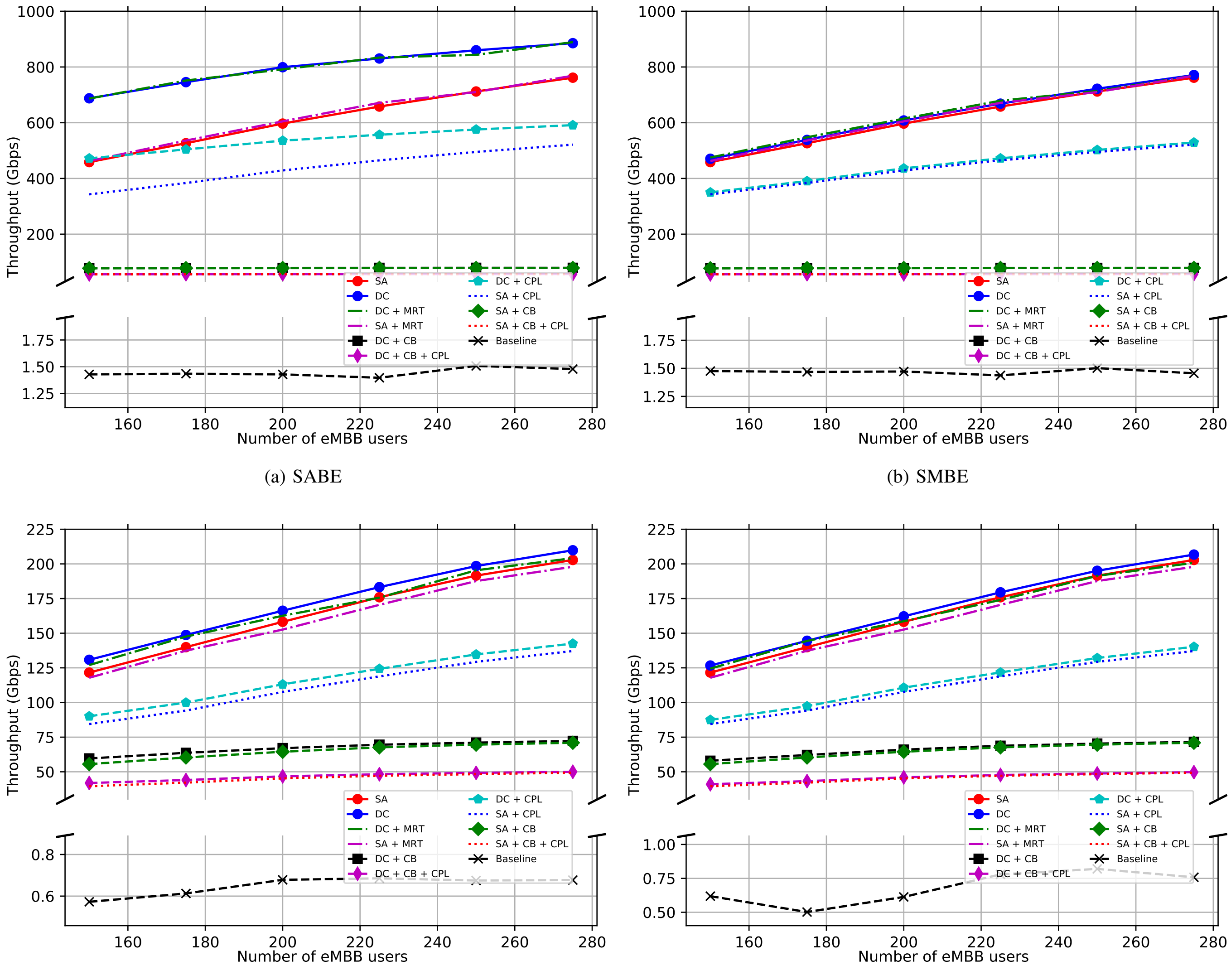

(a) SABE

(b) SMBE

(c) SAIE

(d) SMIE

Fig. 4. Total Network Throughput for multiple combination of constraints being employed on (a) SABE, (b) SMBE, (c) SAIE and (d) SMIE scenarios.

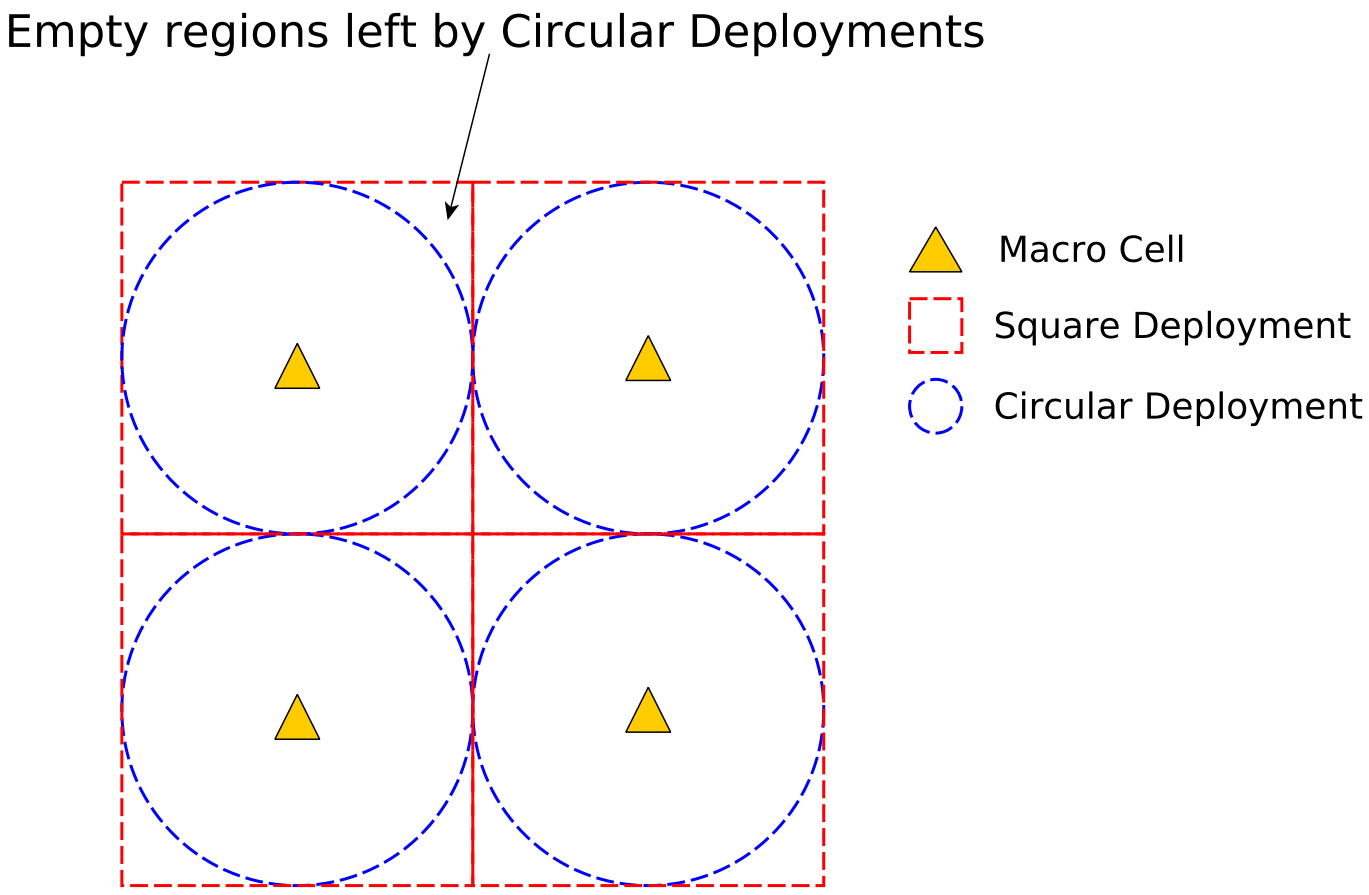

Fig. 5. Circular and Square deployment characteristics for SCs around MCs.

framework outperforms the baseline scenario for all set of constraints and scenarios. Next, the DC scenarios outperform the corresponding SA scenarios when *AnyDC* is employed (Fig. 3(a)). However, when *MCSC* is employed the gains are not as significant (Fig. 3(b)), because with DC in *MCSC* the UEs are connected to one MC and can additionally connect to at most one SC. Further, in SA, due to the nature of our optimization methodology being to maximize the total sum rate, the UEs are associated mainly to the SCs. Hence, the gains for DC scenarios in the *MCSC* setup are not as significant as those in *AnyDC*. Moreover, from Figs. 3(a) and (b), it can be observed that constrained path latency (CPL) and backhaul (CB) severely impact the overall network throughput. The reason being that the APs with the best available SINR, and hence capacity, might not be able to satisfy these latency and backhaul constraints.

We then consider the interference limited regime based scenarios, i.e. *CAIE* and *CMIE*, in Figs. 3(c) and (d). Immediately we can observe a significant reduction in the total network throughput as compared to that in the beamformed regime (Figs. 3(a) and (b)). This is inline with our expectations, since the SINR in an interference limited scenario will be significantly degraded as compared to that observed in the beamformed regime. Further, due to this interference limited nature of the scenarios, in Figs. 3(c) and (d) we do not observe a significant gain in performance from *AnyDC* over *MCSC*.

Next, we analyze the square deployment based scenarios, i.e. *SABE*, *SMBE*, *SAIE* and *SMIE*, in Figs. 4(a)-(d). To reiterate, in square deployment based scenarios, the SCs are distributed in a square geometry around each MC to which they have a backhaul link. Given these scenarios, from Figs. 4(a)-(d), the generic trend of observations does not change compared to the circular deployment. Concretely, we observe that the AURA-5G framework, given any set of constraint combinations and scenarios always outperforms the baseline scenario (beamformed regime scenarios perform better than their interference limited regime counterparts (Figs. 4(a)-(b) and 4(c)-(d)); *AnyDC* based DC scenarios have a significant performance gain over SA scenarios, which is not the case with *MCSC* based DC scenarios (Figs. 4(a)-(b) and 4(c)-(d)); and, latency and backhaul constraints significantly reduce the total network throughput (Figs. 4(a)-(d)).

However, in addition to these aforesaid observations, the square deployment based scenarios provision approximately 6% increase in total network throughput across all the constraint combinations explored, except when backhaul capacity constraints are applied. The reason being:

- In a circular deployment based scenario the SCs are deployed around the MCs such that there will be blind spots, i.e. there will be areas where there is weak or no SC coverage at all, since circular geometries leave empty spaces where their edges meet. However, with a square deployment scenario the probability that there are such blind spots is less, as square geometries do not leave any empty spaces like their circular counterparts. This consequently gives users in these geographically transitional areas a better opportunity to connect with an SC. An illustration to highlight the aforesaid characteristic of circular and square deployments has been shown in Fig. 5.
- A square deployment configuration however means that the probability that SCs are further away from the MC is higher. This consequently increases the probability of employing a wired backhaul from SC to MC which, as can be seen from our emulation framework in Section V, might have lower capacity than the mmWave based wireless backhaul. Hence, this reduces the overall data carrying capacity of the network which corresponds to the stated reduced total network throughput as compared to the circular deployment when backhaul capacity constraints are employed.

*2) mMTC with eMBB services based scenarios:* For the scenarios where both mMTC and eMBB services co-exist, we firstly re-iterate the fact that the mMTC devices only consume backhaul resources (see Section V). Hence, the main total network throughput characteristics stay similar to those observed for the scenarios where only eMBB services exist. However, certain scenarios where we take into account the backhaul capacity constraints show interesting observations. Consequently, in Figs. 6(a)-(d), for the *CABE*, *CABEm*, *CAIE* and *CAIEm* scenarios we illustrate the total network throughput only when CB and CPL (see Table IV for descriptions) constraints have been imposed. Note that, the above mentioned scenarios are all circular deployment based.

Concretely from Figs. 6(a)-(d), we observe that the presence of mMTC devices leads to a reduction in the overall network throughput experienced by the eMBB services. This is along expected lines, given that the mMTC devices, as stated before, consume a portion of the available backhaul capacity. In addition, from Figs. 6(a) and (c), it can be seen that the total network throughput for the beamformed regime (Fig. 6(a)) is much higher than that observed in an interference limited regime (Fig. 6(c)).

Next we consider, through Figs. 7(a) and (b), scenarios *CMBEm* and *CMBE*, where *MCSC* configuration for DC modes is utilized. We observe that, for the scenarios where mMTC services also exist along side the eMBB services, the total network throughput achieved by the eMBB services is lower. The reasoning, as has already been stated before in this section, is that the mMTC devices consume a portion of the available backhaul capacity thus reducing the overall achievable throughput for the remaining services in the system. We further present the Square Deployment scenarios in an interference limited regime in Figs. 7(c) and (d), and compare them with their circular deployment counterparts presented in Figs. 6(c) and (d). We deduce that, as compared to the circular deployment scenarios in Figs. 6(c) and (d), the total network throughput observed for the square deployment scenarios is lower when CB and CPL constraints are considered. The reason being, and we re-iterate, that a square deployment configuration leads to a higher probability of the SCs being further away from the MC. As a consequence, this increases the probability of employing a wired backhaul from SC to MC which might have lower capacity than the mmWave based wireless backhaul. Hence, this reduces the overall data carrying capacity of the network which corresponds to the stated reduced total network throughput as compared to the circular deployment.

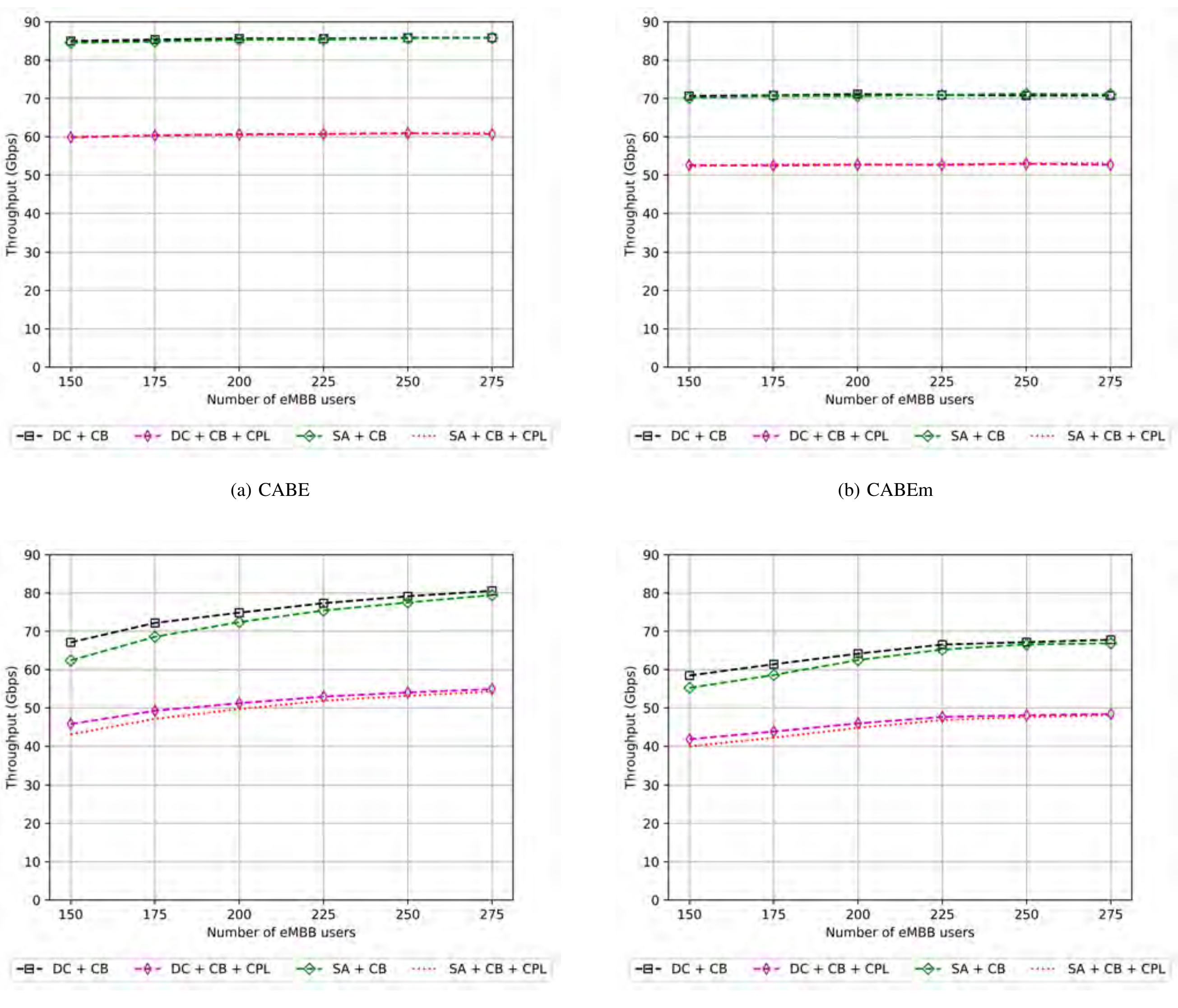

(a) CABE

(b) CABEm

(c) CAIE

(d) CAIEm

Fig. 6. Total Network Throughput for multiple combination of constraints being employed on (a) CABE, (b) CABEm, (c) CAIE and (d) CAIEm scenarios.

## B. System Fairness

We analyze the fairness of our optimization based framework through the Jain's Fairness Index [51] for the scenarios explored in this work. The fairness index is computed for the constraint combinations and scenarios that have been discussed in Section VI.A, and then a detailed discussion has been provided. It is important to state here that the objective of evaluating the AURA-5G framework for fairness measure is, *to be able to study the impact of various constraints and scenarios, prevalent in the 5G networks, on the fairness offered by the system, given the objective function of total sum rate maximization* (Section III).

*1) eMBB service based scenarios:* In Figs. 8(a)-(d), the Jain's fairness index for the scenarios *CABE*, *CMBE*, *CAIE*

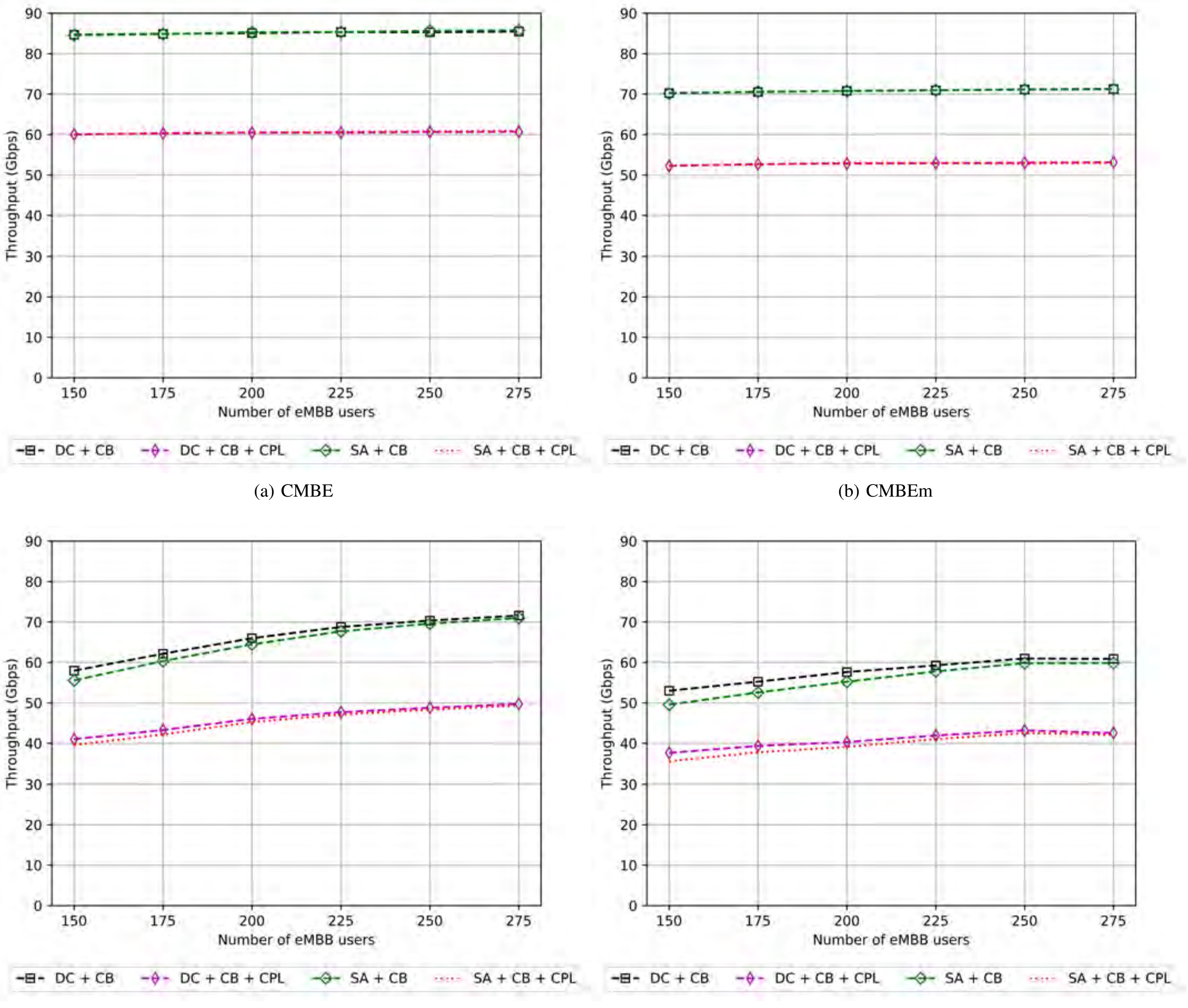

(a) CMBE

(b) CMBEm

(c) SAIE

(d) SAIEm

Fig. 7. Total Network Throughput for multiple combination of constraints being employed on (a) CMBE, (b) CMBEm, (c) SAIE and (d) SAIEm scenarios.

and *CMIE* with the different constraint combinations have been illustrated. Specifically, from Fig. 8(a) we observe that the AURA-5G framework in the Single Association (SA) setup provisions a higher fairness as compared to the Dual Connectivity setup, except when backhaul capacity constraints are considered. The reason being, since SA allows the UEs to connect to at most one AP, network resources are more evenly distributed and hence more users are able to connect and reserve resources in the network. However, since in DC the UEs have the possibility of connecting to two APs, the amount of resources available per user in the network is significantly less. Hence, the disparity in the amount of resources reserved by the users in DC modes is much higher. This, as a consequence, results in the aforementioned fairness characteristic. However, as can be seen in Fig. 8(a), when we consider the Minimum Rate (MRT) requirement constraint there is a slight improvement in the fairness, because the algorithm tries to find a solution wherein each user is allocated at least 100 Mbps. This forces the system to allocate resources more fairly so as to satisfy the minimum rate constraint and thus results in the marginal increase in fairness, as observed.

Further, in Fig. 8(a), we observe that the imposition of path latency (CPL) and backhaul capacity (CB) constraints results in significant lowering of the overall system fairness. This is as a consequence of only a small subset of APs and backhaul paths being able to satisfy the set constraints. Hence, in order to satisfy these requirements the UEs share the limited available resources on these candidate APs and backhaul links. Moreover, given that the algorithm aims to maximize the total sum rate of the system, UEs with better SINR are allocated better bandwidth resources from the limited available system resources. Hence, this creates significant disparities between the throughput of different users, thus leading to a reduction in system fairness. Lastly, DC scenarios provision better fairness than the SA scenarios when backhaul capacity (CB) constraint is applied. This is so because, the users have the opportunity to compensate for the lack of backhaul capacity resources in one link by acquiring bandwidth resources in the other link connected to the second AP selected. However, in SA, the lack of backhaul capacity resources combined with the nature of the optimization algorithm to maximize the total sum rate leads to a significant disparity in how the system resources are allocated to the users.

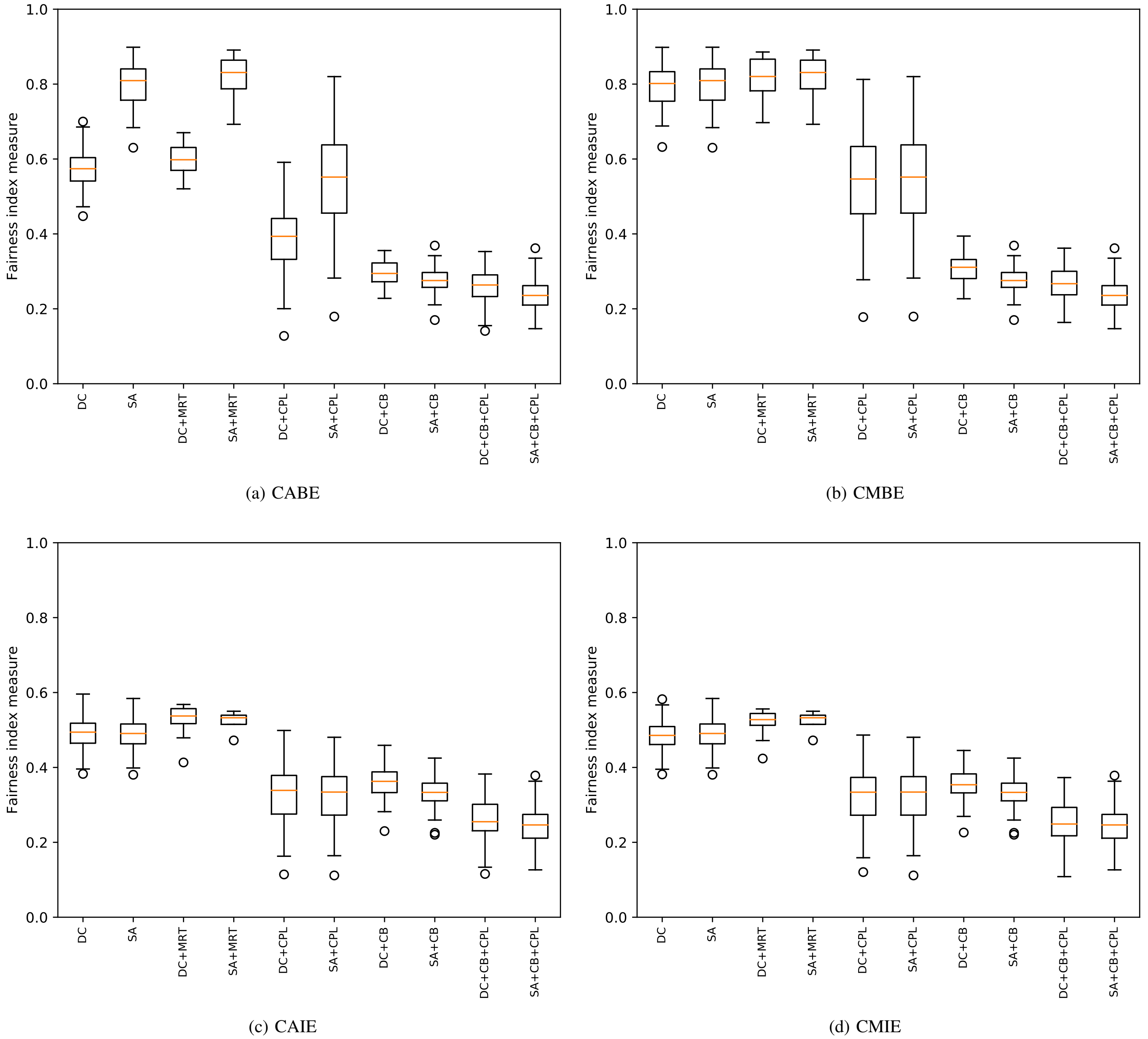

(a) CABE

(b) CMBE

(c) CAIE

(d) CMIE

Fig. 8. Jain's Fairness index deviation measure for multiple combination of constraints being employed on (a) CABE, (b) CMBE, (c) CAIE and (d) CMIE scenarios.

Next, in Fig. 8(b), for the *CMBE* scenario wherein *MCSC* setup is utilized, an overall improvement in the system fairness in the DC modes is observed. This is as a result of the fact that the users are now forced to select one MC amongst the two APs they choose. Hence, this relieves resources from the SCs which are the main drivers for maximizing the total sum rate of the system. However, this was not the case in the *AnyDC* scenario, wherein users could select even two SCs. As a consequence, MCSC provides more users with the opportunity to select an SC and maximize their possible data rate, which leads to the improvement in the system fairness, as stated before. Moreover, and as expected, for the interference limited regime scenarios shown in Figs. 8(c) and (d), the fairness measures are significantly lower as compared to the beamformed based scenarios (Figs. 8(a) and (b)). Given the severe interference and the objective of maximizing sum rate, only a select few users will have a good SINR, which as a consequence will receive the maximum share of the network resources. Hence, this leads to the severe disparity in the achievable rate per user, which subsequently explains the drop in the system fairness. Note that, rest of the trends in system fairness measures for the interference limited regime scenarios follow those already observed for the beamformed regime scenarios.

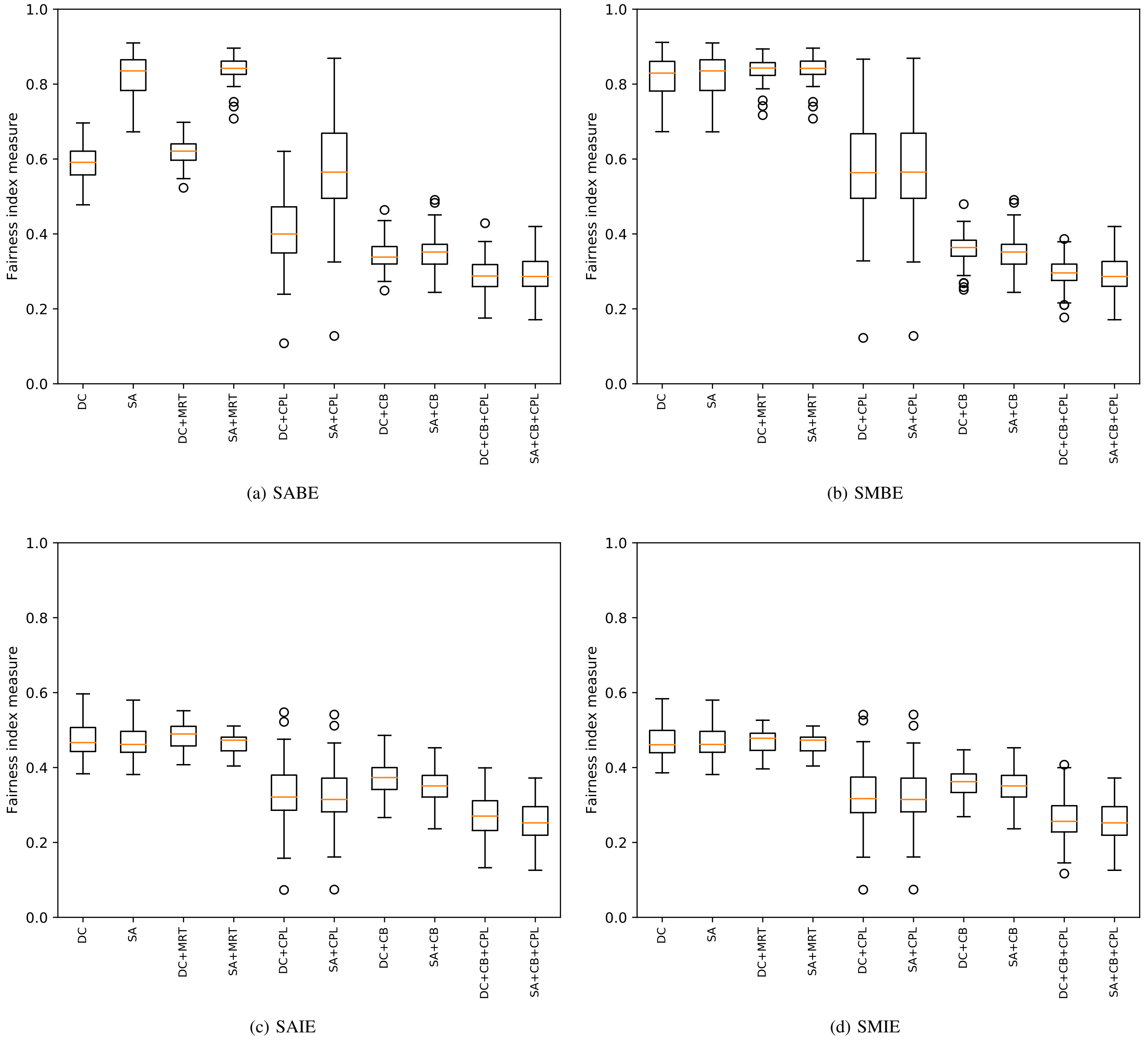

Fig. 9. Jain's Fairness index deviation measure for multiple combination of constraints being employed on (a) SABE, (b) SMBE, (c) SAIE and (d) SMIE scenarios.

Following the discussion for circular deployment based scenarios, we next consider the square deployment based scenarios, i.e. *SABE*, *SMBE*, *SAIE* and *SMIE*, in Figs. 9(a)-(d). From these figures, we observe that the generic trend for the fairness measure is similar to those observed for the circular deployment scenarios (discussed above). However, the square deployment for certain constraint combinations and scenarios enhances the overall system fairness. An example being the SABE scenario, wherein for all constraint combinations we observe between 5-6% improvement in system fairness. This is because of the reasons we have already elaborated in Section VI.A.2, i.e. square deployments result in less blind spots within the deployment, hence resulting in a fairer allocation of resources to the users as compared to the circular deployment.

*2) mMTC with eMBB based scenarios:* For the scenarios where mMTC and eMBB services are considered together, we present our observations through Fig. 10. It can be seen from Figs. 10(a)-(d) that the fairness index does not change significantly as compared to the fairness measure observed in eMBB only scenarios, even though we consider the mMTC devices within our framework. The reason for such a behavior is two-folds. Firstly, the fairness is computed by utilizing the throughput experienced by each individual eMBB user in the system, which is a function of the access network resources. And secondly, the mMTC users operate in the guard band, thus not consuming any access resources from eMBB users. Henceforth, there are very slight variations in the fairness index measure as mMTC users only impact the system performance through backhaul resource consumption.

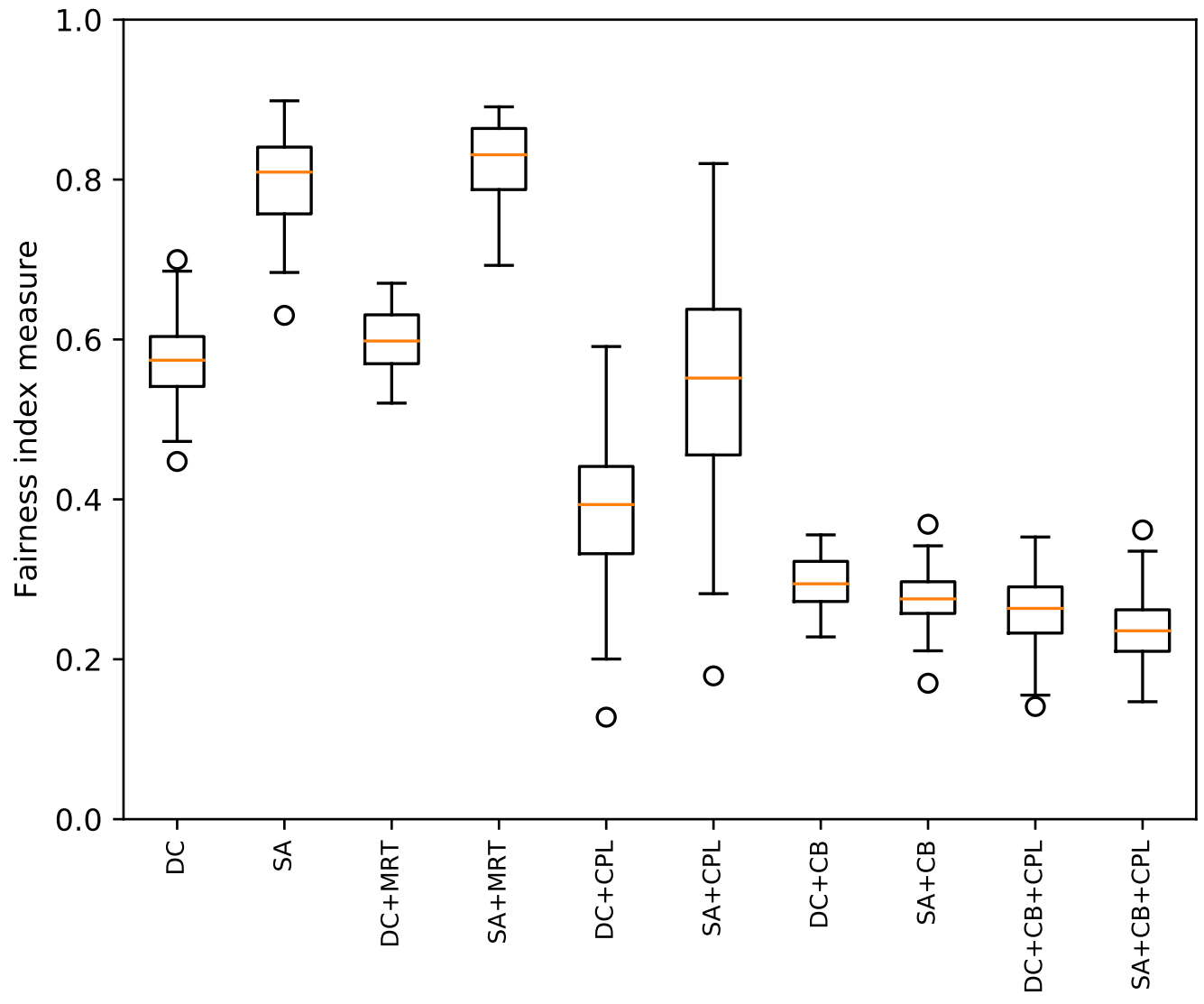

(a) CABE

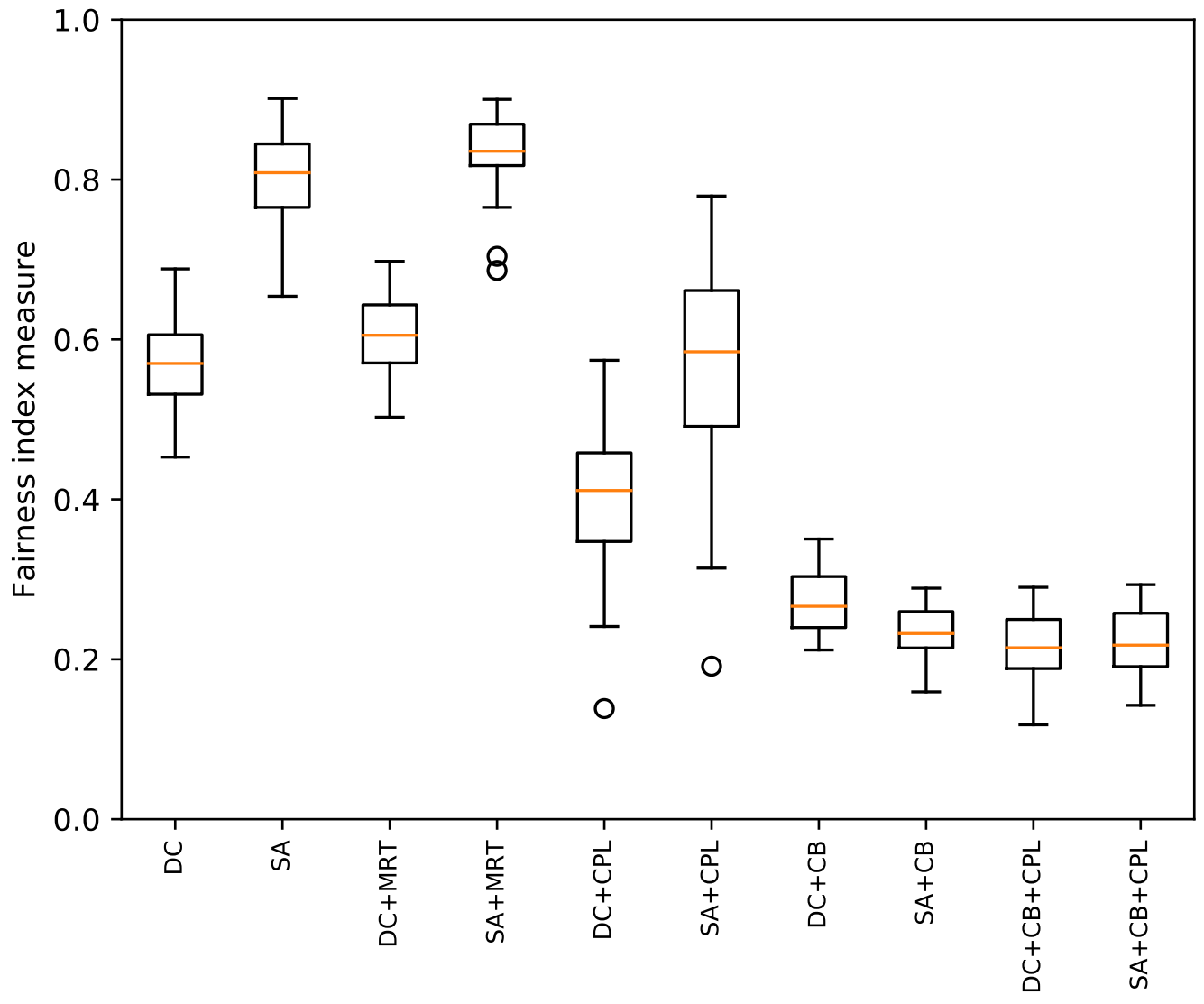

(b) CABEm

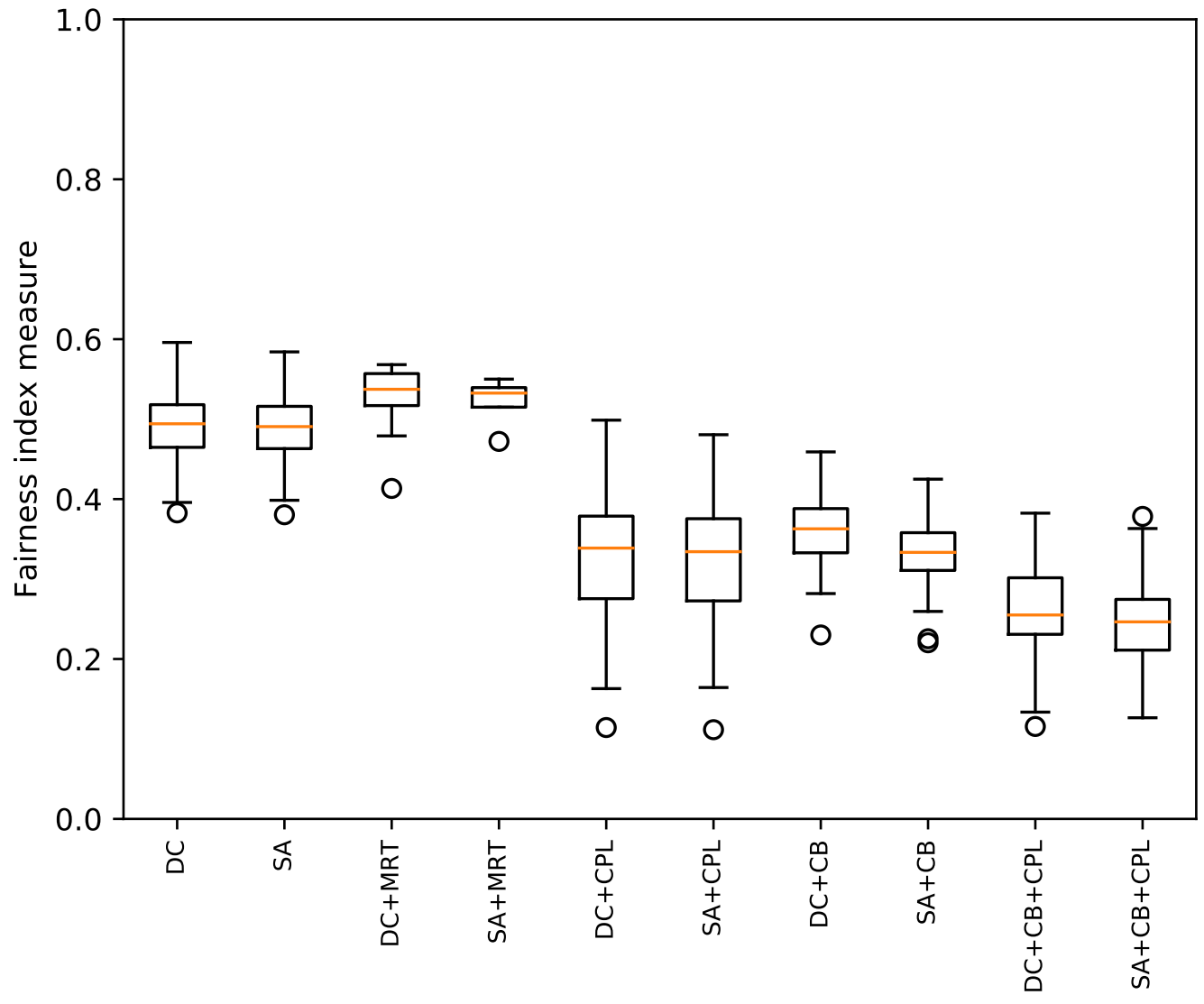

(c) CAIE

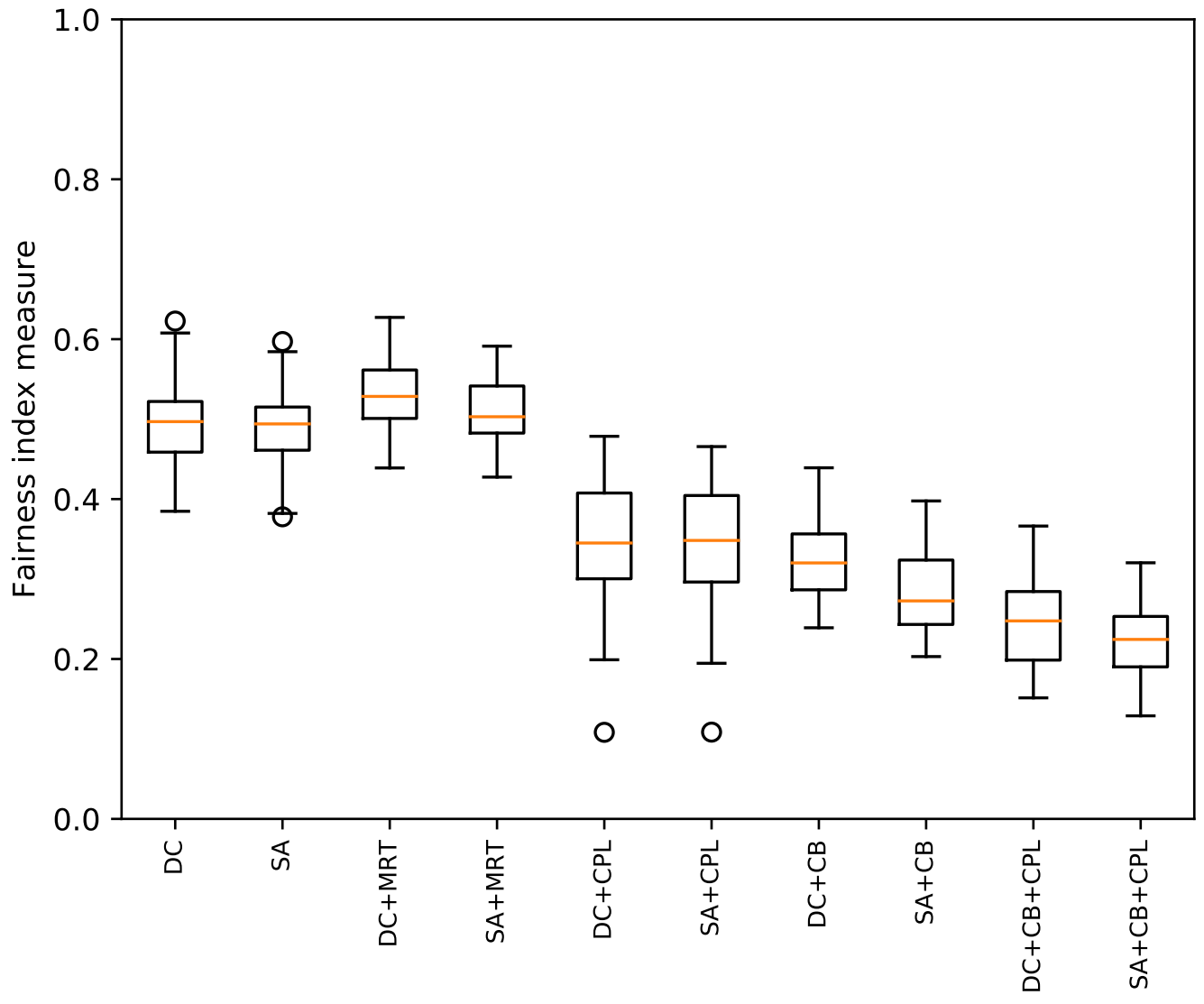

(d) CAIEm

Fig. 10. Jain's Fairness measure for multiple combination of constraints being employed on (a) CABE, (b) CABEm, (c) CAIE and (d) CAIEm scenarios.

Further, we also considered scenarios with square deployment and circular deployment along side the *MCSC* setup. And similar to the aforesaid deductions, we observed negligible change in the fairness index when mMTC and eMBB services are considered together as compared to when only eMBB devices are considered. Note that, for the sake of brevity, we do not present the illustrations for any other scenarios except those presented in Fig. 10.

### C. Backhaul Utilization

The primary goal for analyzing the backhaul utilization after the AURA-5G framework has been implemented on certain scenarios is to determine if the existing backhaul setup, wherein we consider a combination of wired and wireless backhaul links with the wired links having capacities of 1Gbps and 10Gbps (Section V), is a bottleneck. Further, we also utilize this analysis to understand the compliance of the AURA-5G framework with the backhaul capacity constraints. For the sake of brevity, and as we have done throughout this paper, for this analysis we select a subset of representative scenarios. Hence, through Figs. 11 and 12, we depict the backhaul utilization as observed for CABE, CMBE, CAIE,

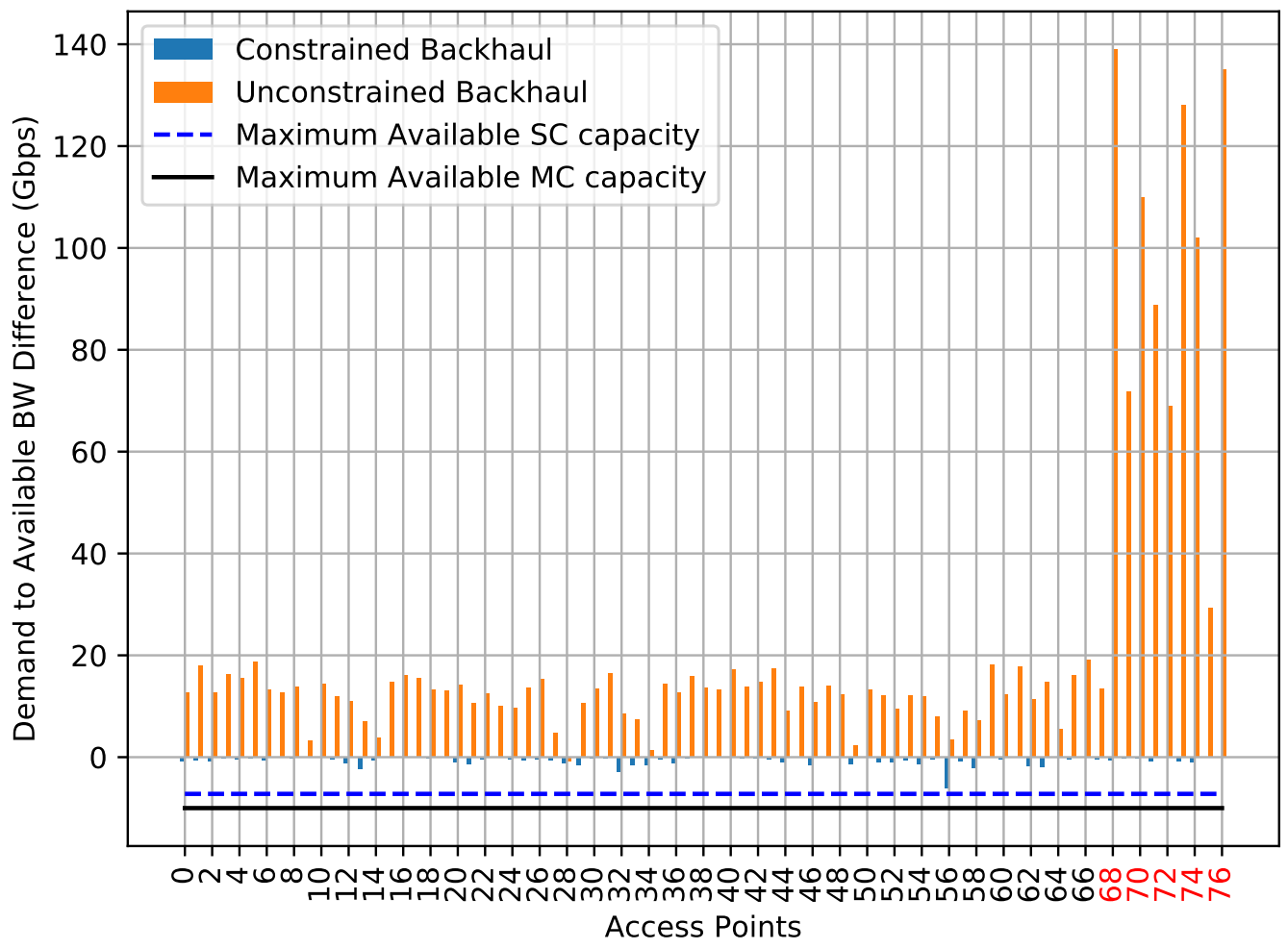

(a) CABE

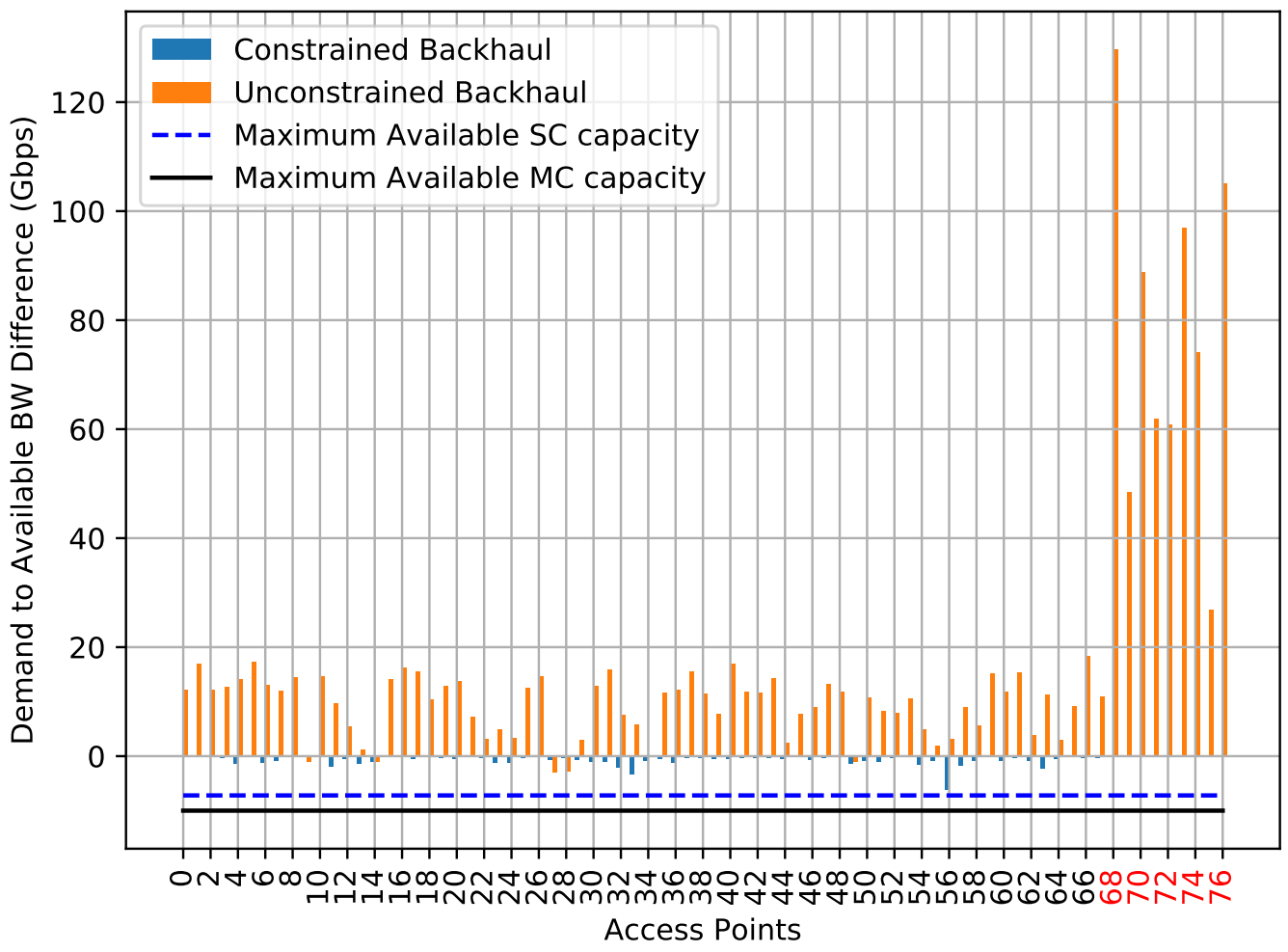

(b) CMBE

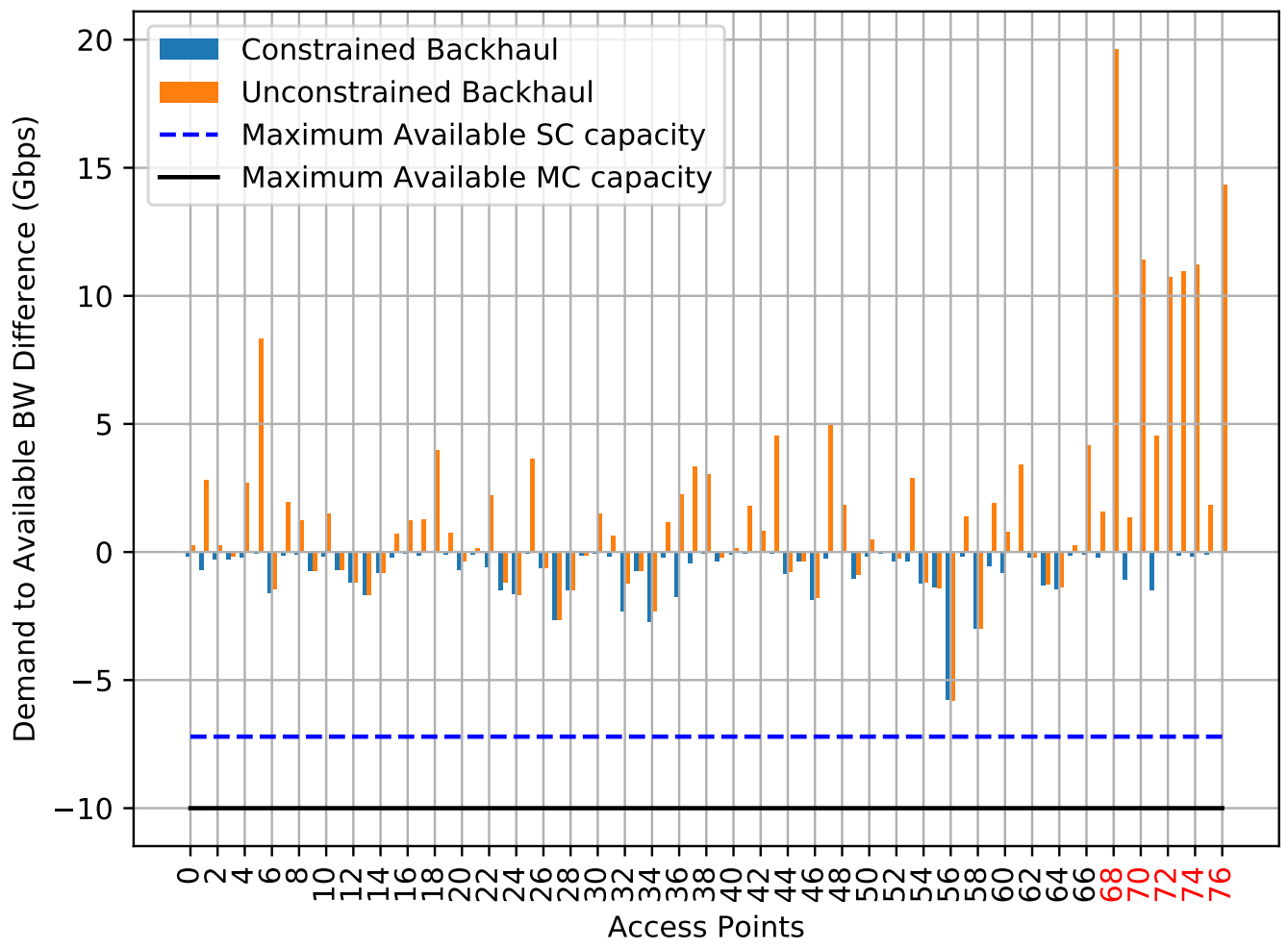

(c) CAIE

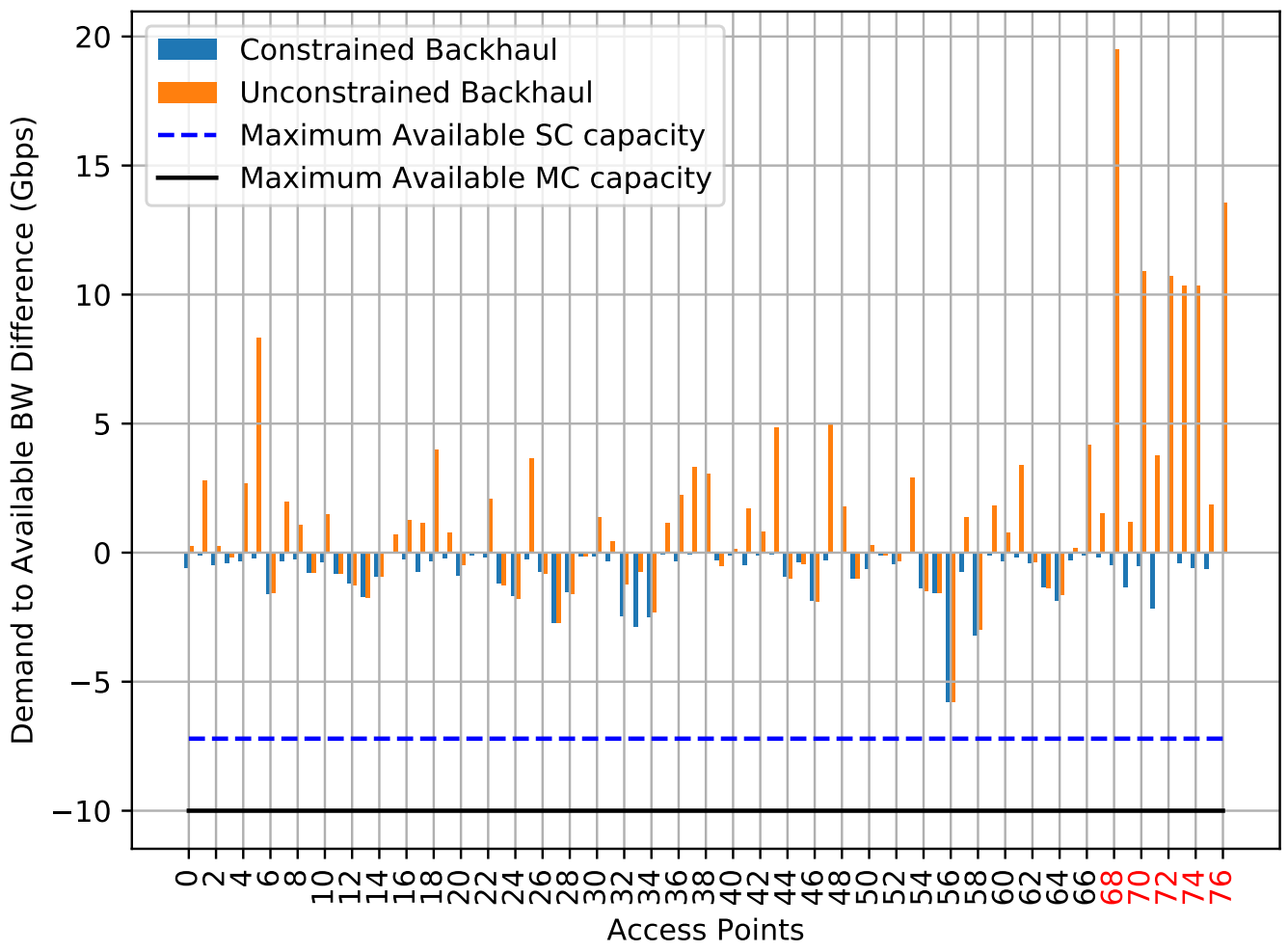

(d) CMIE

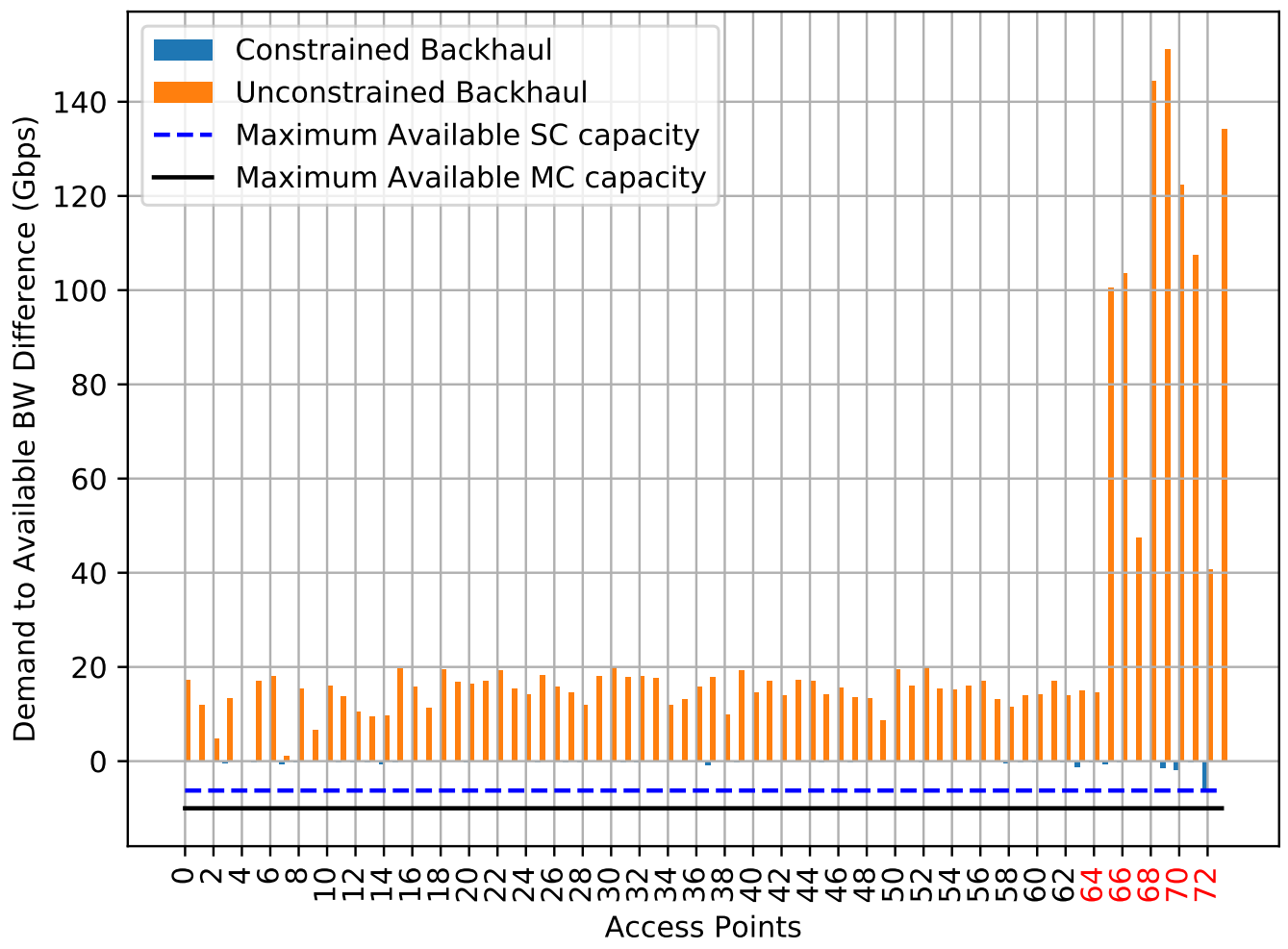

(e) SABE

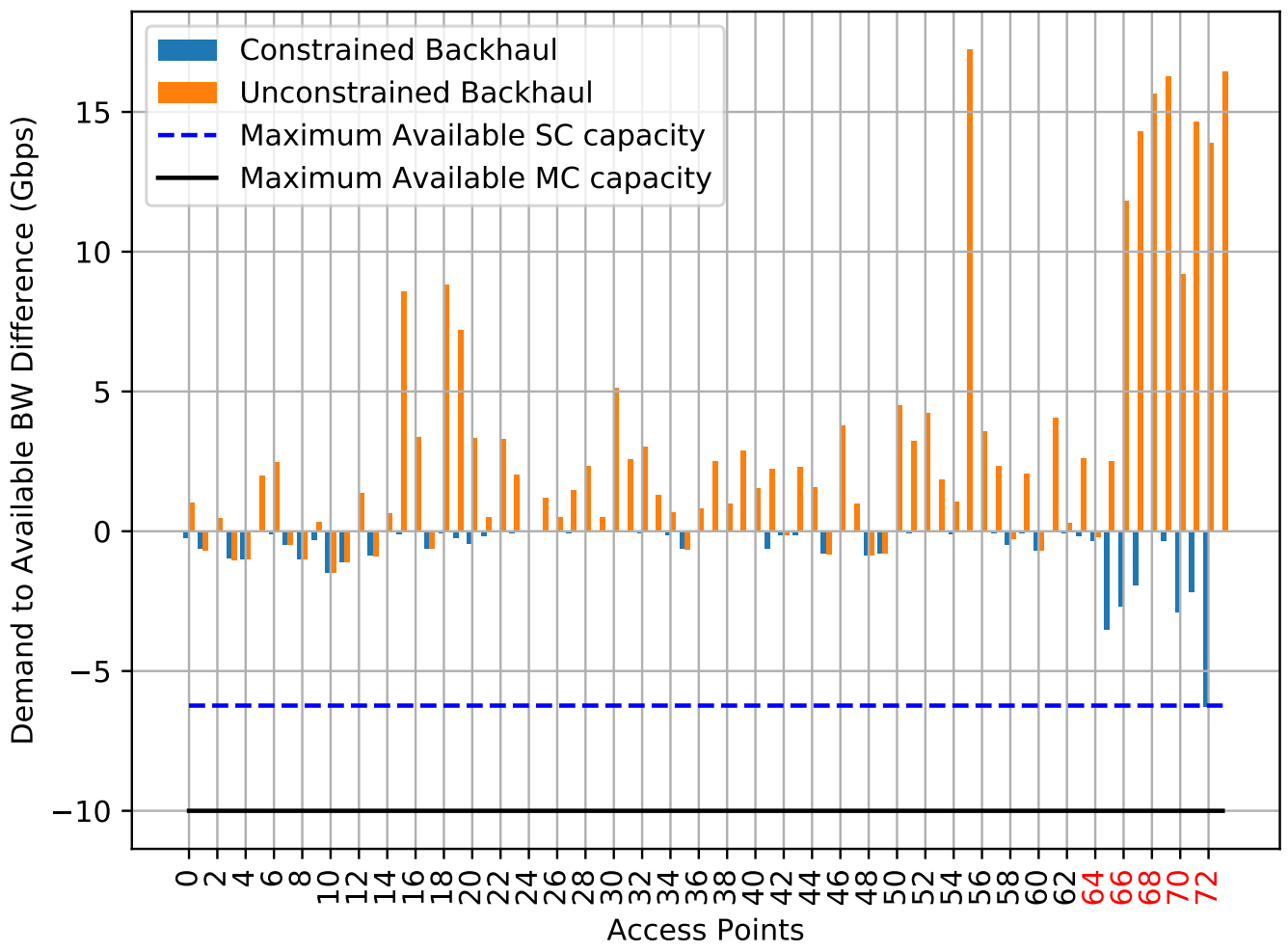

(f) SAIE

Fig. 11. Backhaul Utilization for Dual Connectivity (DC) and DC with Backhaul Capacity constraints in (a) CABE, (b) CMBE, (c) CAIE and (d) CMIE scenarios. Red colored AP indices are for MCs and the rest for SCs.

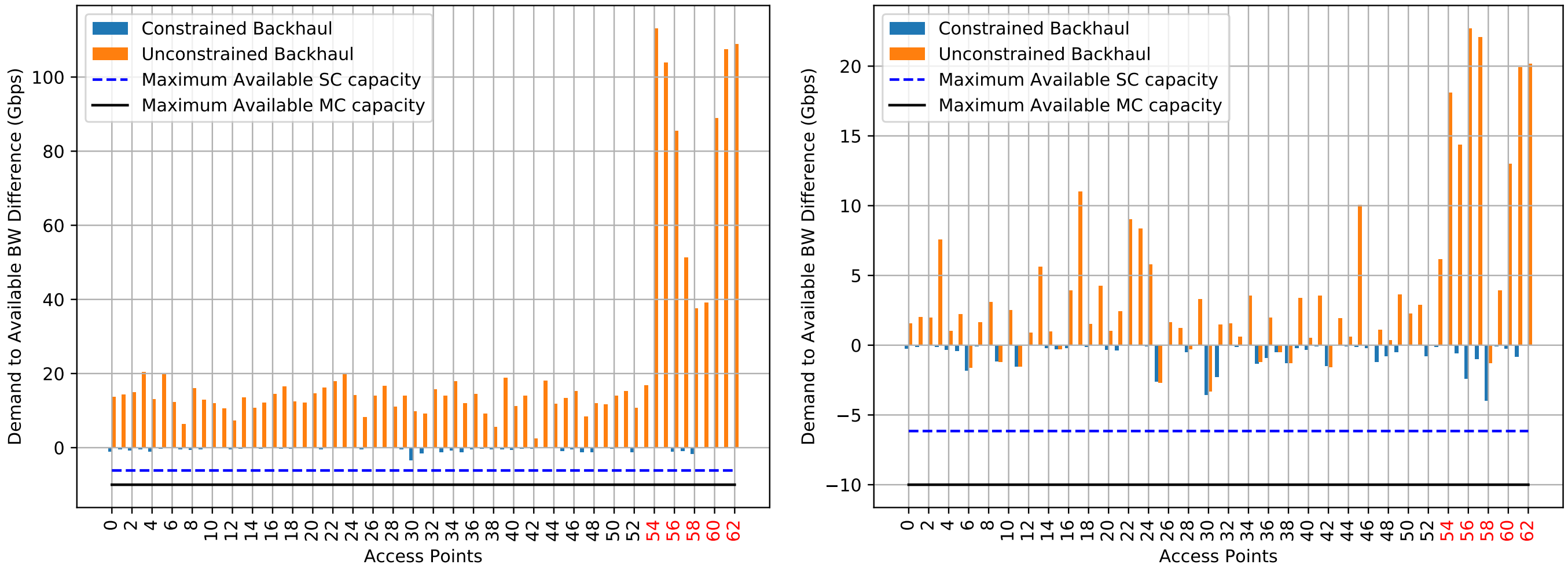

(a) CABEm

(b) CAIEm

Fig. 12. Backhaul Utilization for Dual Connectivity (DC) and DC with Backhaul Capacity constraints in (a) CABEm and (b) CAIEm scenarios. Red colored AP indices are for MCs and the rest for SCs.

CMIE, SABE, SAIE, CABEm and CAIEm scenarios. The choice of the aforesaid scenarios stems from the fact that these selected scenarios include the *MCSC* and *AnyDC* setup, beamformed and interference limited regime, square deployment setups, as well as the mMTC and eMBB services together in the beamformed and interference limited scenarios alongside *AnyDC* setup. This set of scenarios owing to their diversity and challenging nature give the necessary and sufficient idea with regards to the backhaul utilization characteristics.

From Figs. 11(a)-(f) and 12(a)-(b), we firstly observe that the AURA-5G framework is successful in satisfying the backhaul capacity constraints as and when they have been imposed. Here by backhaul capacity constraints we mean that the wired backhaul links are capped by their designated capacities as stated in Table IV. Further, the wireless backhaul links are constrained by the capacity computed utilizing the *Shannon-Hartley* formula based on their corresponding SINR value. It is also important to state here that on the vertical axis in all the subplots presented in Figs. 11 and 12, we represent the difference between the demand and the available capacity. Hence, a negative value on the vertical axis indicates that the backhaul resources on the corresponding AP have not been fully utilized, whilst a positive value indicates over-utilization by the corresponding amount. And so we can see that for the unconstrained scenarios the backhaul resources are always over-utilized. However, for the backhaul capacity constrained (CB) scenarios, we observe that our framework succeeds in finding an optimal solution without over-utilizing the total available backhaul resources. This significant difference in backhaul utilization also reflects the greedy nature of the optimization framework, whose objective is to maximize the total network throughput. Note that we have also indicated the maximum available backhaul capacity amongst the SCs and MCs. This assists the readers in understanding the maximum data carrying capacity that the set of SCs and MCs in the network have, as well as in exemplifying the fidelity of the AURA-5G framework. Concretely, the maximum capacity on an SC should not exceed that on an MC. This is so because, all SCs, in our framework, route their traffic through the MC. Hence, a higher maximum bandwidth availability on the SC as compared to an MC would be equivalent to trying to balance a big box on a thin needle. Additionally, and for the reasons as stated above, from the backhaul utilization values for the unconstrained setup we observe that the backhaul through the MCs is significantly over-utilized as compared to the SCs.

Next, we observe that scenarios wherein beamforming has been applied, i.e., Figs. 11(a), (b), (e) and 12(a), are severely limited for backhaul resources. The reason being that, the blue bars (darker bars if viewed in black and white), which indicate the available backhaul capacity in the constrained setup, are extremely small. This indicates that nearly all of the available capacity has been utilized, even without the Minimum Rate constraints being applied. The reason being that, beamforming results in an improved SINR measure within the system. This consequently enables the users to achieve a better throughput, and hence, the aforementioned backhaul utilization characteristic. Thus, an important insight for network operators, suggesting a requirement for network re-dimensioning (Section VII), can also be drawn from these observations. Further, in Figs. 11(c), (d), (f) and 12(b), wherein the interference limited regime has been adopted, the overall backhaul utilization in the unconstrained setup is much lower than that observed for the scenarios involving beamformed regime. This is as a result of the severe interference causing a significant loss in SINR, and hence, per user throughput. This claim is also corroborated by the reduction in network throughput observed in Section VI.A for interference limited scenarios.

Lastly, through Figs. 12(a) and (b), wherein the mMTC services have been considered alongside the eMBB services, it can be observed that the AURA-5G framework is able to provision optimal user-AP associations whilst adhering to the backhaul capacity constraints. Furthermore, as compared to the corresponding scenarios where only eMBB services are present, i.e., CABE and CAIE, the backhaul utilization for the constrained backhaul case in CABEm (Fig. 12(a)) and CAIEm (Fig. 12(b)) scenarios is slightly higher. This is so because, in addition to the eMBB services, the mMTC services also consume a portion of the backhaul resources. Hence, the overall increase in backhaul utilization.

### *D. Latency Requirement Compliance*

As part of our fidelity analysis for the AURA-5G framework, we delve into how it satisfies the specified service latency requirements through Figs. 13(a)-(f). It is important to state here that, the latency (or the downlink delay which we define as latency in our work) is governed by the number of hops the data has to traverse to reach the user from the core network. Hence, we consider certain representative scenarios such as *CABE*, *CMBE*, *SAIE*, *SMIE*, *CMIEm* and *CAIEm* for our analysis. These scenarios encompass the *AnyDC* and *MCSC* setup, the beamformed and interference limited regimes, as well as the eMBB only and eMBB with mMTC services based setups. Note that, we do not include the last wireless hop, i.e. MC or SC to the UE, in our optimization framework as it does not induce any variability within the scenario given that it will be omnipresent whatever the given association be. Hence, we focus on the number of hops that the data has to traverse from the CN to the AP.

From Figs. 13(a), (c) and (e), wherein the *AnyDC* setup is employed, we observe that the density of users with $3ms$ latency is higher as compared to the ones where *MCSC* setup is employed, i.e. in Figs. 13(b), (d) and (f). This is so because, in *AnyDC* to maximize the total sum rate our algorithm tries to find SCs first for all the users. However, in the *MCSC* setup, we force our algorithm to find atleast one MC for each user. Hence, given the fact that an SC is connected to the CN through a corresponding MC, the latency incurred by the users in the *MCSC* scenarios is comparatively less as compared to the *AnyDC* scenario. The operators can utilize this insight to design access point selection schemes wherein for services that can tolerate higher delays the *AnyDC* setup maybe employed, whereas for services with extreme latency constraints, an *MCSC* setup could be employed.

Further, for the square deployment scenarios (Figs. 13(c) and (d)) and mMTC based scenarios (Figs. 13(e) and (f) the trend for latency compliance follows that of the *CABE* (Fig. 13(a)) and *CMBE* (Fig. 13(b)) scenarios, as discussed above. Hence, through this analysis we reinforce fidelity of the AURA-5G framework towards the joint optimization problem that we explore in this paper.

### *E. Convergence Time Distribution*

Next, we study the convergence time to the optimal solution for the AURA-5G framework. This will be critical for real time implementation, and hence is of immense value to not only the academic but also to the industrial community. The reason being, network scenarios with the combination of constraints discussed in this work, will be prevalent in 5G networks. And given that there will be a central controller within a local area of these networks [2], [3], [52], the newly designed mobility management algorithms, such as the AURA-5G framework, will be placed on these controllers to enhance the QoE for the users. Consequently, through Figs. 14 and 15, we evaluate the convergence time parameter for the various constraint combinations imposed on the myriad scenarios explored in this paper. From the CDFs presented in Figs. 14 and 15, a probability measure of 1 indicates that all the 100 iterations (monte carlo trials) for the specific constraint combination over the scenario under study have converged. On the other hand, a probability measure of 0 indicates that none of the iterations converge. Note that the simulations were performed on a commodity server with 20 cores (with each being an i9-7900x at 3.3GHz core), Ubuntu 16.04 LTS OS, and 64GB of RAM.

From Figs. 14(a)-(d) and 15(a)-(b) we observe that for all scenarios and most constraint combinations, the AURA-5G framework is able to determine an optimal solution. It is worth mentioning that, the AURA-5G framework is able to provision an optimal solution in an acceptable time frame, given the density and heterogeneity of 5G networks. This is of extreme importance for real-time implementation because of the elevated level of dynamics in 5G networks as compared to its predecessors.

Next, we observe that for the Single Association (SA) scenarios the time required for obtaining an optimal solution is significantly less as compared to the Dual Connectivity (DC) mode scenarios. This is so because, the solution space for an SA scenario will be much smaller than that for a DC scenario, hence the time required to search for the optimal solution is correspondingly also reduced. Next, we observe that as constraints are imposed the amount of time required to search for the optimal solution increases. This is inline with our intuition, since addition of constraints adds extra dimensions to the search space. Most notably scenarios with the Minimum Rate (MRT) constraints for both the SA and DC modes do not converge to an optimal solution in the given timeframe (we set a 600 seconds cutoff time) for all the Monte Carlo trials carried out. This reflects the complexity introduced by the MRT constraint and a possible requirement to re-dimension the network so as to be able to accomodate the rate requirements for any given topology. We refer the reader to Section VII for further details on the network re-dimensioning aspects.

Further, in Figs. 14(a)-(d) and Figs. 15(a)-(b), we also highlight an exception to the generic trend stated above. The Path latency (CPL) constraint when imposed on SA and DC leads to a faster search time as compared to their respective SA and DC counterparts in most scenarios. This is due to the fact that while most constraint combinations in our work lead to an increasingly complex search space, and hence an increased convergence time as corroborated by our results in Figs. 14 and 15, the addition of path latency constraint creates a cut in the solution hyperspace that reduces the overall complexity of the search space and consequently the convergence time. This is also indicative of the fact that very few APs in the topology are actually able to satisfy the path latency constraint

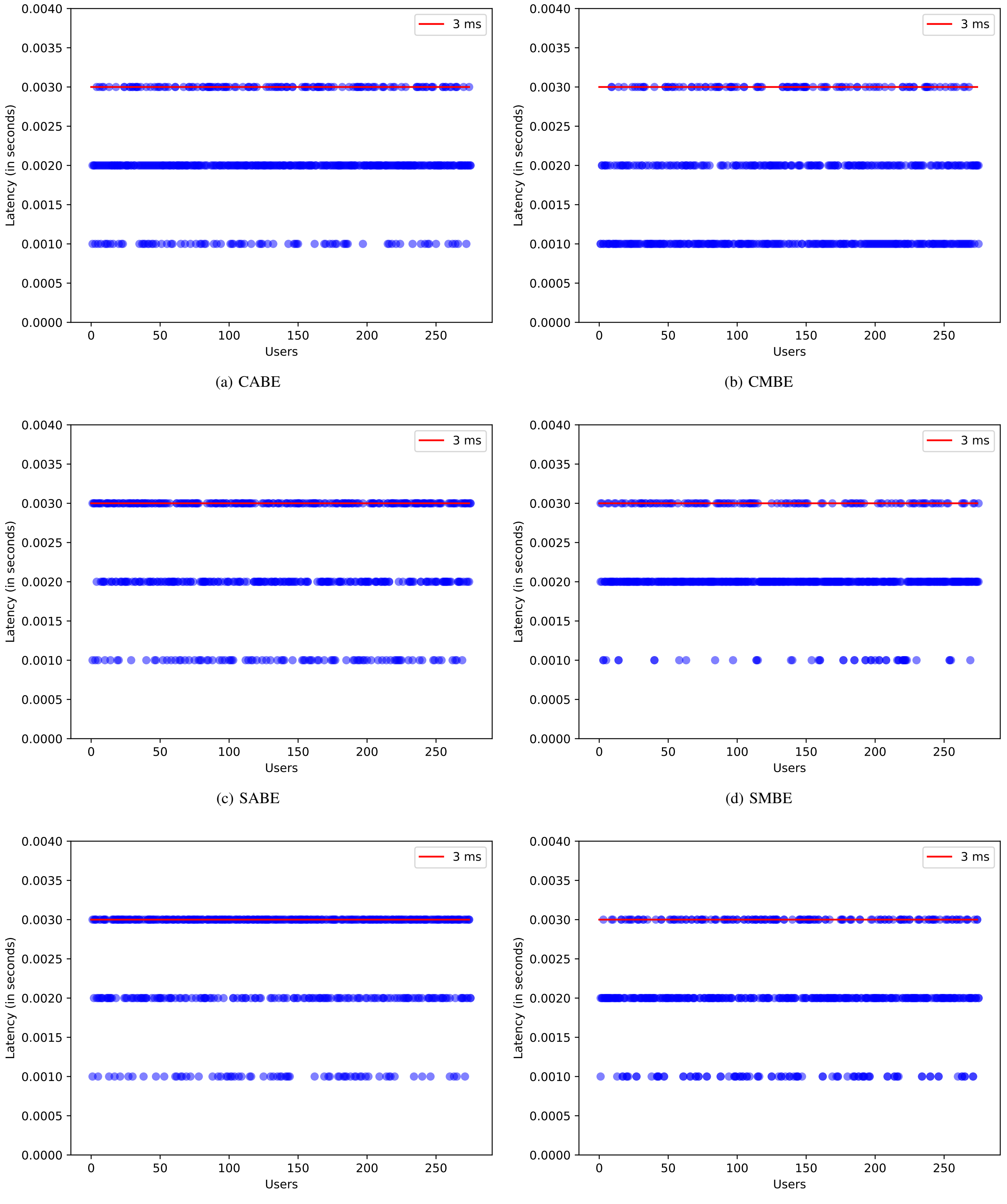

(a) CABE (b) CMBE

(c) SABE (d) SMBE

(e) CAIEm (f) CMIEm

Fig. 13. Observed Latency for (a) CABE, (b) CMBE, (c) SABE, (d) SAIE, (e) CAIEm and (f) CMIEm scenarios.

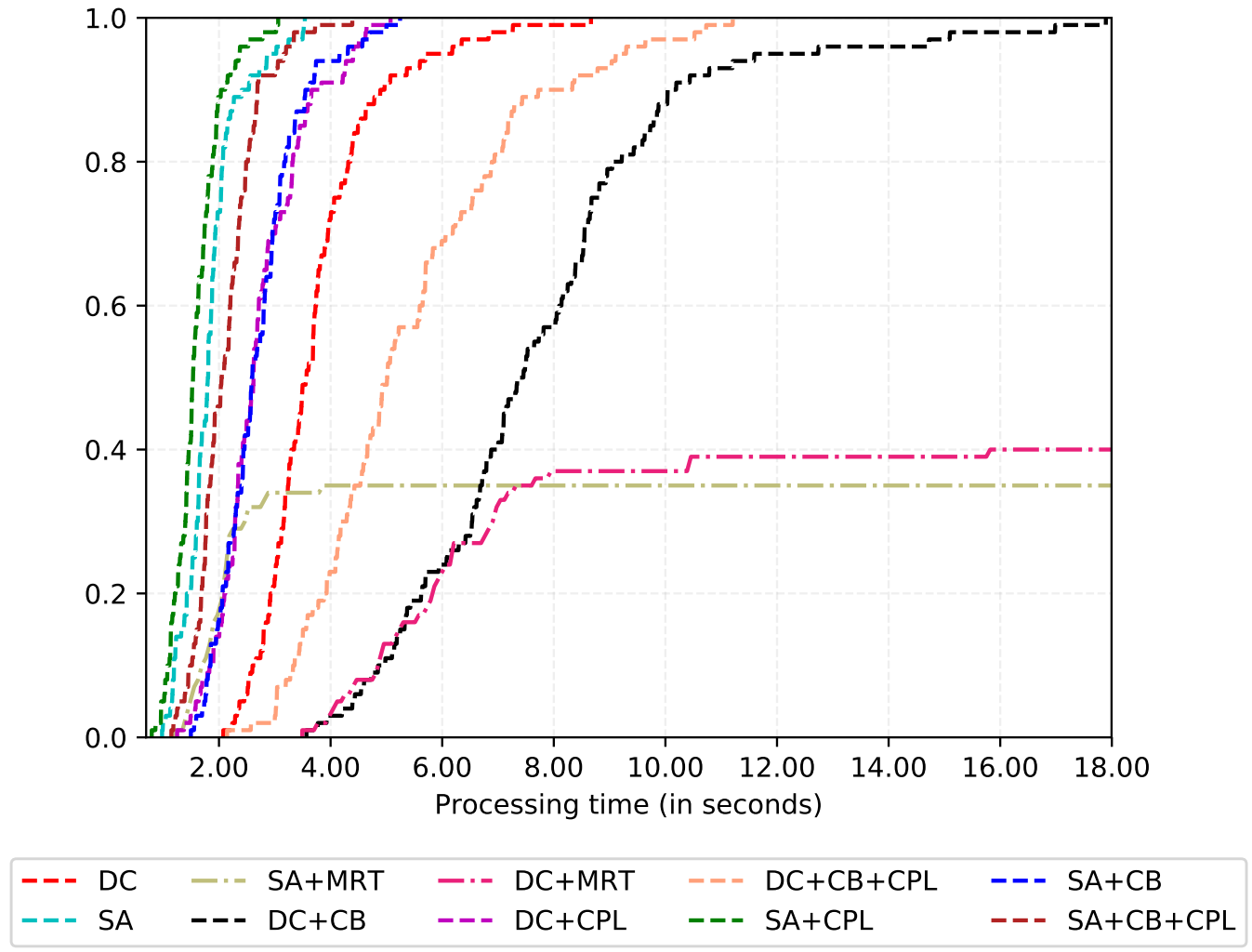

(a) CABE

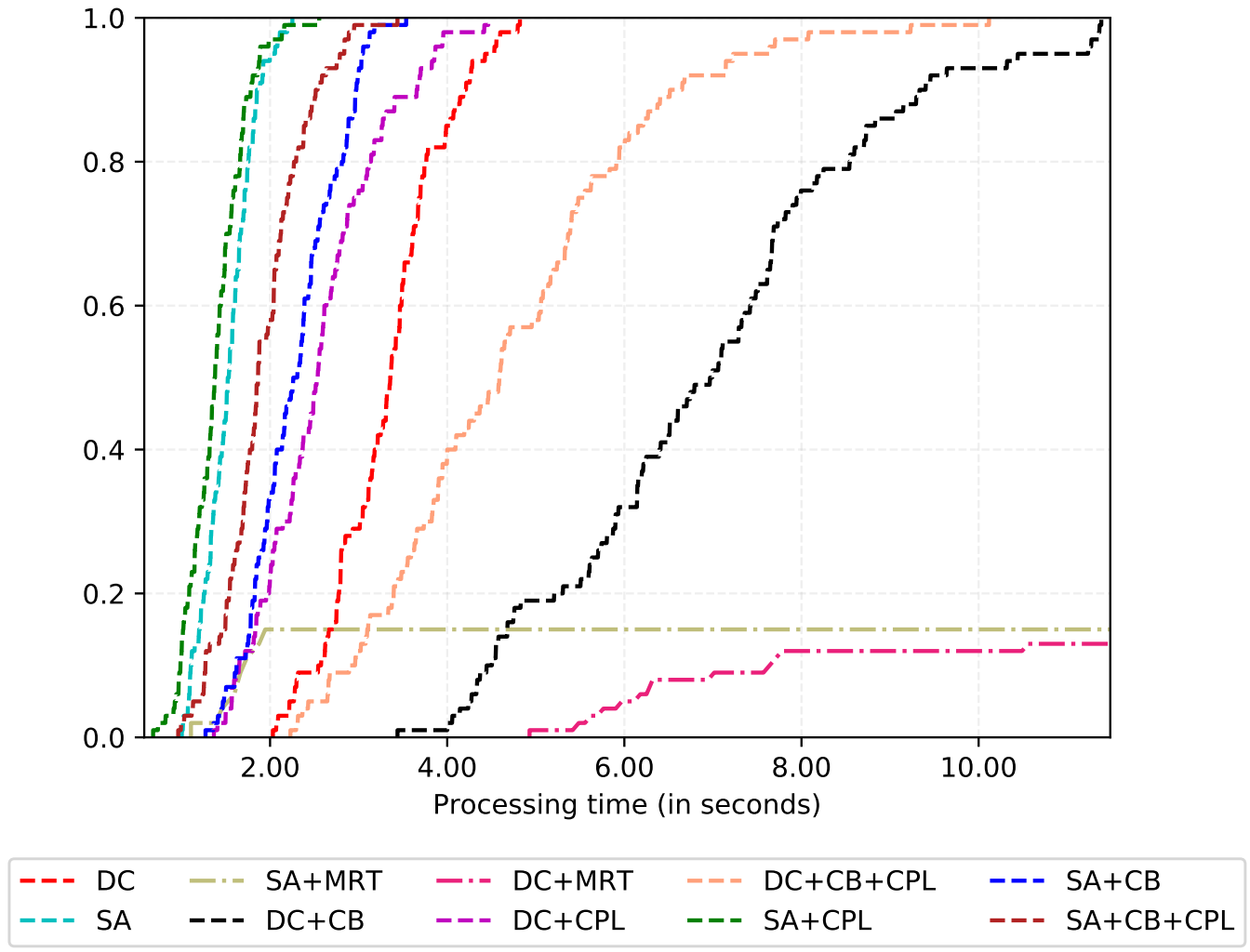

(b) CAIE

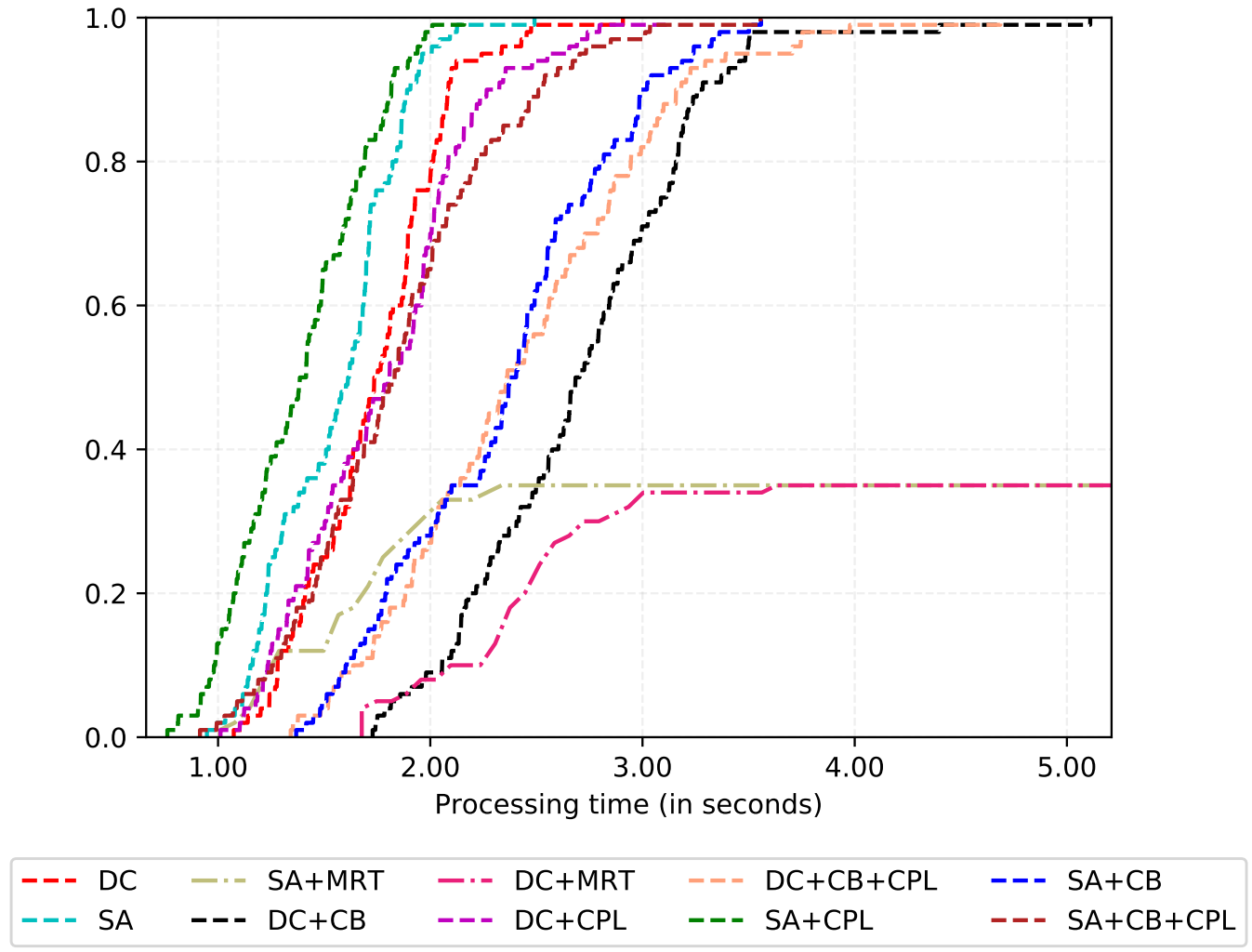

(c) CMBE

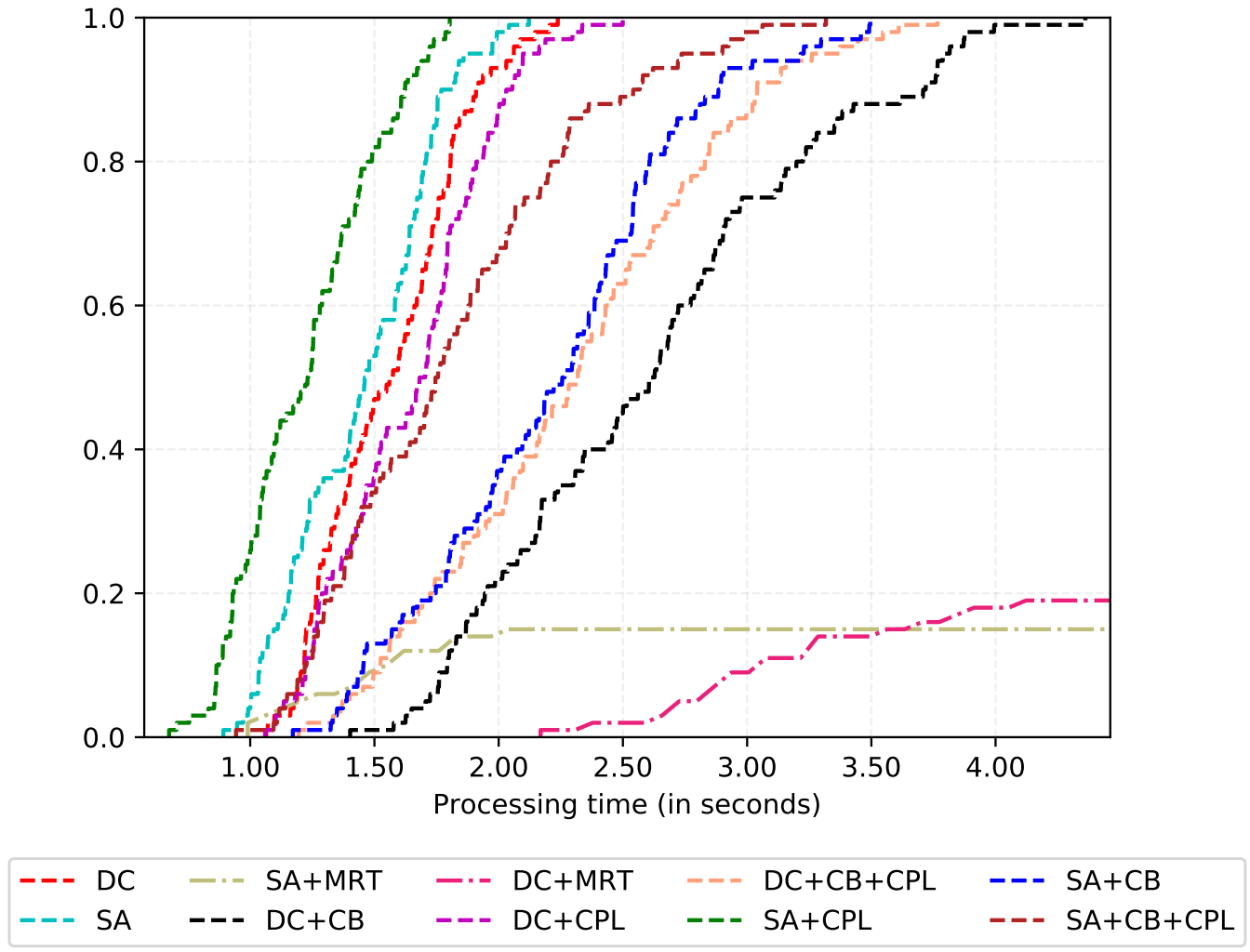

(d) CMIE

Fig. 14. Convergence time CDF (Empirical) for (a) CABE, (b) CAIE, (c) CMBE and (d) CMIE scenarios.

when imposed in combination with other network constraints. Thus, this gives an insight into the contrasting behavior of different constraints, and their overall impact on the system performance.

Lastly from Figs. 15(c) and (d) wherein the mMTC services are considered as well, it can be observed that most of the iterations for the scenarios, in which the backhaul capacity is constrained, do not converge to an optimal solution in the stipulated time. This is so because the mMTC services place an additional burden on the backhaul by consuming a portion of their available capacity. This, as a result, leads to a more challenging scenario for the AURA-5G framework to determine an optimal solution as the eMBB services have less amount of available backhaul capacity. Consequently, we observe the non-convergent behavior of the scenarios with the backhaul capacity constraint.

### *F. Solvability Analysis*

In Section VI.E we observed that certain scenarios with backhaul capacity and minimum rate constraints do not converge to an optimal solution in the cutoff time period of 600 seconds, as defined in our evaluation framework. However, it might very well be possible that either a solution does not exist or the optimizer got timed out, i.e. it might or might not have a feasible solution, but, it was not able to determine the same up until the 600 seconds timeframe. Hence, with this background, in this section we undertake a solvability analysis with the specified time limit parameters and aim to understand the bottleneck constraints for the AURA-5G framework, given the various scenarios we have studied.

For the solvability analysis we introduce Figs. 16(a)-(f), wherein we have also provisioned an analysis for the most complex combination of constraints, i.e. CB + MRT, CPL + MRT and CB + MRT + CPL (see Table V for description). It must be highlighted that the analysis provisioned here is for the case when we consider 275 eMBB users in the system, i.e. the maximum number of users evaluated within our evaluation framework. From Figs. 16(a)-(f) we observe that for all the scenarios explored, the Minimum Rate (MRT) constraint behaves as a bottleneck constraint for the optimizer in the AURA-5G framework. This is also reflected from the time convergence plots in Figs. 14 and 15. The reason being that there is limited access bandwidth available. In addition, given the nature of the scenario, i.e. if its beamformed or interference limited, the SINR characteristics transform and subsequently impact the decision of the optimization framework. Such system based variability in SINR also limits the achievable per user rate, hence, rendering the MRT constraint as a bottleneck.

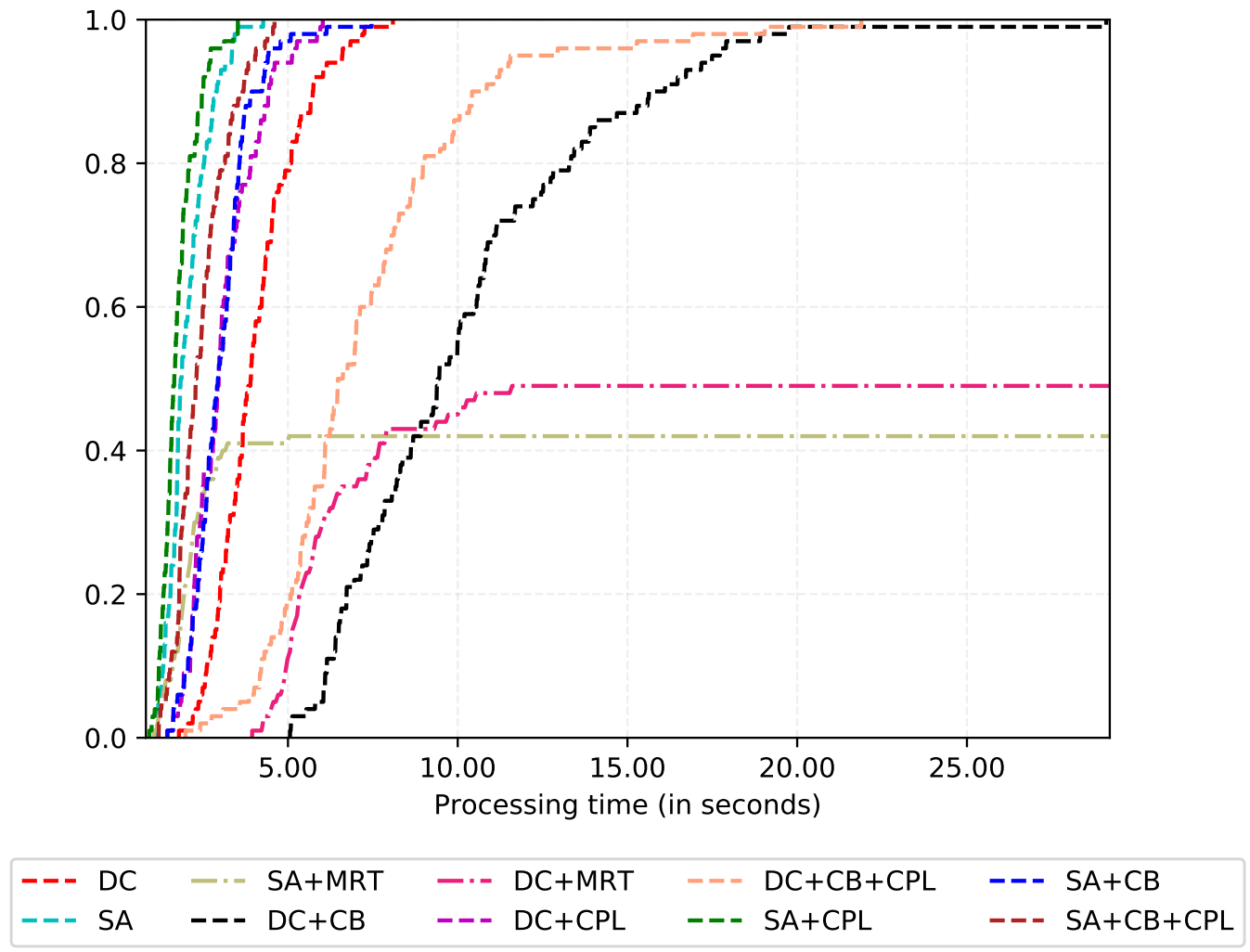

(a) SABE

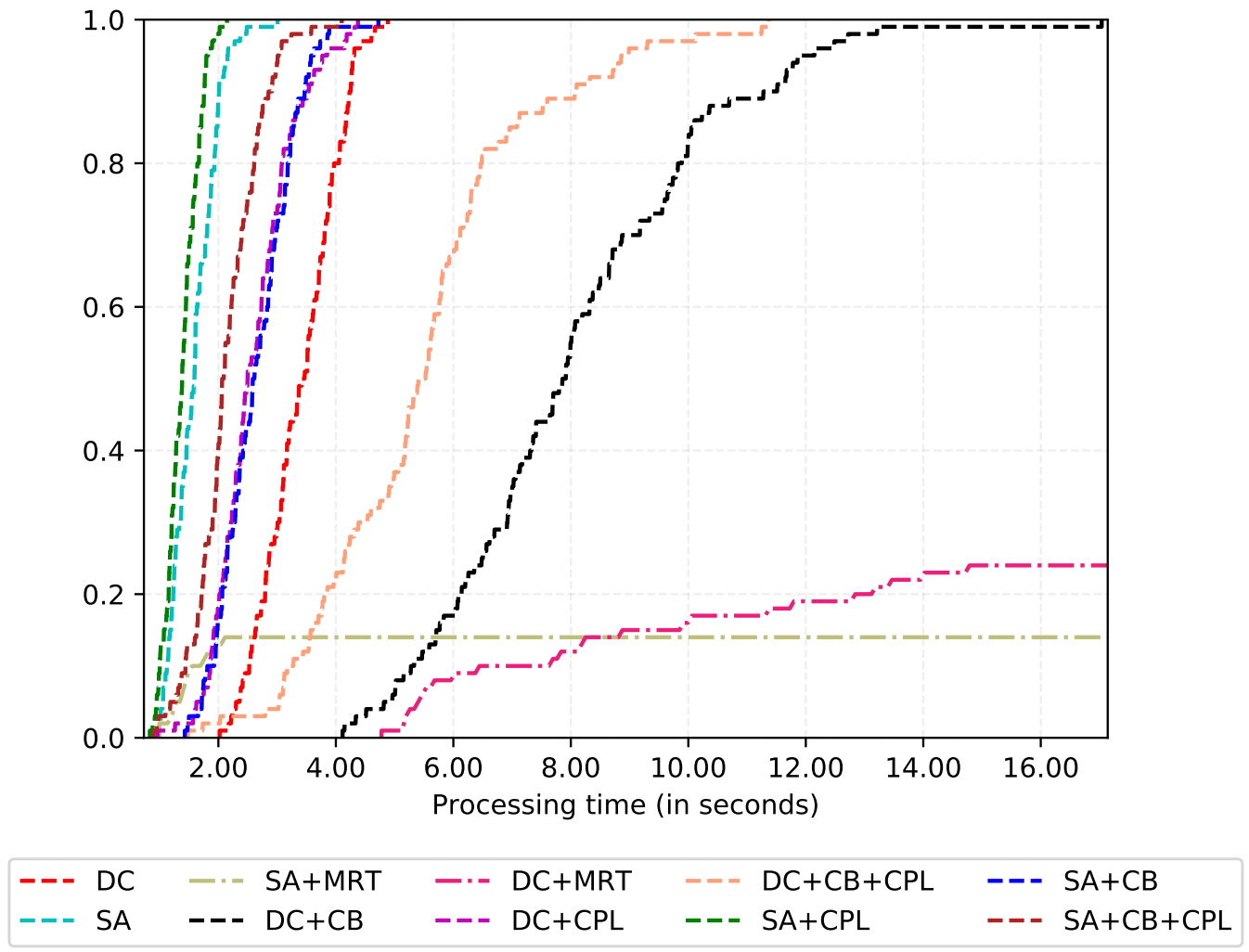

(b) SAIE

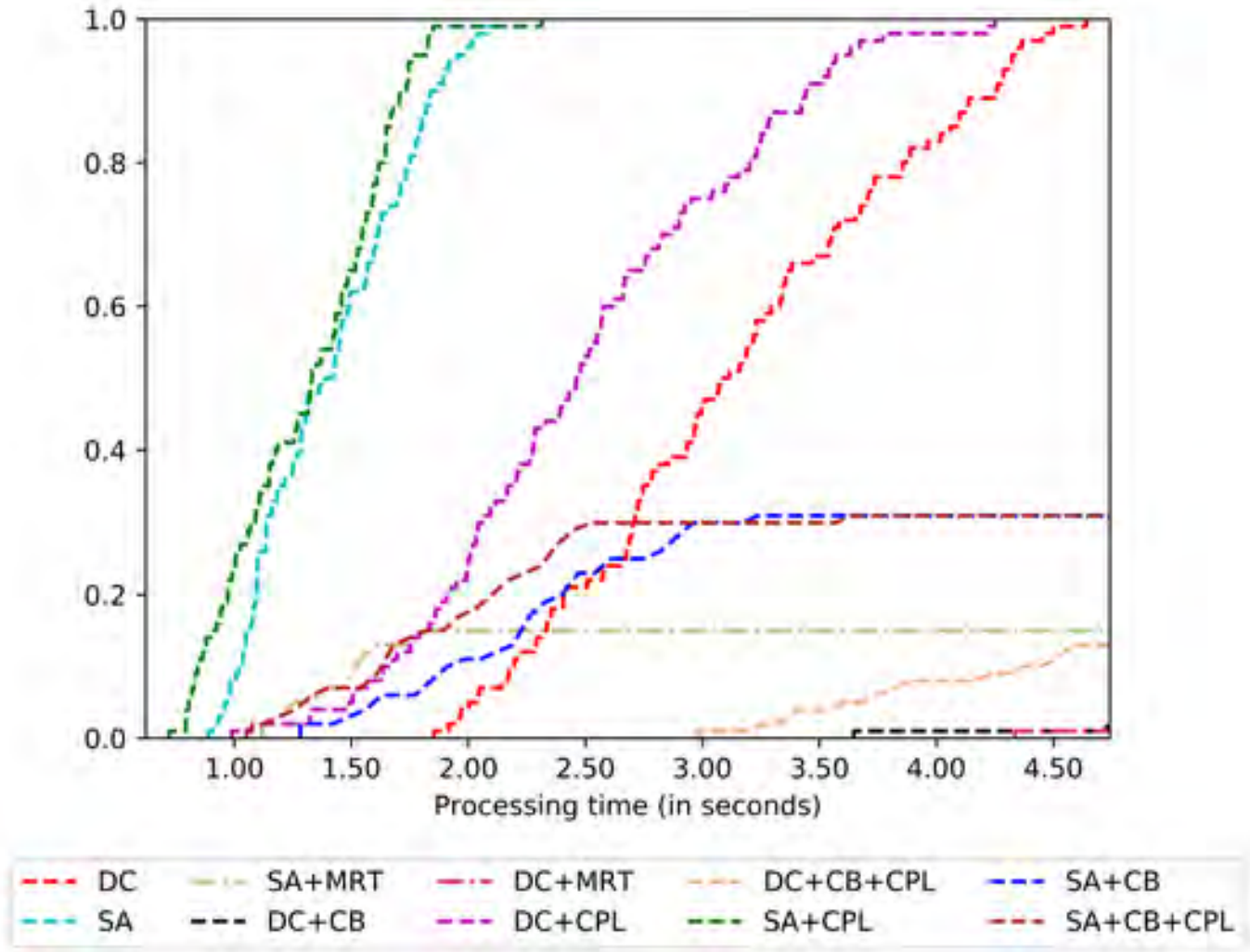

(c) CAIEm

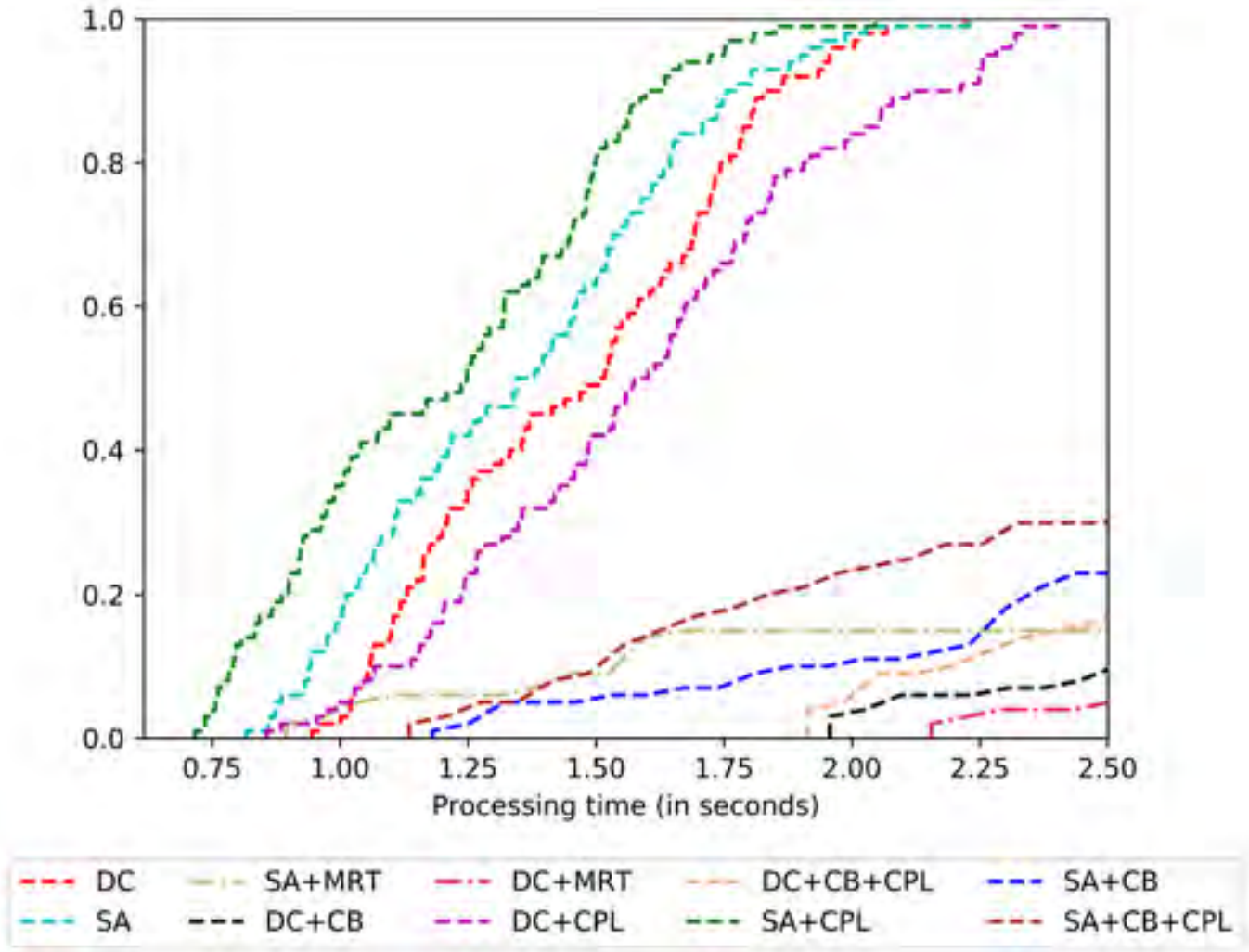

(d) CMIEm

Fig. 15. Convergence time CDF (Empirical) for (a) SABE, (b) SAIE, (c) CAIEm and (d) CMIEm scenarios.

Further, from Figs. 16(a)-(f) we see that in the interference limited regime the optimizer performance is much more severely affected as compared to the beamformed scenario, which is in line with the rest of our analysis so far. Moreover, for the square deployment scenario (Figs. 16(e) and (f)) the backhaul constraints are even more restrictive given the fact that the probability of the SC APs being more distant from the MC is higher. Hence, the probability of having a wired backhaul, which has a 1 Gbps capacity and is in most cases much lower than what is offered by the mmWave wireless backhaul, is also subsequently higher. As a consequence, from Figs. 16(a), (c) and (e) it can be deduced that for the square deployment scenario with CB and MRT constraint, the system performance is impacted severely with at least 10 more iterations without a solution compared to those in the circular deployment scenario.

Next, when we observe the scenarios wherein both the mMTC and eMBB services have been considered (Figs. 17(a) and (b)), it can be seen that the backhaul capacity constraint also emerges as a bottleneck. This is corroborated from the time convergence curves in Figs. 15(c) and (d), where the scenarios with the CB constraints do not illustrate convergence for all the iterations. This is because of the fact that the

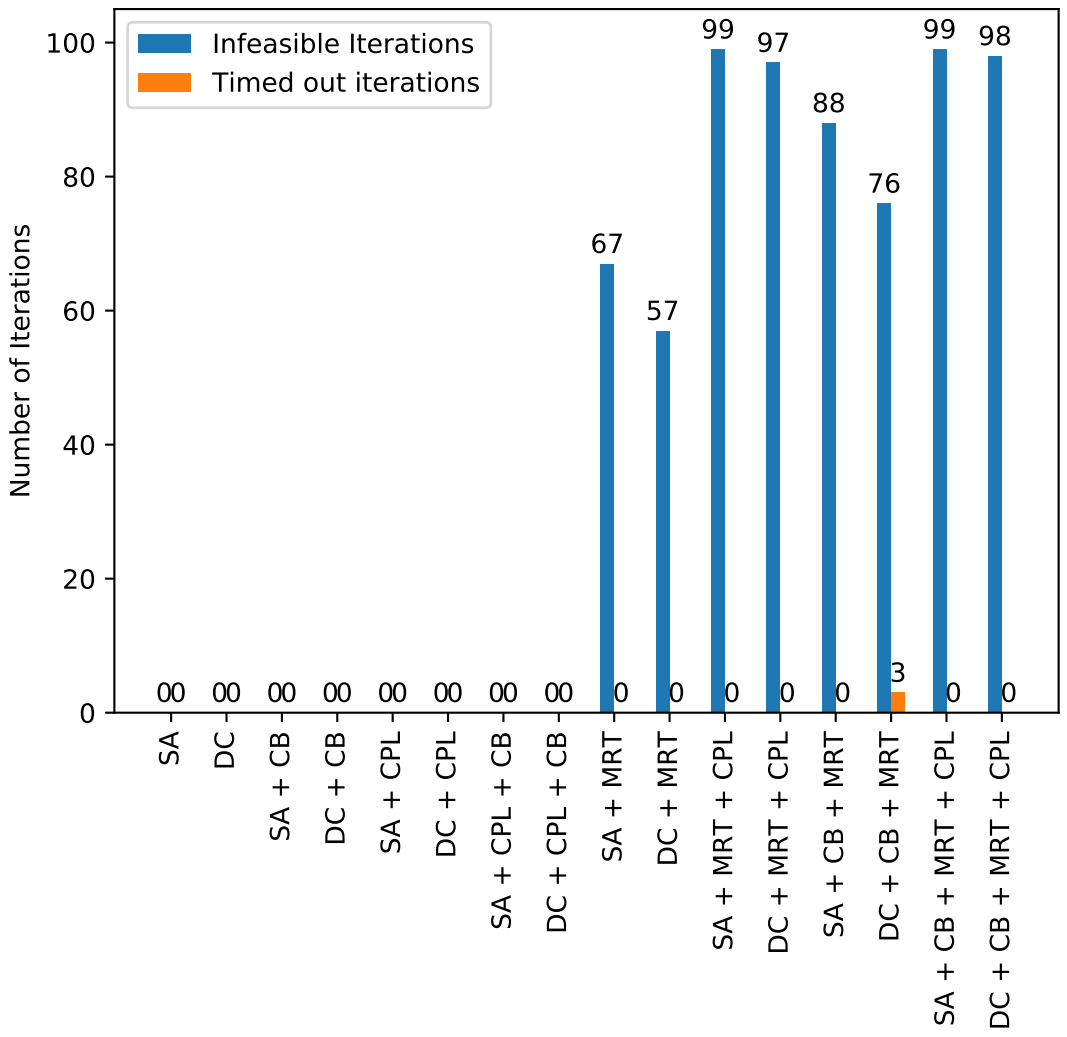

(a) CABE

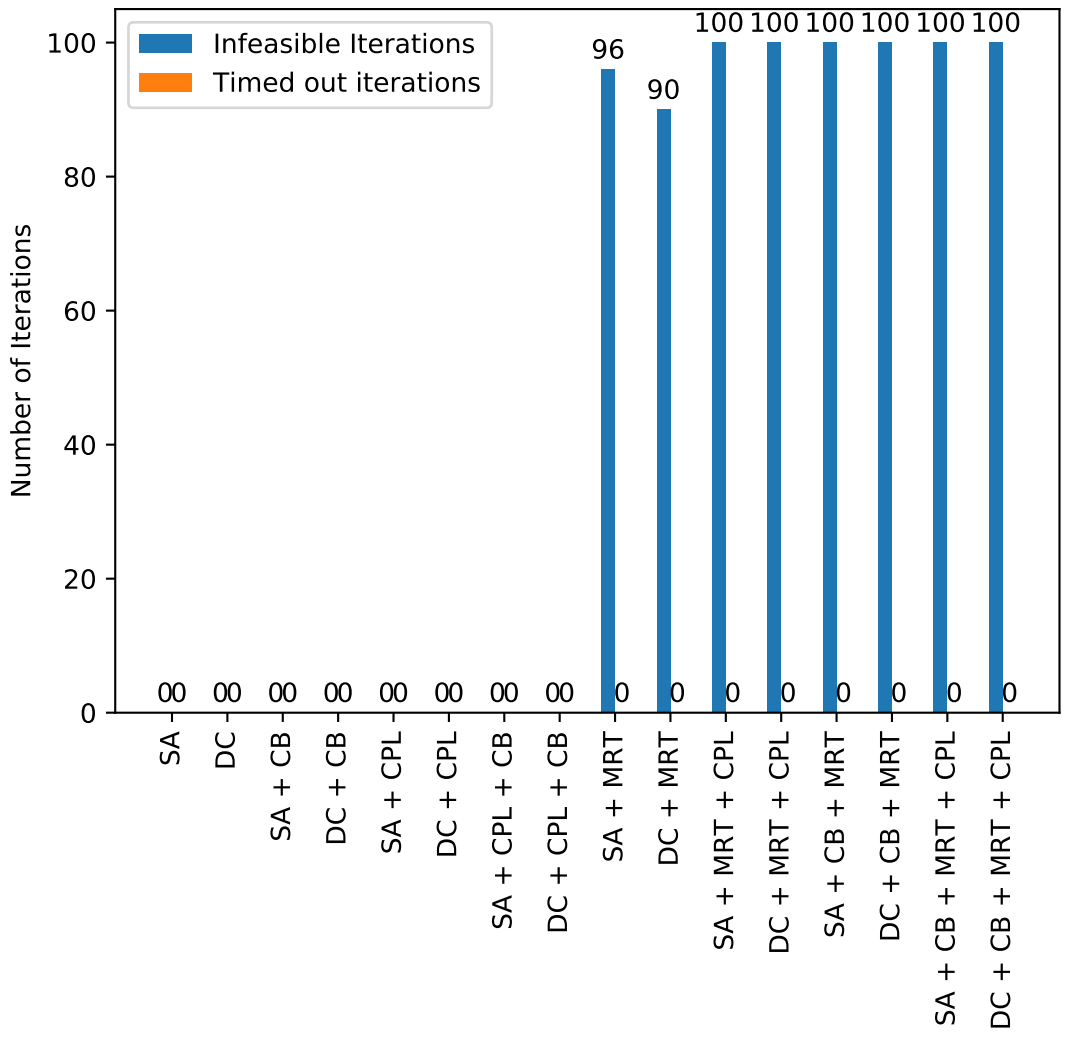

(b) CAIE

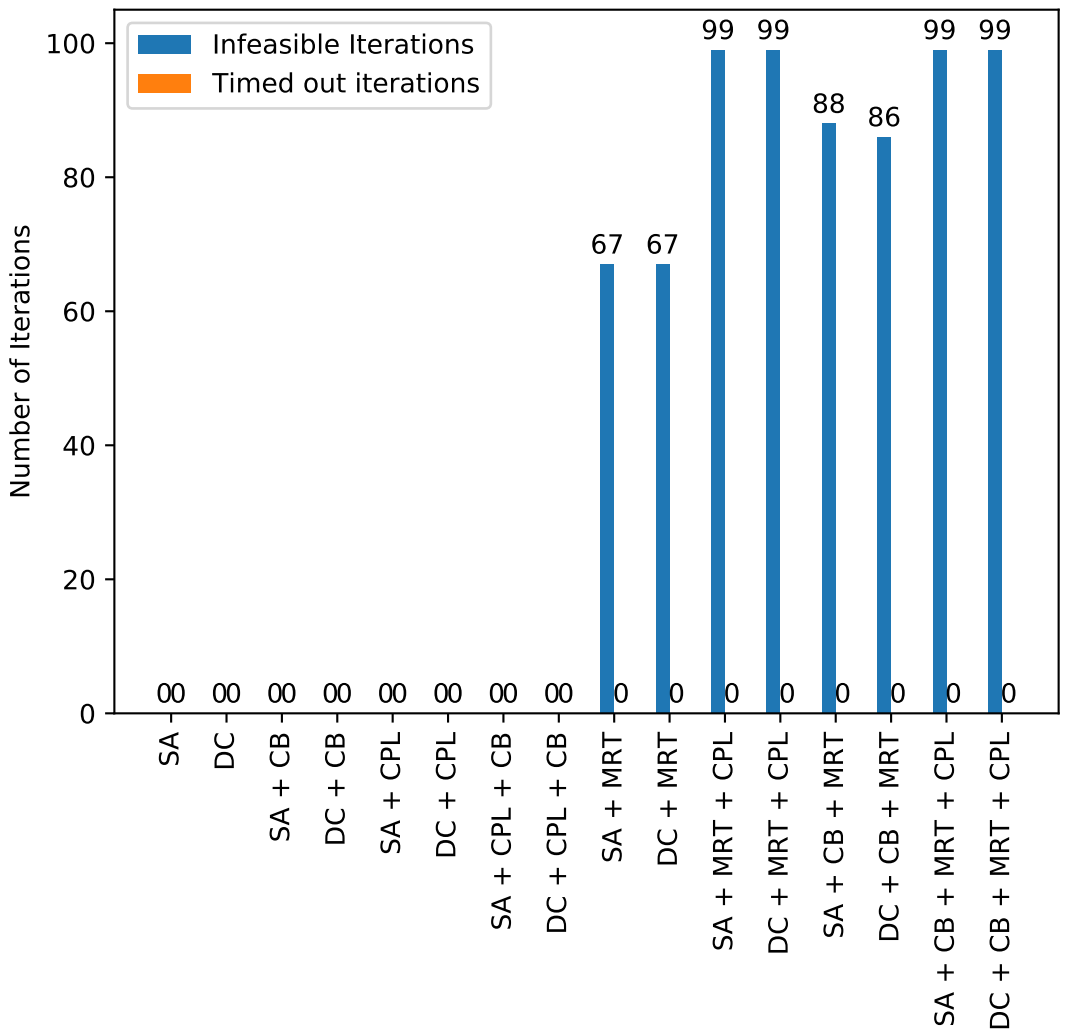

(c) CMBE

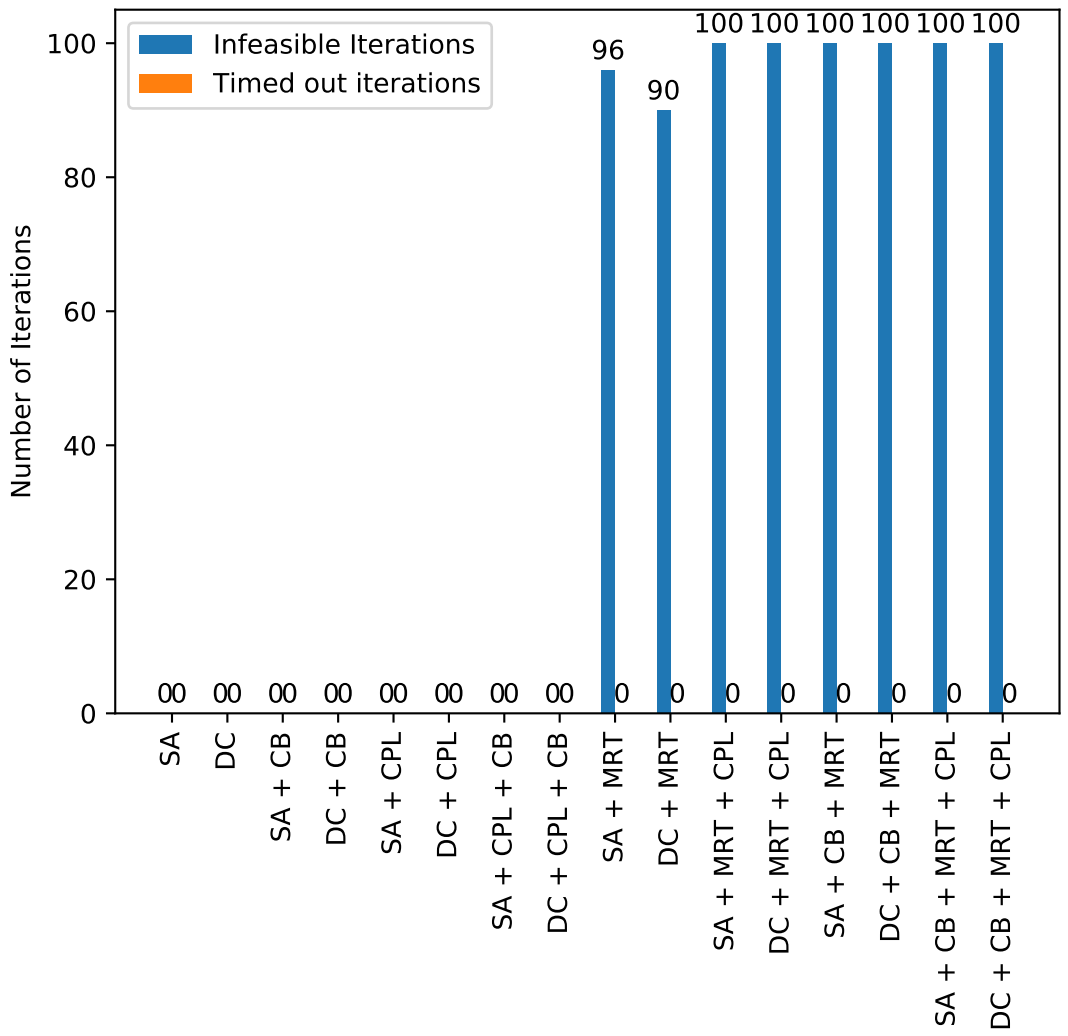

(d) CMIE

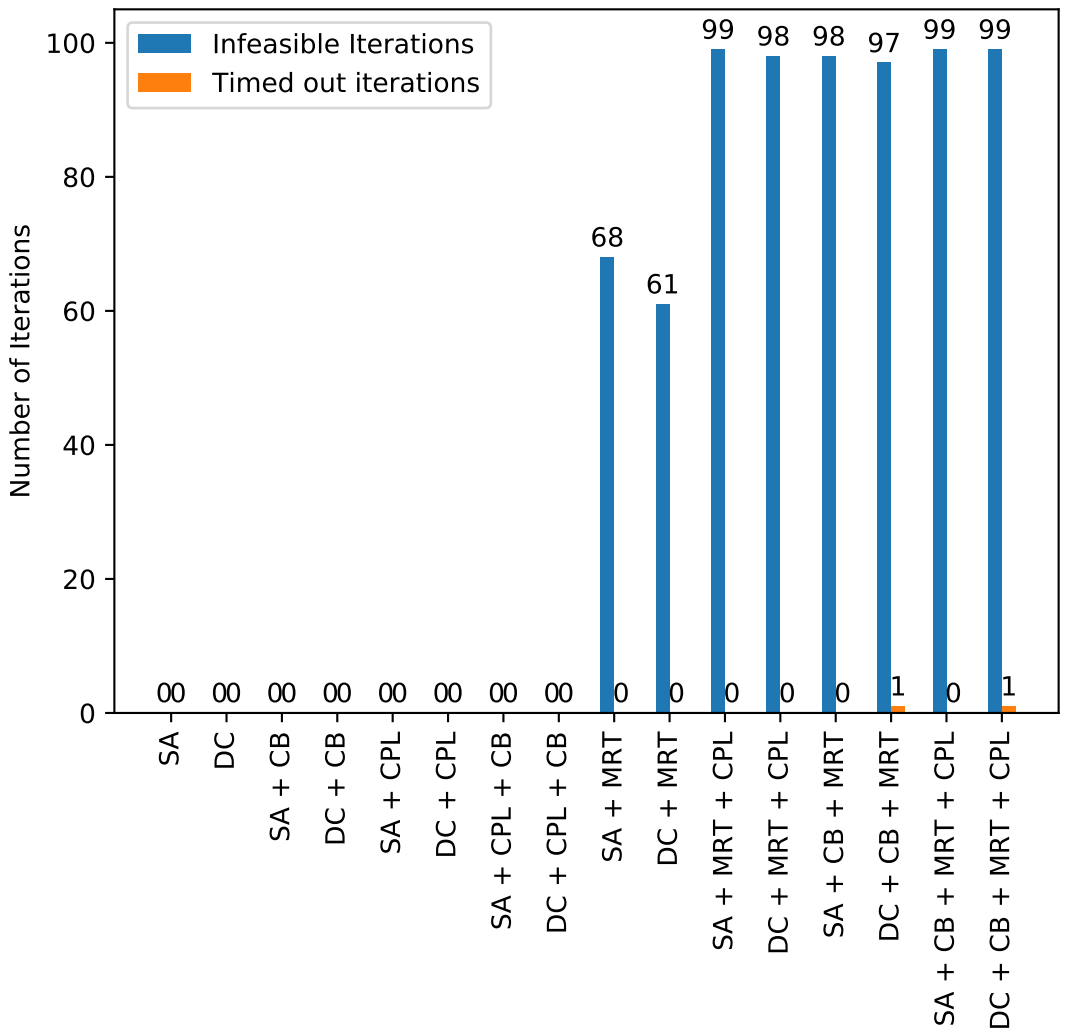

(e) SABE

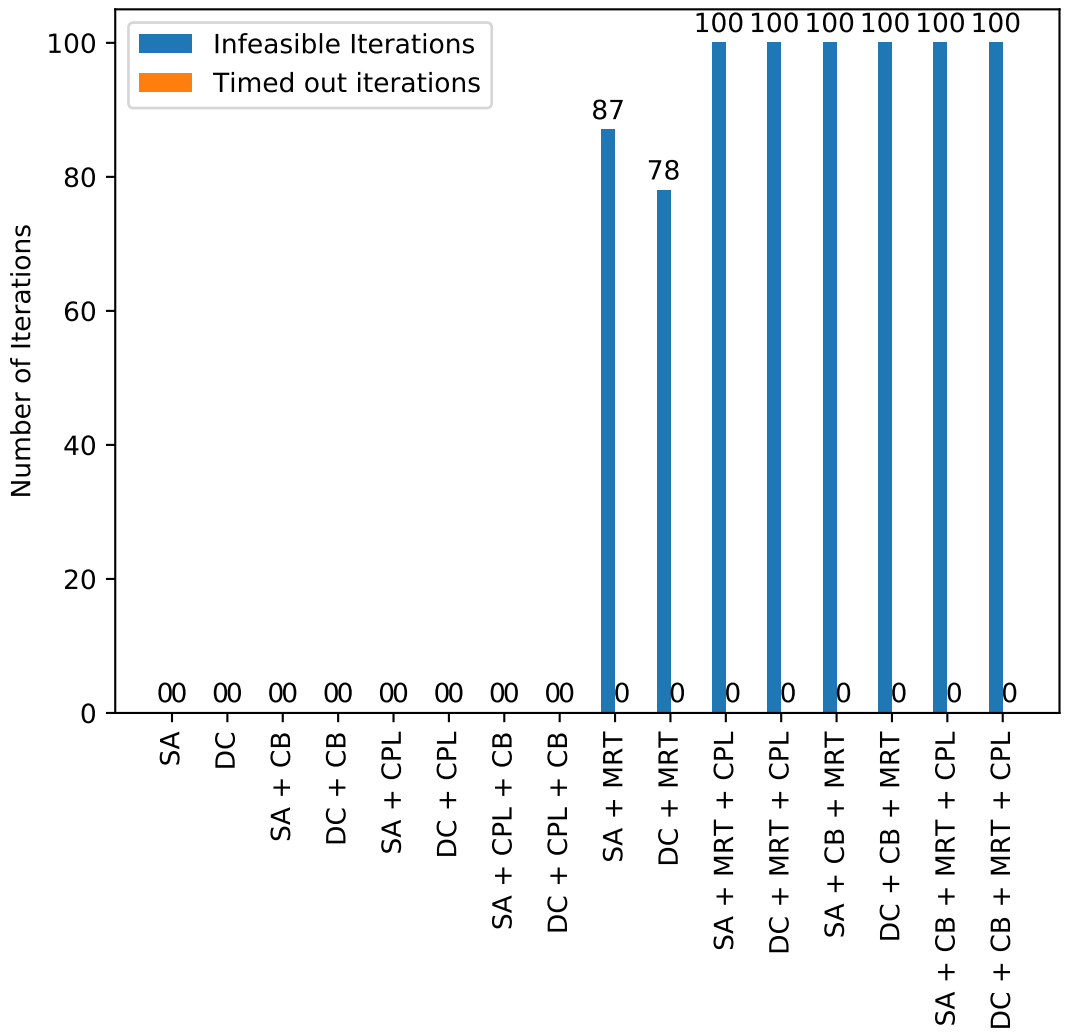

(f) SAIE

Fig. 16. Optimizer Status for (a) CABE, (b) CAIE, (c) CMBE, (d) CMIE, (e) SABE and (f) SAIE scenarios with 275 eMBB users.

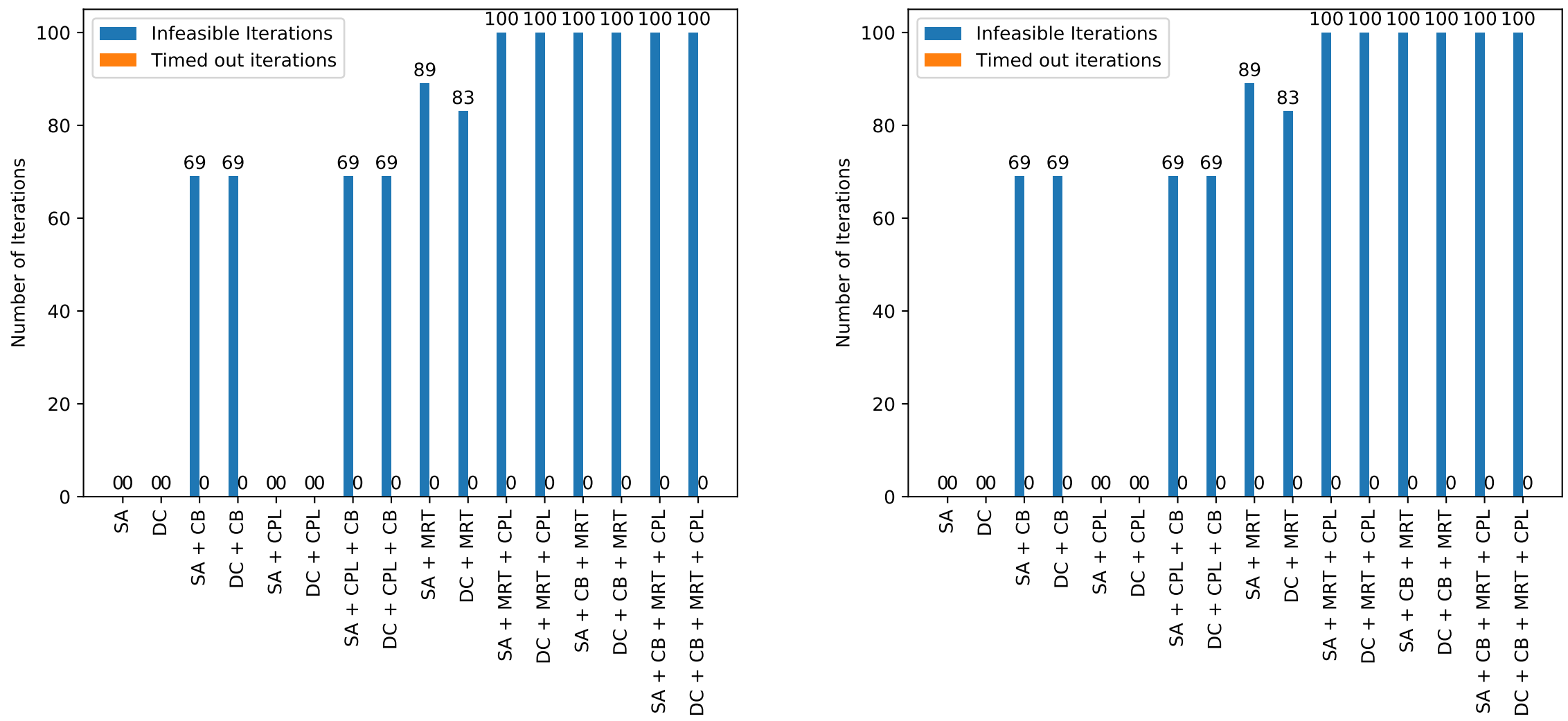

(a) CAIEm

(b) CMIEm

Fig. 17. Optimizer Status for (a) CAIEm and (b) CMIEm scenarios with 275 eMBB users.

mMTC devices consume a portion of the available backhaul capacity. Consequently, due to the reduced backhaul capacity, the optimizer finds itself in a situation wherein these very backhaul resources transform into a bottleneck.

Lastly, from Figs. 16(a)-(f) and 17(a)-(b) we deduce that as the complexity of the constraint combinations increases, the AURA-5G framework finds it increasingly challenging to determine the optimal solution. In particular, the MRT and CPL constraints appear to be fairly challenging for our user association methodology. Further, as has already been stated above, for the scenarios with mMTC services included, constraint combinations with CB also transform into being extremely challenging ones to satisfy. Consequently, in the next section we explore certain network re-dimensioning methodologies that assist the optimizer to determine an optimal association.

## VII. Network Re-dimensioning

From the analysis presented in Sections VI.E and VI.F, we have observed that certain constraint combinations for the scenarios analyzed prove to be significantly difficult for the MILP framework to satisfy. These insights can be a useful network designing tool for the operators, and subsequently they can re-dimension/upgrade their networks to meet the demands. Hence, in this section through Figs. 18-29, we discuss certain network re-dimensioning options and their corresponding results. We present the fact that re-evaluating and re-defining appropriate network parameters results in an improved performance by the AURA-5G framework. However, for our analysis and also the sake of brevity, we consider only SABEm and CABEm scenarios in this section as they encompass all the complexities of the scenarios that we have studied in this paper.

And so, the analysis presented thus far has led to the conclusion that one of the constraints that has proven to be extremely difficult to satisfy, especially when mMTC and eMBB services are considered together in the system, is the backhaul capacity constraint. Moreover, scenarios wherein *beamforming* and *AnyDC* modes have been utilized will prove to be particularly challenging, given the lack of backhaul resources and the throughput maximization nature of the optimizer. In addition, due to the lack of access resources as well as the prevailing SINR characteristics, the MRT constraint also imposes a severe challenge for the AURA-5G framework. Hence, through Figs. 18-23, we analyzed scenarios with a re-dimensioned backhaul and access network wherein both mMTC and eMBB users are considered alongside the circular and square deployment, and the beamformed and AnyDC regime.

For the re-dimensioning we firstly calculated the average amount of backhaul utilized in all the SCs when no re-dimensioning is done for the scenario under study. Next, we increase the backhaul capacity of all the SCs in the system by a percentage of this average consumption. For the percentage increment we utilized four quantized levels, i.e. 30%, 50%, 80% and 100%. Subsequently, and to account for the worst case scenario, we increment the capacity of the backhaul for each MC by 10 times the aforementioned average SC backhaul utilization. The factor of 10 arises from the fact that in our evaluation framework the maximum number of supported SCs by an MC is also 10. Next, we re-dimension the access network by increasing the average number of SCs per MC in the topology from 6-7 (uniform distribution of 3 to 10 SCs per MC) to 8 (uniform distribution of 6 to 10 SCs per MC). This, automatically provisions more access network resources, in terms of bandwidth, as well as increases the likelihood for a user to find an SC in close proximity. It is important to state here that, we maintain the receive beam angle and beamforming gain. Whilst, these can be exploited to improve the performance of the system further, it might lead to increased capital/operating expenditure for the operator given required infrastructure overhaul, such as antenna replacement, etc. Hence, we leave the discussion on this aspect for a sequel

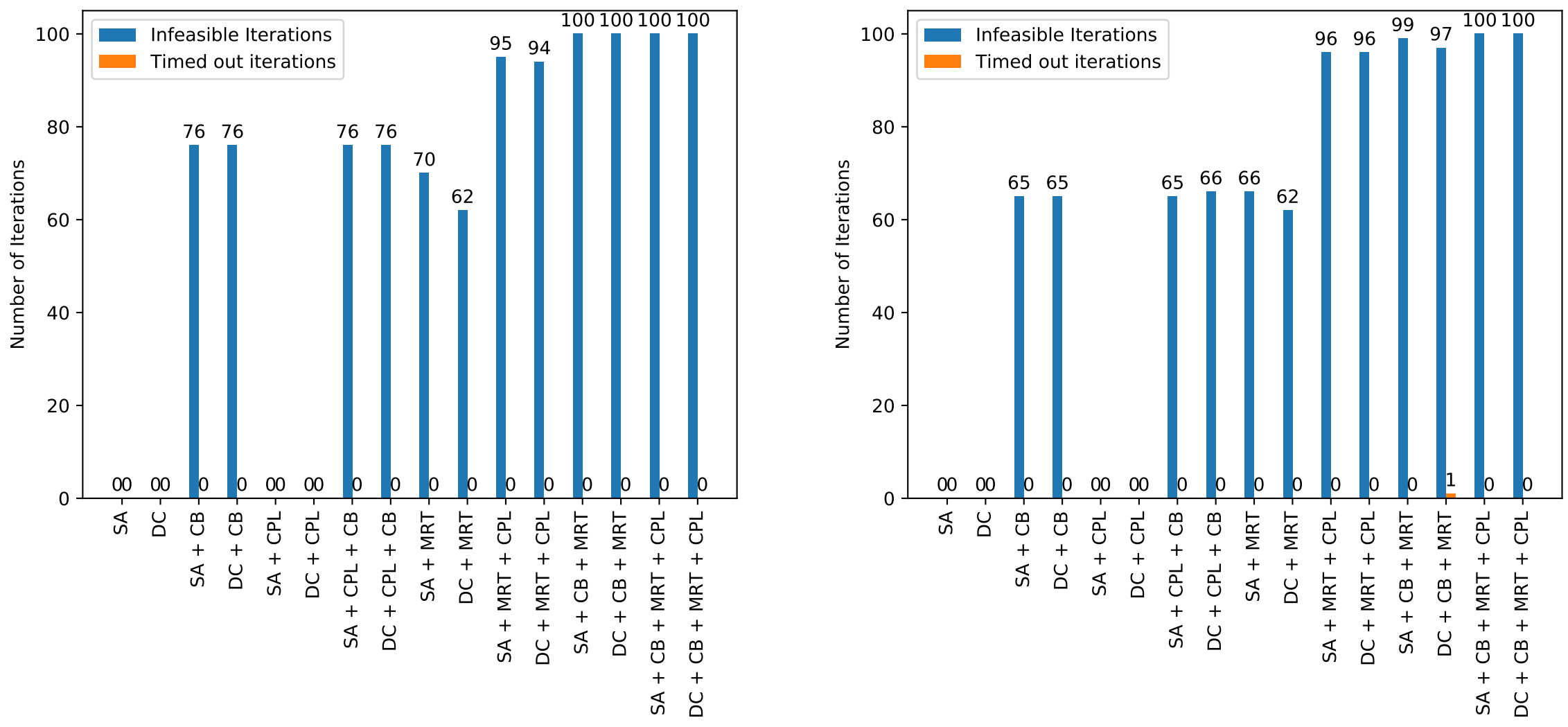

(a) SABEm without Relaxed Backhaul (b) SABEm with Relaxed Backhaul

Fig. 18. Optimizer Status for (a) SABEm without Relaxed Backhaul, and (b) SABEm with Relaxed Backhaul scenarios with 275 eMBB users.

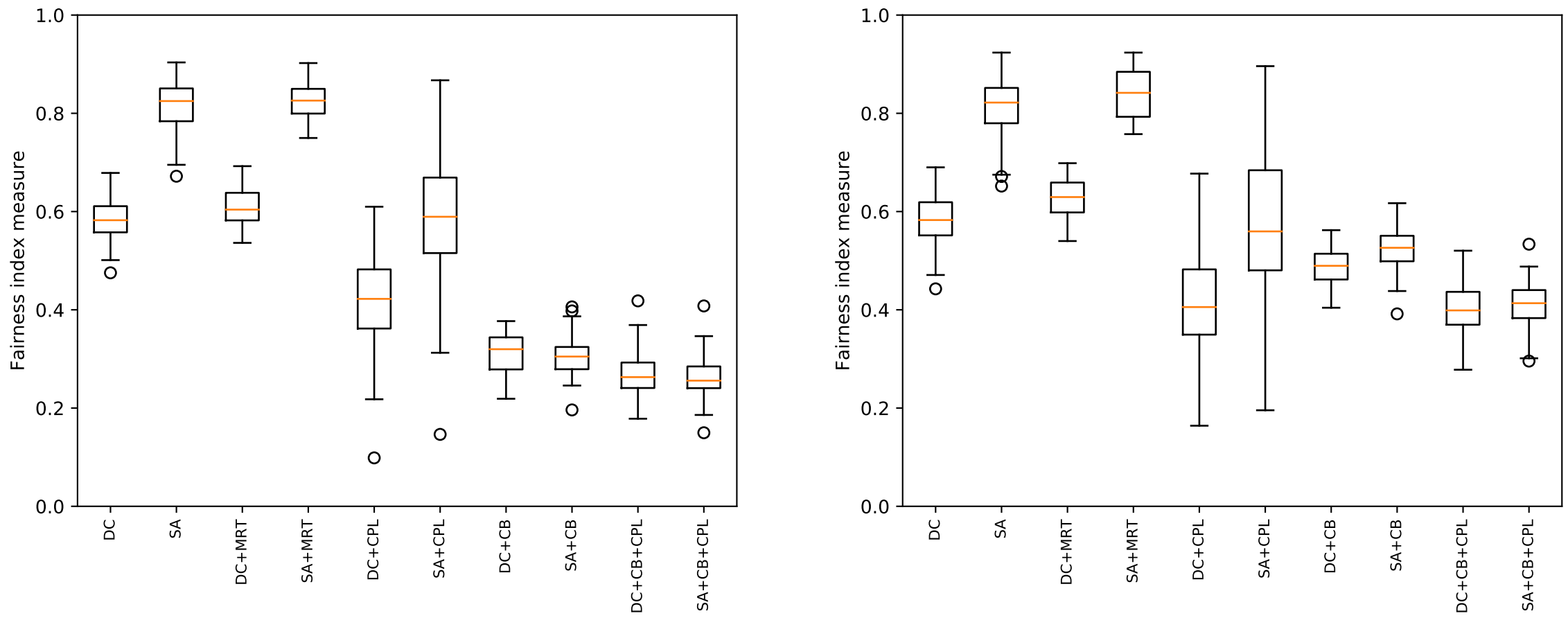

(a) SABEm without Relaxed Backhaul (b) SABEm with Relaxed Backhaul

Fig. 19. System Fairness Measure for (a) SABEm without Relaxed Backhaul, and (b) SABEm with Relaxed Backhaul scenarios with 275 eMBB users.

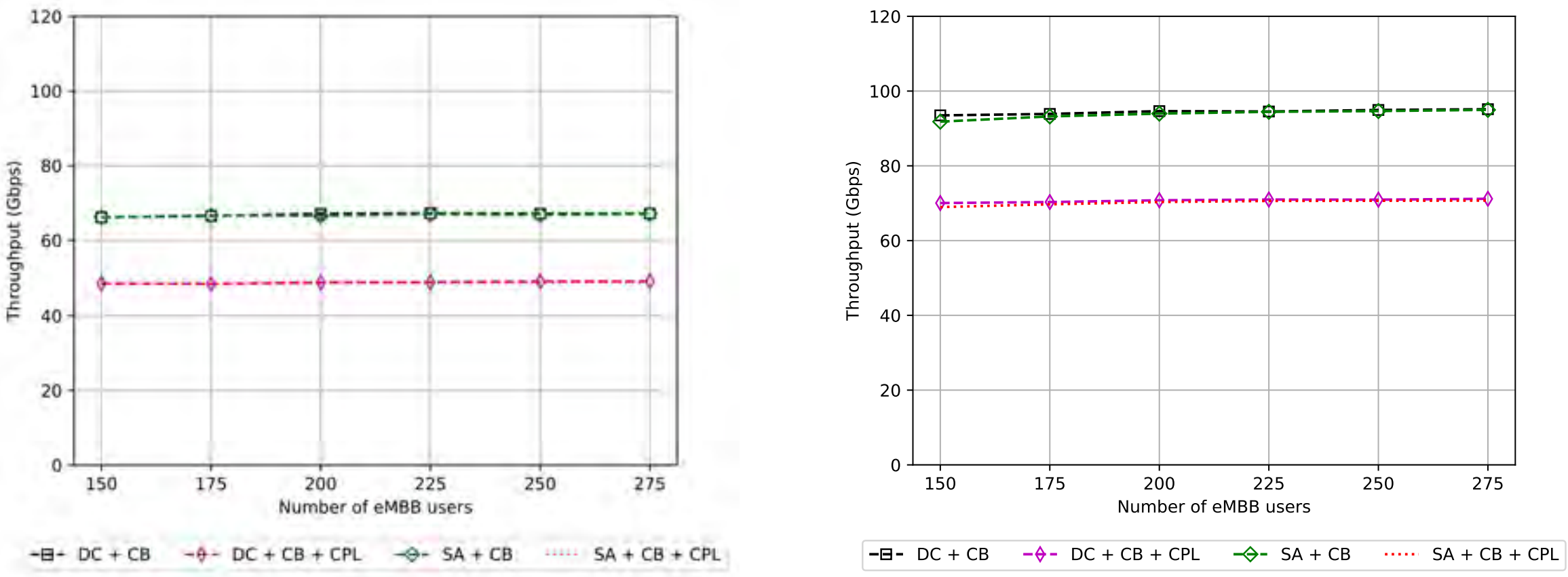

(a) SABEm without Relaxed Backhaul (b) SABEm with Relaxed Backhaul

Fig. 20. Total Network Throughput for (a) SABEm without Relaxed Backhaul, and (b) SABEm with Relaxed Backhaul scenarios with 275 eMBB users.

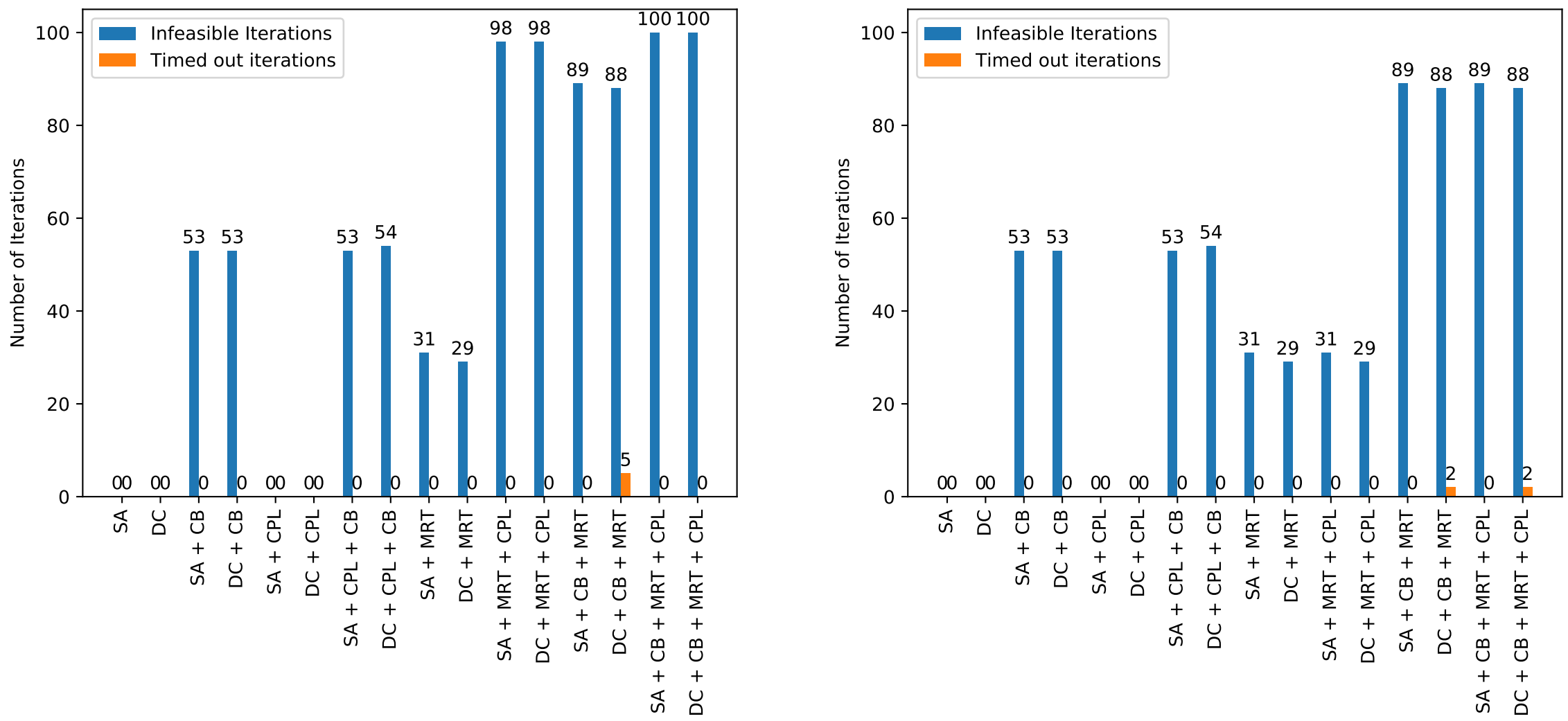

(a) SABEm with Relaxed BH and Increased SC density

(b) SABEm with Relaxed BH, Increased SC density and 5ms latency requirement

Fig. 21. Optimizer Status for (a) SABEm with Relaxed Backhaul and Increased SC density scenario with 275 eMBB users, and (b) SABEm scenario with Relaxed Backhaul, Increased SC density, 5 ms downlink latency requirement and 275 eMBB users.

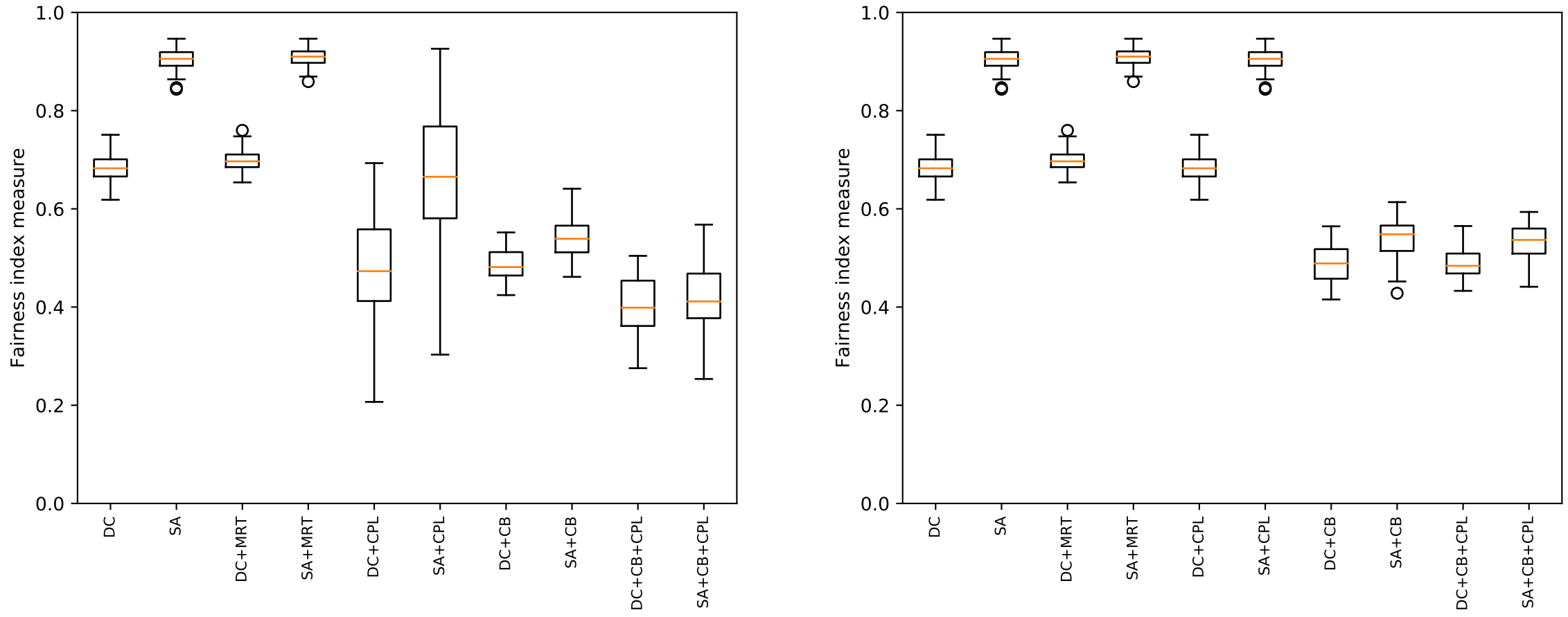

(a) SABEm with Relaxed BH and Increased SC density

(b) SABEm with Relaxed BH, Increased SC density and 5ms latency requirement

Fig. 22. System Fairness Measure for (a) SABEm with Relaxed Backhaul and Increased SC density scenario with 275 eMBB users, and (b) SABEm scenario with Relaxed Backhaul, Increased SC density, 5 ms downlink latency requirement and 275 eMBB users.

work.

Consequently from the analytical results in Figs. 18(a)-(b) and 21(a), we observe that when the backhaul capacities are enhanced by the methodology explained above, the scenarios where backhaul was a constraint ceases to be so anymore. For example, in Fig. 18(a), constraint combinations *only* CB, and CB + CPL highlight the fact that the backhaul capacity is a constraint for the scenario under study, i.e. SABEm. Hence, by the re-dimensioning employed as specified in our work, through Fig. 18(b), we observe that the number of iterations that converge for the *only* CB and CB + CPL constraint combinations increases by 14.47%. Furthermore, when we employ the increased SC density framework to provision more access network resources, the percentage improvements in the number of converged iterations, as can be seen in Fig. 21(a), for constraint combinations CB, CB + CPL, MRT and CB + MRT are at 30.2%, 30%, 55% and 11%, respectively . As a result, to a great extent the re-dimensioning performed, according to the guidelines specified above, helps in alleviating the bottlenecks that hampered the AURA-5G framework earlier.

In addition to the solvability analysis, discussed above, we see that the re-dimensioning efforts result in an increase in the system fairness. This is more prominent for constraint combinations CB and CB + CPL, as understood from Figs. 19(a) and (b), while from Fig. 22(a) we deduce that a re-dimensioned access network topology leads to an across the board positive effect on the system fairness. The positive effects of network re-dimensioning are also prevalent in the

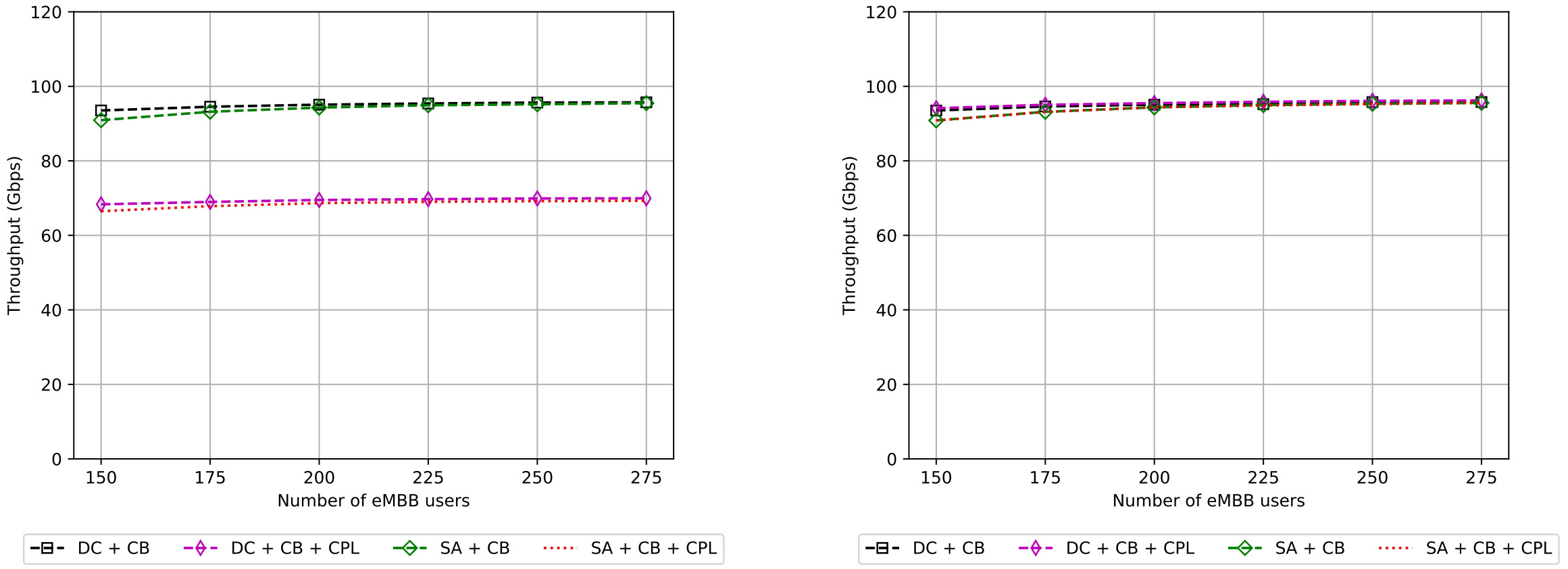

(a) SABEm with Relaxed BH and Increased SC density

(b) SABEm with Relaxed BH, Increased SC density and 5ms latency requirement

Fig. 23. Total Network Throughput for (a) SABEm with Relaxed Backhaul and Increased SC density scenario with 275 eMBB users, and (b) SABEm scenario with Relaxed Backhaul, Increased SC density, 5 ms downlink latency requirement and 275 eMBB users.

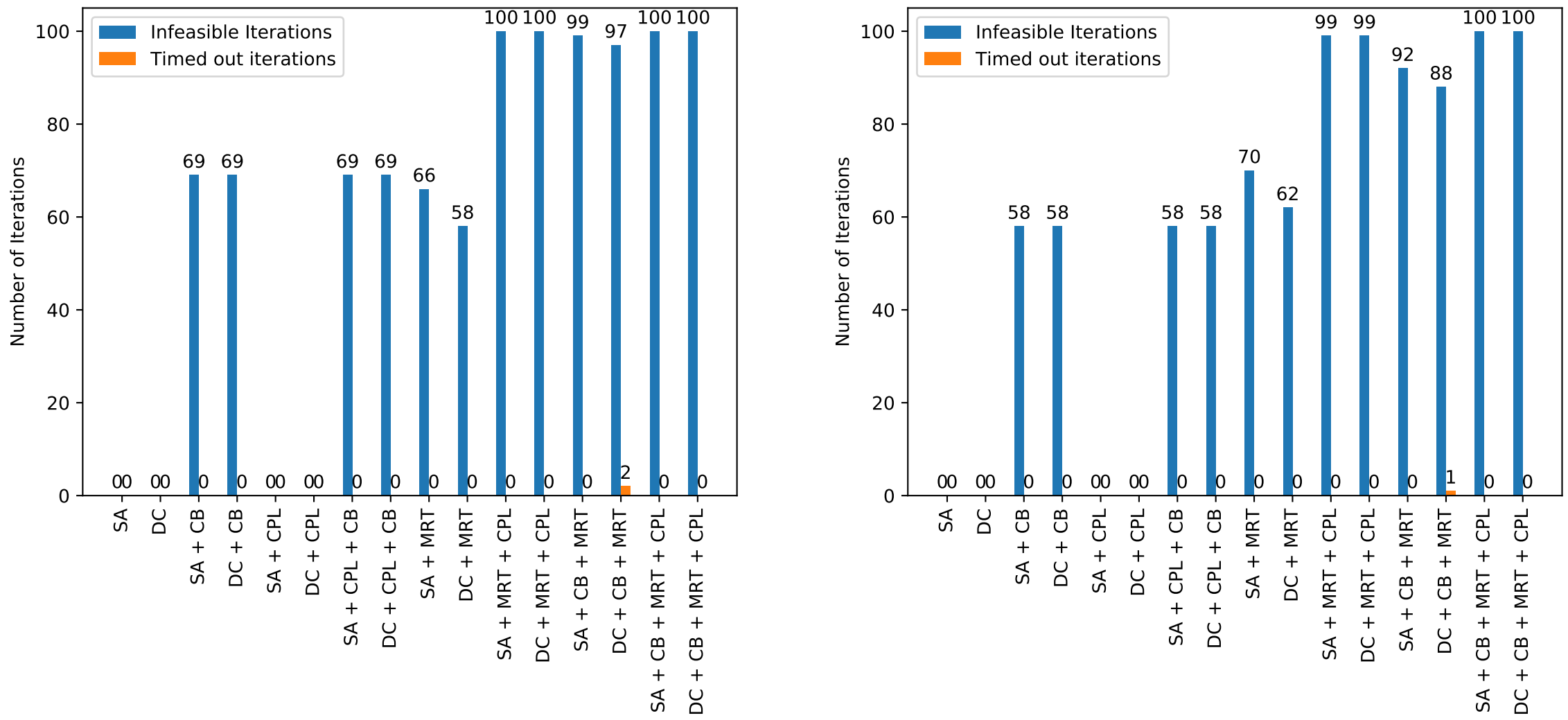

(a) CABEm without Relaxed Backhaul

(b) CABEm with Relaxed Backhaul

Fig. 24. Optimizer Status for (a) CABEm without Relaxed Backhaul, and (b) CABEm with Relaxed Backhaul scenarios with 275 eMBB users.

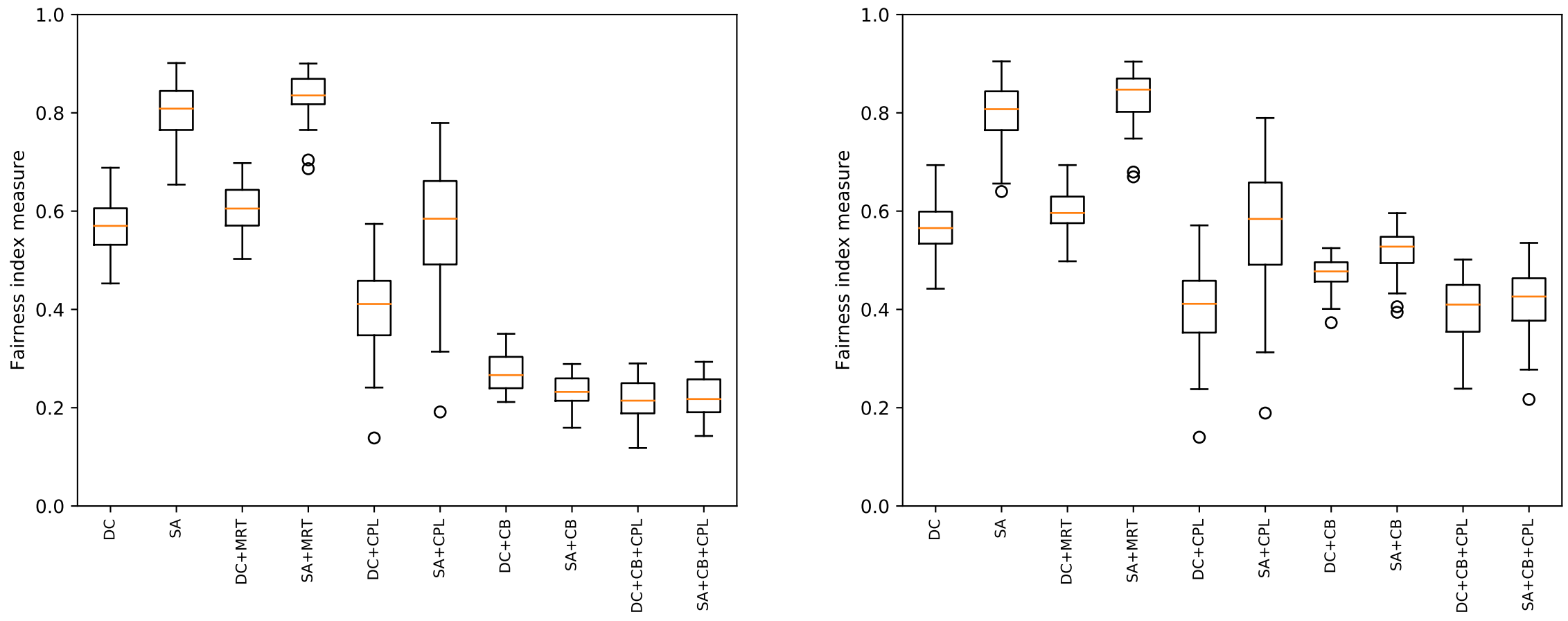

(a) CABEm without Relaxed Backhaul

(b) CABEm with Relaxed Backhaul

Fig. 25. System Fairness Measure for (a) CABEm without Relaxed Backhaul, and (b) CABEm with Relaxed Backhaul scenarios with 275 eMBB users.

network throughput plots in Figs. 20(a)-(b) and 23(a). From these results we observe that the network re-dimensioning enables approximately 35% and 40% increase in the total network throughput for the CB and CB + CPL constraint combinations, respectively.

However, as can be seen from Fig. 21(a), path latency still remains a bottleneck constraint in scenarios where MRT+CPL and CB+MRT+CPL constraint combinations are imposed. Moreover, when path latency is imposed as the only constraint our optimizer converges to a solution in each of the 100 Monte Carlo trials. Hence, in a network wherein there can be multiple constraint combinations, such as MRT+CPL and MRT+CPL+CB, the operators should be careful when they sign the service level agreements (SLAs). These SLAs should not be overly restrictive, such as the one we have here where the 3ms downlink latency cannot be guaranteed in most topologies. As a consequence, we present our observations for the case when this downlink latency requirement is relaxed to 5ms. Immediately, through Fig. 21(b) we observe that the optimizer is able to determine an optimal association in 68.3% more iterations for the CB+MRT constraint scenario, and in 11% more iterations for the CB+MRT+CPL. Further, the fairness and the total network throughput in the presence of MRT+CPL and MRT+CPL+CB constraints are also improved as seen through Figs. 22(b) and 23(b). In addition to the relaxation in the SLAs, edge clouds, through appropriate placement [53], [54], can also provision great improvements in system performance. This is so because, they bring the services closer to the users which reduces the total round trip time, and hence the downlink delay as well.

Also, from Figs. 24-29, wherein the circular deployment is considered, we observe a similar trend in results as that in Figs. 18-23. Concretely, from Figs. 24(a) and (b) we notice that the number of iterations that converge to an optimal solution for the CB and CB+CPL constraint combinations increases by 15.9%. For the system fairness, from Figs. 25(a) and (b), it can be observed that for scenarios with CB and CB + CPL, the fairness is also improved. The reason being, the improved backhaul capacity allows the AURA-5G framework to assign resources more equitably to the users in the system. Additionally, as seen from Figs. 26(a) and (b), the improvement in system throughput is nearly 71.4% and 60% for the CB and CB+CPL scenarios, respectively. Furthermore, from Figs. 27-29, it can be deduced that the increase in the average number of SCs per MC from 6 to 8, as well as having less stricter latency requirements, results in resolving to a great extent the bottleneck nature of the MRT and path latency constraint alongside increasing the system fairness and the total network throughput.

Thus, the above observations highlight an important aspect of the AURA-5G framework, wherein the operators can test their network topology and infer its performance as well as the bottlenecks. And although we randomly increased the various network settings, the operators, through the use of the AURA-5G framework, are empowered with the ability to re-dimension their networks according to the inferences they make and inline with their ultimate objectives.

## VIII. Conclusion

In this paper, we have provided a first holistic study in literature with regards to the joint optimization based user association framework for multiple applications in 5G networks. The study entails utilizing a MILP formulation based joint optimization process which takes into account the service classes and accordingly assigns the bandwidth and AP to the users in the system. Further, we organized the entire process into a software suite, wherein the location generation, system model specifics, optimization and data analytics are performed. We refer to this softwarized framework as *AURA-5G*, and present it to the academic and industrial community for analyzing the user association and resource allocation behavior in conjunction with each other.

Next, for the optimizer, we developed a MILP framework wherein the objective is to maximize the overall network throughput, and find the appropriate user-AP-bandwidth association that achieves the same. We established the various constraints that help us perform the multi-dimensional study carried out in this paper. Subsequently, we also establish certain methodologies to resolve the non-linearities introduced by multiplication of binary decision variables. This, assists in reducing the complexity of our optimization problem. Further, we also present a novel discussion on the complexity of SINR calculation and how our computation method, while being a lower bound and sub-optimal, assists in reducing the complexity of the AURA-5G framework.

In addition, as part of this study we presented the performance of the AURA-5G framework for dual connectivity (DC) and single connectivity (SA) modes. For the performance evaluation process, we utilized realistic network scenarios and parameters so as to be able to ensure the efficacy of the AURA-5G framework. Consequently, we showed that the established framework outperforms the baseline scenario in terms of the total network throughput. Further, we presented the performance characteristics of the AURA-5G framework for the DC and SA modes alongside the multiple constraint combinations in terms of system fairness, backhaul utilization, latency compliance, convergence time distribution and solvability. Note that, for DC modes we present a novel analysis for scenarios where the choice of SC does not have to be geo-restricted by the choice of MC and where the user has the opportunity to connect to any two APs, i.e. SC-SC, SC-MC or MC-MC.

We now summarize some of the important findings from our detailed analysis as follows:

1. For the *total network throughput* metric (Figs. 3, 4, 6 and 7), scenarios wherein circular deployment, beamformed, only eMBB services and *AnyDC* setup was considered, showed significant performance gains for dual connectivity as compared to single association. However, with the *MCSC* setup the gains were not as significant. Further, the path latency (CPL) and backhaul (CB) constraints severely impact the overall network throughput for all scenarios. In addition, for the scenarios with the interference limited regime, one can immediately observe a significant reduction in the overall network throughput

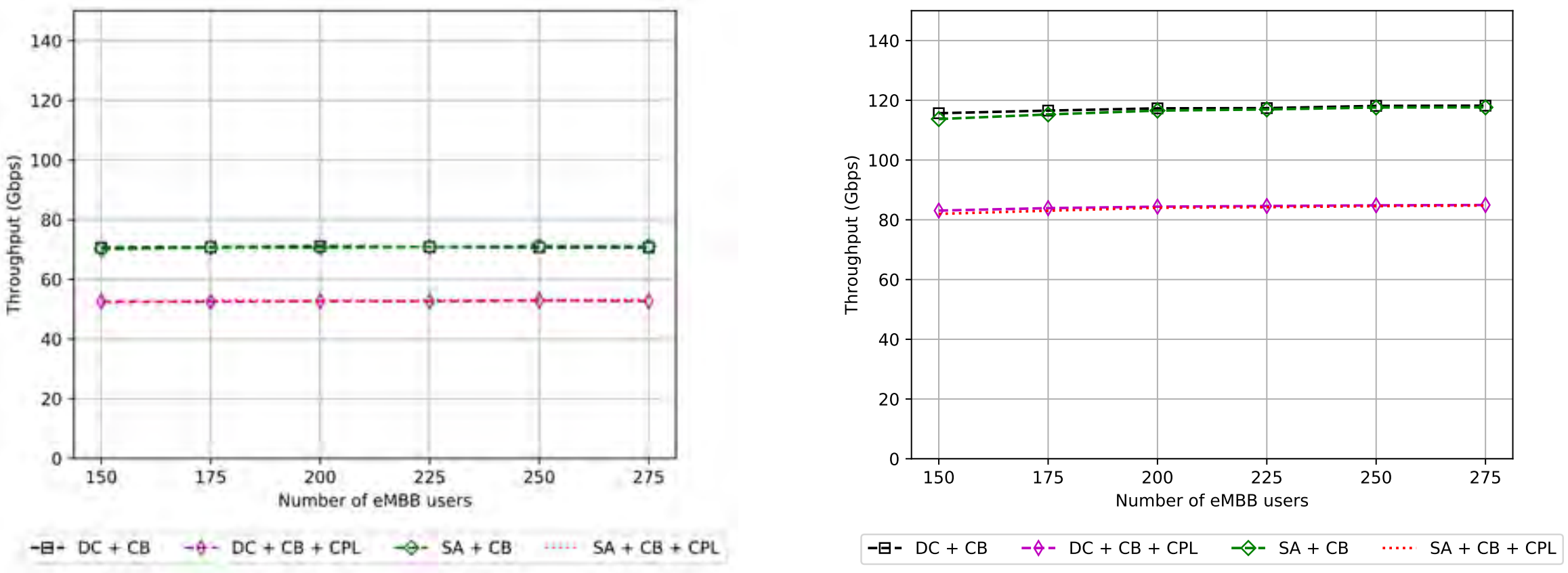

(a) CABEm without Relaxed Backhaul

(b) CABEm with Relaxed Backhaul

Fig. 26. Total Network Throughput for (a) CABEm without Relaxed Backhaul, and (b) CABEm with Relaxed Backhaul scenarios with 275 eMBB users.

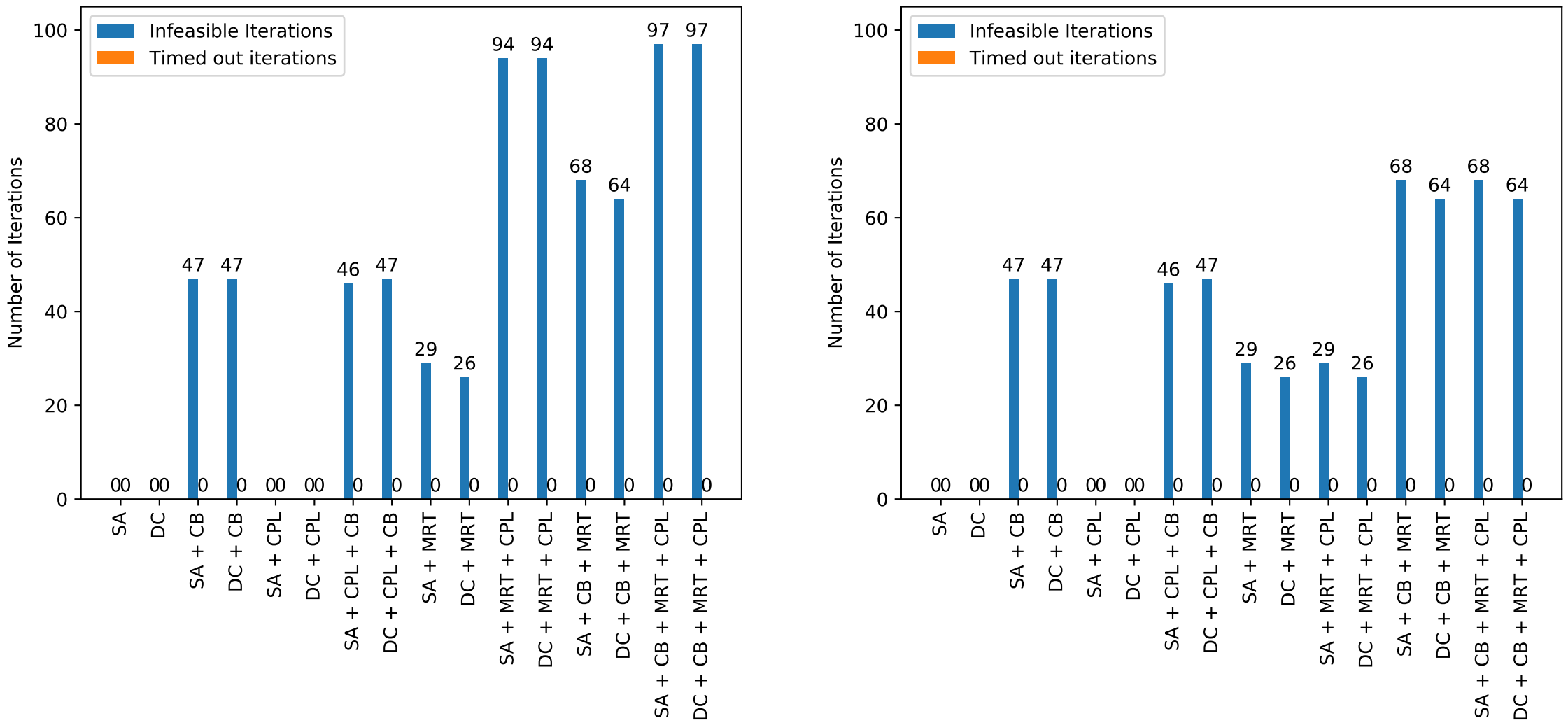

(a) CABEm with Relaxed BH and Increased SC density

(b) CABEm with Relaxed BH, Increased SC density and 5ms latency requirement

Fig. 27. Optimizer Status for (a) CABEm with Relaxed Backhaul and Increased SC density scenario with 275 eMBB users, and (b) CABEm scenario with Relaxed Backhaul, Increased SC density, 5 ms downlink latency requirement and 275 eMBB users.

due to the degradation in SINR. Moreover, for the square deployment scenarios, a gain of nearly 6% in the total network throughput is observed, as compared to the circular deployment scenarios. Finally, for the scenarios with both eMBB and mMTC services, since we consider the mMTC services to operate in the guard band and consume only the backhaul resources, a corresponding reduction in the overall throughput for the eMBB services in the presence of CB constraint was observed.

2. For the *system fairness* metric (Figs. 8-10), scenarios wherein only eMBB services, circular deployment, *AnyDC* setup and beamforming are considered, showed that single association achieved higher fairness than dual connectivity. The Minimum Rate (MRT) constraint however assisted in a slight improvement of system fairness, given the DC setup. Moreover, the CB and CPL constraints resulted in a siginificant lowering of the overall system fairness. Furthermore, for *MCSC* setup, an overall improvement in system fairness for the DC setup was observed. However, for the scenarios where an interference limited regime was considered, a significant drop in system fairness was noticed, given the SINR degradation and the greedy nature of the objective function (Section III, equation 14). Next, the square deployment scenarios showed an overall improvement of 5-6% in system fairness as compared to the circular deployment scenarios. Lastly, we analyzed the scenarios wherein both eMBB and mMTC services co-exist. For these scenarios, we observed that fairness measure is not affected significantly as compared to that noticed for only eMBB scenarios.
3. For the *backhaul utilization* metric (Figs. 11 and 12), we discern that the AURA-5G framework works exceptionally well in being able to adhere to the strict backhaul capacity constraints imposed by the network.

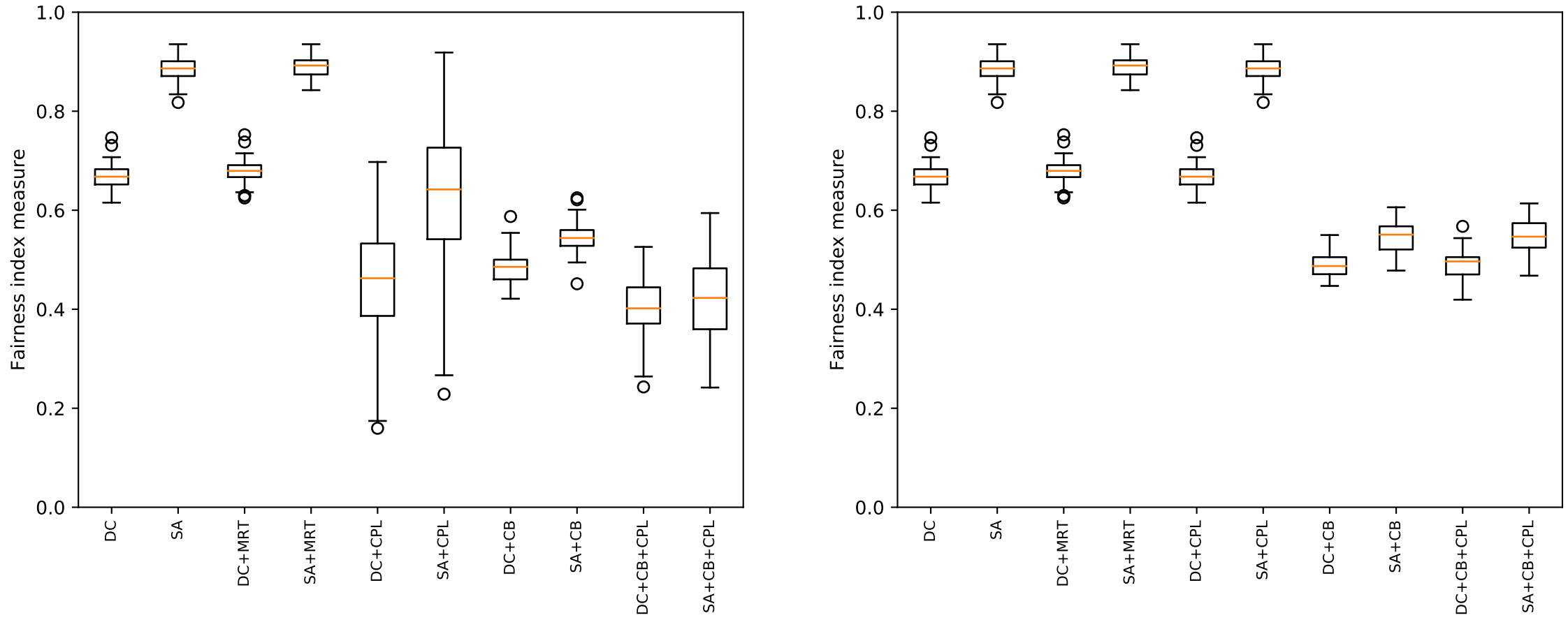

(a) CABEm with Relaxed BH and Increased SC density

(b) CABEm with Relaxed BH, Increased SC density and 5ms latency requirement

Fig. 28. System Fairness Measure for (a) CABEm with Relaxed Backhaul and Increased SC density scenario with 275 eMBB users, and (b) CABEm scenario with Relaxed Backhaul, Increased SC density, 5 ms downlink latency requirement and 275 eMBB users.

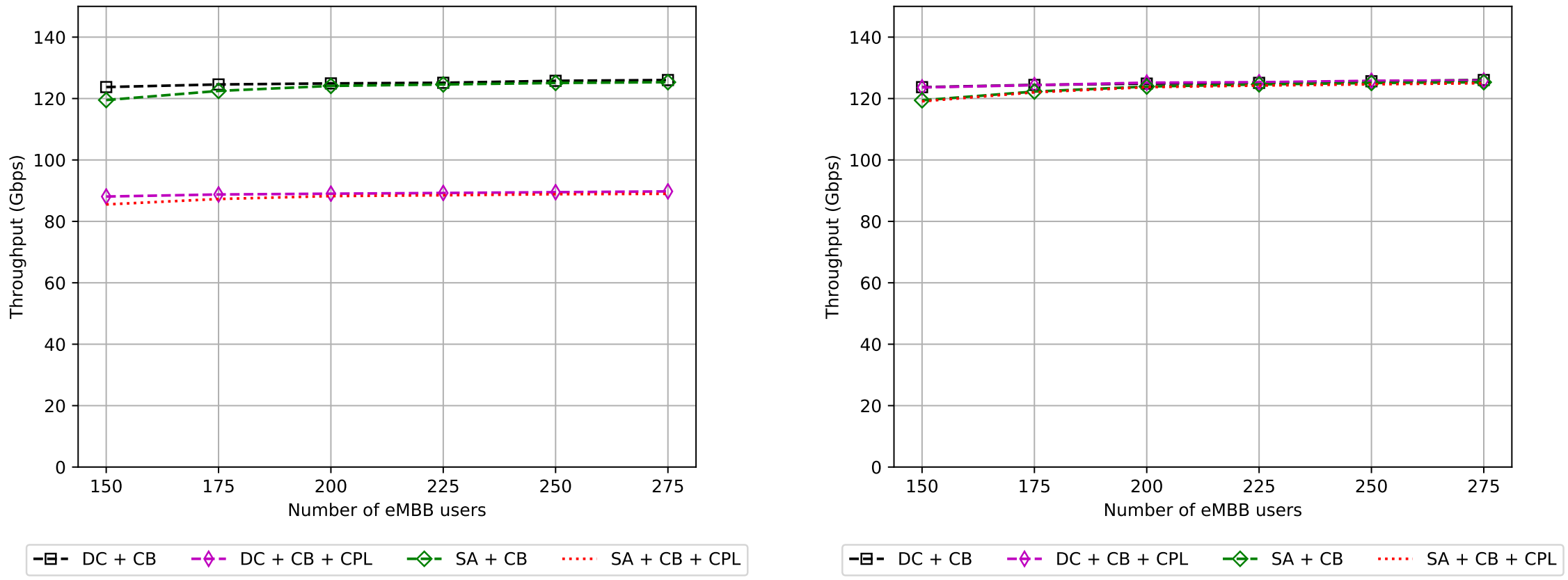

(a) CABEm with Relaxed BH and Increased SC density

(b) CABEm with Relaxed BH, Increased SC density and 5ms latency requirement

Fig. 29. Total Network Throughput for (a) CABEm with Relaxed Backhaul and Increased SC density scenario with 275 eMBB users, and (b) CABEm scenario with Relaxed Backhaul, Increased SC density, 5 ms downlink latency requirement and 275 eMBB users.

Further, for the scenarios with beamforming we observe that the backhaul capacity is almost completely utilized as compared to that in the scenarios with interference limited regime. Additionally, for scenarios wherein both eMBB and mMTC devices are considered, it was observed that the overall backhaul utilization by the eMBB devices is lower than that in scenarios where only eMBB devices exist.

4. For the *latency compliance* metric (Fig. 13), we observed again that the AURA-5G framework is able to determine user-AP-bandwidth associations such that the latency constraints are satisfied. It was observed that, while in *AnyDC* setups the users accessed SCs more than MCs, for the *MCSC* setups, a higher density of users was observed to have access to MCs and thus a reduced latency.
5. Through our novel convergence time distribution and solvability analysis, it was observed that certain constraint combinations are very restrictive and hence, the network requires re-dimensioning. It is imperative to state that such insights will be significantly important for the operators in network planning.
6. We presented, in Section VII, an analysis of certain challenging scenarios wherein the network re-dimensioning was carried out on both the access and backhaul network. We showed that, a simple re-dimensioning process, wherein the SC density was increased from 6 to 8 SCs on average per MC, the backhaul capacity was increased and less restrictive SLAs were agreed to, resulted in significant improvement in system performance, thus alleviating the bottleneck constraints concern.

Lastly, as part of our future work we will be analyzing scenarios wherein the URLLC services will be considered and serviceability of requests through edge servers will also be taken into account.